\documentclass[epj]{svjour}
\usepackage{graphics}
\usepackage{bm}% bold math
\usepackage{amsmath}
\usepackage{graphicx}
\usepackage{multirow}
\usepackage{rotating}
\begin{document}
\title{Photoproduction of $\bm{\eta}$ mesons from the neutron: 
cross sections and double polarization observable $\bm{E}$}
\author{
  L.~Witthauer\inst{1},
  M.~Dieterle\inst{1},
  F.~Afzal\inst{2}, 
  A.V.~Anisovich\inst{2,4},
  B.~Bantes\inst{3}, 
  D.~Bayadilov\inst{2,4},
  R.~Beck\inst{2}, 
  M.~Bichow\inst{5},
  K.-T.~Brinkmann\inst{2,7},
  S.~B\"ose\inst{2}, 
  Th.~Challand\inst{1},
  V.~Crede\inst{6}, 
  H.~Dutz\inst{3}, 
  H.~Eberhardt\inst{3}, 
  D.~Elsner\inst{3}, 
  R.~Ewald\inst{3}, 
  K.~Fornet-Ponse\inst{3},
  St.~Friedrich\inst{7}, 
  F.~Frommberger\inst{3}, 
  Ch.~Funke\inst{2}, 
  St.~Goertz\inst{3}, 
  M.~Gottschall\inst{2}, 
  A.~Gridnev\inst{4}, 
  M.~Gr\"uner\inst{2}, 
  E.~Gutz\inst{2,7},
  D.~Hammann\inst{3}, 
  Ch.~Hammann\inst{2}, 
  J.~Hannappel\inst{3}, 
  J.~Hartmann\inst{2}, 
  W.~Hillert\inst{3}, 
  Ph.~Hoffmeister\inst{2}, 
  Ch.~Honisch\inst{2}, 
  T.~Jude\inst{3},
  D.~Kaiser\inst{2}, 
  H.~Kalinowsky\inst{2}, 
  F.~Kalischewski\inst{2}, 
  S.~Kammer\inst{3},
  A.~K{\"a}ser\inst{1}, 
  I.~Keshelashvili\inst{1}, 
  P.~Klassen\inst{2}, 
  V.~Kleber\inst{3}, 
  F.~Klein\inst{3},
  K.~Koop\inst{2}, 
  B.~Krusche\inst{1}, 
  M.~Lang\inst{2}, 
  I.~Lopatin\inst{4}, 
  Ph.~Mahlberg\inst{2}, 
  K.~Makonyi\inst{7}, 
  V.~Metag\inst{7},
  W.~Meyer\inst{5},
  J.~M\"uller\inst{2}, 
  J.~M\"ullers\inst{2}, 
  M.~Nanova\inst{7},
  V.~Nikonov\inst{2,4}, 
  D.~Piontek\inst{2},
  G.~Reicherz\inst{5},
  T.~Rostomyan\inst{1},
  A.~Sarantsev\inst{2,4},
  Ch.~Schmidt\inst{2}, 
  H.~Schmieden\inst{3},
  T.~Seifen\inst{2}, 
  V.~Sokhoyan\inst{2}, 
  K.~Spieker\inst{2}, 
  A.~Thiel\inst{2}, 
  U.~Thoma\inst{2}, 
  M.~Urban\inst{2}, 
  H.~van Pee\inst{2}, 
  N.K.~Walford\inst{1},
  D.~Walther\inst{2}, 
  Ch.~Wendel\inst{2}, 
  D.~Werthm\"uller\inst{1},
  A.~Wilson\inst{2,6},
  and A.~Winnebeck\inst{2}
\newline(The CBELSA/TAPS collaboration)
\mail{B. Krusche, Klingelbergstrasse 82, CH-4056 Basel, Switzerland,
\email{Bernd.Krusche@unibas.ch}}
}
\institute{Department of Physics, University of Basel, CH-4056 Basel, Switzerland
  \and Helmholtz-Institut f\"ur Strahlen- und Kernphysik der Universit\"at Bonn, Germany
  \and Physikalisches Institut, Universit\"at Bonn, Germany
  \and National Research Centre "Kurchatov Institute", Petersburg Nuclear Physics Institute, Gatchina, Russia
  \and Institut f\"ur Experimentalphysik I, Ruhr-Universit\"at Bochum, Germany
  \and Department of Physics, Florida State University, Tallahassee, USA
  \and II. Physikalisches Institut, Universit\"at Giessen, Germany
}
\date{Received: date / Revised version: date}

\authorrunning{L. Witthauer et al.}
\titlerunning{Photoproduction of $\eta$ mesons from neutrons}
\abstract{Results from measurements of the photoproduction of $\eta$ mesons from quasifree protons and 
neutrons are summarized. The experiments were performed with the CBELSA/TAPS detector at the electron 
accelerator ELSA in Bonn using the $\eta\to3\pi^{0}\to6\gamma$ decay. A liquid deuterium target was used 
for the measurement of total cross sections and angular distributions. The results confirm earlier 
measurements from Bonn and the MAMI facility in Mainz about the existence of a narrow structure in 
the excitation function of $\gamma n\rightarrow n\eta$. The current angular distributions show a 
forward-backward asymmetry, which was previously not seen, but was predicted by model calculations including 
an additional narrow $P_{11}$ state. Furthermore, data obtained with a longitudinally polarized,
deuterated butanol target and a circularly polarized photon beam were analyzed to determine the double 
polarization observable $E$. Both data sets together were also used to extract the helicity dependent cross 
sections $\sigma_{1/2}$ and $\sigma_{3/2}$. The narrow structure in the excitation function of 
$\gamma n\rightarrow n\eta$ appears associated with the helicity-1/2 component of the reaction. 
}
 % end of PACS codes

\maketitle
\section{Introduction}
\label{intro}
The excitation spectrum of nucleons is one of the most important testing grounds of our understanding
of the strong interaction in the non-perturbative regime \cite{Klempt_10,Crede_13}. While in the past,
most experimental information came from hadron induced reactions such as elastic and inelastic pion
scattering (profiting from the large reaction cross sections), the last two decades have seen a huge 
world-wide effort to study the electromagnetic excitations of baryons with photon induced reactions
\cite{Krusche_03,Burkert_04} and with electron scattering \cite{Aznauryan_12,Aznauryan_13}. 

This program has two central parts. The study of many different final states should eliminate
experimental bias and the measurement of several types of polarization observables in addition to
differential cross sections should provide data sets that allow for (almost) uniquely determined
partial wave analyses. Since the electromagnetic excitation of nucleon resonances depends on
isospin, it is not sufficient to study only photon induced reactions on protons, but the neutron target 
must also be investigated. This part of the experimental program is still much less advanced than
the measurements with free proton targets. It involves experiments with quasifree neutrons
(mostly bound in the deuteron) with the difficulties from the coincident detection of
the recoil nucleons, Fermi motion of the bound nucleons, and effects from final state
interactions (FSI), complicating the interpretation of the results. However, recent progress
was encouraging (see e.g. \cite{Krusche_11,Dieterle_14}), and currently such experiments are
in the focus at the Bonn ELSA and Mainz MAMI facilities.

Photoproduction of $\eta$ mesons is a selective reaction and due to the isoscalar nature of
this meson, only $N^{\star}$ resonances can decay to the nucleon ground state via $\eta$ emission
(higher lying $\Delta$ states can emit $\eta$ mesons in decays to $\Delta(1232)\eta$ resulting in
a $N\eta\pi$ final state \cite{Kaeser_15}). The main features of the $\gamma p\rightarrow p\eta$
reaction at low incident photon energies (threshold is at $E_{\gamma}\approx707$~MeV) are well 
understood. This reaction is dominated by the excitation of the $N(1535)1/2^-$ and $N(1650)1/2^-$
states \cite{Krusche_95,Krusche_97} and a very small contribution from the $N(1520)3/2^-$ state
has been identified mainly from the $S_{11}-D_{13}$ interference term in the photon beam asymmetry 
\cite{Ajaka_98,Elsner_07,Bartalini_07}. Angular distributions for $\gamma p\rightarrow p\eta$
from threshold up to photon energies of 2.5~GeV were measured by several experiments, 
in particular at CLAS \cite{Dugger_02,Williams_09}, ELSA \cite{Crede_05,Bartholomy_07,Crede_09}, 
GRAAL \cite{Renard_02}, LNS \cite{Nakabayashi_06}, and at MAMI \cite{Krusche_95,McNicoll_10}.
Recently, also results for the transverse target asymmetry $T$ and the beam-target asymmetry 
$F$ were published from a MAMI experiment \cite{Akondi_14} and results for the double
polarization observable $E$ became available from CLAS \cite{Senderovich_16}. 
The status of $\eta$ production from nucleons and nuclei with hadron and photon induced reactions 
was recently reviewed in \cite{Krusche_15}.    
      
Photoproduction of $\eta$ mesons off neutrons is an intriguing example of why measurements with neutron 
targets are important. Measurements with deuterium and helium targets in the threshold region 
\cite{Krusche_95a,Hoffmann_97,Weiss_01,Weiss_03,Hejny_99,Hejny_02,Pfeiffer_04,Pheron_12}
determined the isospin structure of the electromagnetic excitation of the dominant $N(1535)1/2^-$   
resonance \cite{Krusche_03}. But a real surprise came from the first studies of the 
$\gamma n\rightarrow n\eta$ reaction at somewhat higher energies, which revealed an unexpected,
narrow structure in the excitation function around incident photon energies of 1~GeV 
($W\approx1.67$~GeV). This structure was first reported by the GRAAL collaboration 
\cite{Kuznetsov_07} and confirmed by the CBELSA/TAPS \cite{Jaegle_08,Jaegle_11} collaboration 
at ELSA and also at LNS in Sendai \cite{Miyahara_07}. Recent high-statistics measurements 
at MAMI using a deuterium and $^3$He target \cite{Werthmueller_13,Witthauer_13,Werthmueller_14} 
provided much more precise results for the excitation functions and the angular distributions. 
These measurements determined the position of the structure at a final state invariant mass of 
$W = (1670\pm5)$~MeV and a width of only $\Gamma~=~(30\pm15)$ MeV. The nature of this structure 
is not yet understood. Some analyses suggested a contribution from a new narrow nucleon resonance 
\cite{Polyakov_03,Arndt_04,Choi_06,Fix_07,Shrestha_12}. In its 2014 edition, the Review of Particle 
Physics (RPP) \cite{PDG_14} lists this as a tentative $N(1685)$ state with unknown quantum numbers,
but from the 2016 version \cite{PDG_16} it was removed again. 
However, other models use coupled-channel effects of known nucleon resonances \cite{Shklyar_07,Shyam_08} 
or intermediate strangeness states \cite{Doering_10}. A fit \cite{Anisovich_15} to the high statistics 
A2 deuteron data \cite{Werthmueller_13,Werthmueller_14} by the BnGa group suggests an interference 
in the $J^P = 1/2^-$ $S$ partial wave between contributions of the well-known $N(1535)$ and $N(1650)$ 
resonances \cite{Anisovich_15}. This solution requires a sign change of the electromagnetic 
coupling of the $N(1650)$ for the neutron compared to the results listed in the RPP \cite{PDG_16}. 
Alternative fits with a narrow $P_{11}$-like $N(1685)$ resonance were seen as inferior since they 
yielded a larger $\chi^2$ \cite{Anisovich_15}.

In the meantime, the situation became even more complicated. A footprint of this narrow structure 
was also found in the beam asymmetry of Compton scattering from protons \cite{Kuznetsov_15}.  
A further narrow structure at an invariant mass of $W\approx1.726$~GeV in Compton scattering 
\cite{Kuznetsov_15} was also identified in the $\gamma n\rightarrow n\eta$ reaction 
\cite{Werthmueller_15}. There are also pronounced structures in this energy range in the Legendre
expansion coefficients of the beam asymmetry for $\eta$ production off the proton \cite{Krusche_15}.
The appearance of such structures in quite different reactions makes explanations with intricate 
interference effects less probable. It is obvious that the $\gamma n\rightarrow n\eta$ reaction 
deserves a lot of attention in this energy range.

The results as a function of the incident photon energy for the total cross sections and 
angular distributions measured at ELSA in Bonn \cite{Jaegle_08,Jaegle_11} and at MAMI in Mainz 
\cite{Werthmueller_13,Werthmueller_14} agreed. However, there was some discrepancy in the absolute scale 
of the cross sections when the effective $W$ was reconstructed from the final state kinematics.
It was already discussed in \cite{Krusche_15} that there is some internal inconsistency between the results
as function of $E_{\gamma}$ and as function of $W$ (although just inside quoted systematic uncertainties) 
in \cite{Jaegle_11}. The results as function of reconstructed $W$ from the current work, measured with a
very similar setup as the results from \cite{Jaegle_11} but with a much finer binning in the angular 
distributions than in \cite{Jaegle_11} supersede the previous ones from ELSA.  

More information about the interesting structures in the $\gamma n\rightarrow n\eta$ reaction can be
obtained with the measurement of polarization observables. So far, only the beam asymmetry $\Sigma$
has been investigated with the GRAAL experiment \cite{Fantini_08}. The particularly interesting double 
polarization observable $E$ can be measured with a circularly polarized photon beam and a longitudinally 
polarized target. It is defined as:
\begin{equation}
E= \frac{\sigma_{1/2}-\sigma_{3/2}}{\sigma_{1/2}+\sigma_{3/2}} =  
\frac{\sigma_{1/2}-\sigma_{3/2}}{2\sigma_0}\, ,
\label{eq:E}
\end{equation}
where $\sigma_{1/2}$ and $\sigma_{3/2}$ are the helicity dependent cross sections with photon and
nucleon spin antiparallel or parallel, respectively, and $\sigma_0$ is the unpolarized cross section.
The helicity dependent cross sections, which can be easily computed when $E$ and $\sigma_0$ are known,
are especially interesting, since they allow direct access to the spin structure. Resonances 
with spin $J=1/2$ (like $S_{11}$ and $P_{11}$) contribute only to $\sigma_{1/2}$. States with spin 
$J\geq 3/2$ can also appear in $\sigma_{3/2}$ and very likely will. There are no known 
examples of $J\geq 3/2$ states which contribute only to $\sigma_{1/2}$, most of them have larger 
contributions to $\sigma_{3/2}$ \cite{PDG_16}. Very recently, results for the quasifree proton and 
neutron were published by the A2 collaboration at MAMI \cite{Witthauer_16}. Here, we will report results 
from a measurement of this observable at ELSA. 

\begin{figure*}[t]
\centerline{
\raisebox{-.09\height}{\resizebox{0.7\textwidth}{!}{\includegraphics{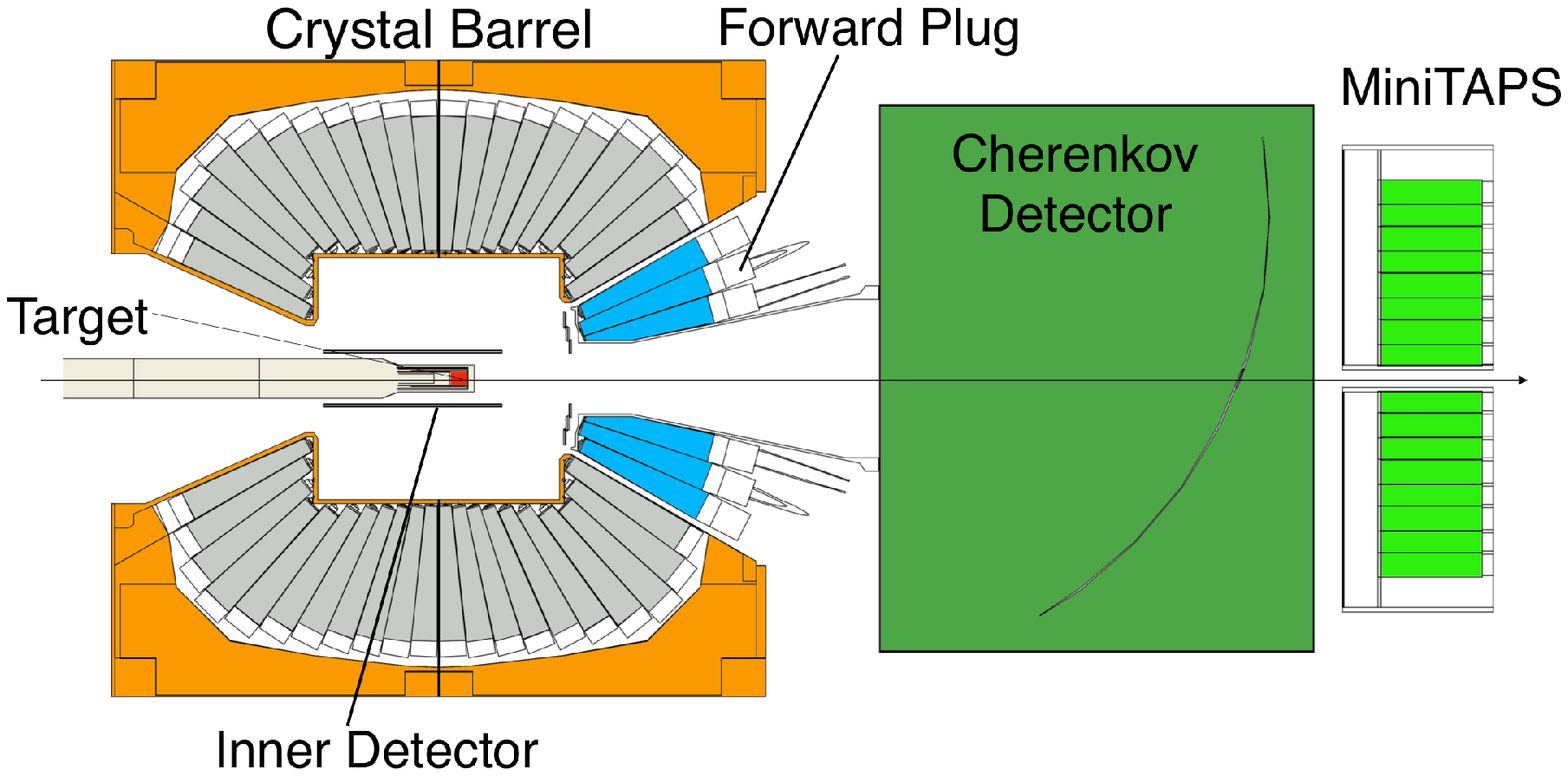}}}
\resizebox{0.3\textwidth}{!}{\includegraphics{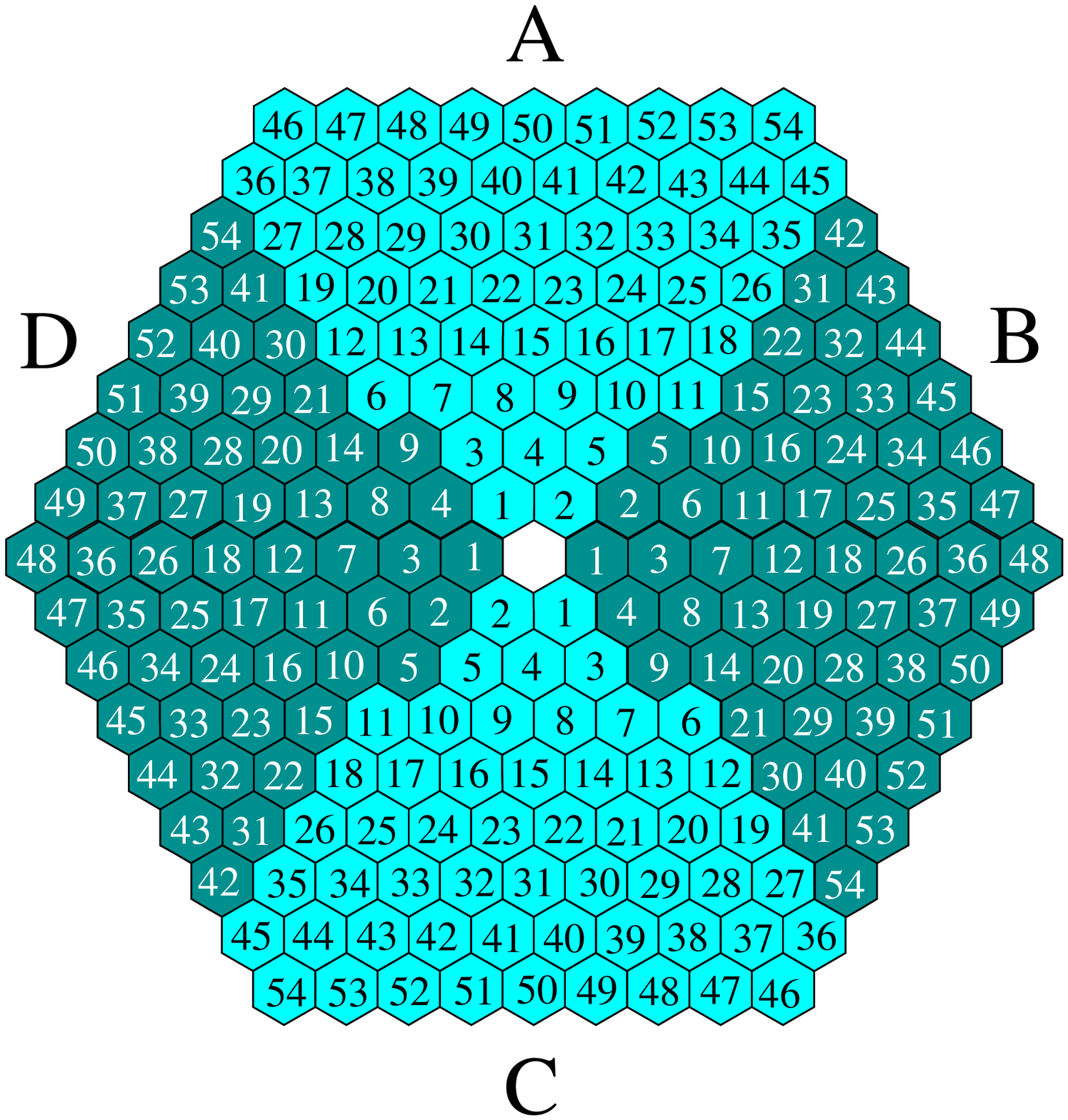}
}}
\caption{Left hand side: detector setup of the CBELSA/TAPS experiment at ELSA in Bonn with the main 
detectors Crystal Barrel and MiniTAPS and the Cerenkov detector in between. Right hand side:
front view of the MiniTAPS detector with the four logical sectors.}
\label{fig:Setup}       
\end{figure*}

\section{Experimental Setup}
\label{sec:1}

The measurements for the unpolarized cross sections and the double polarization observable $E$ were 
carried out by the CBELSA/TAPS collaboration at the electron stretcher accelerator ELSA \cite{Hillert_06} 
in Bonn, Germany. The accelerator provided an electron beam with an energy of $E_0=2.35$~GeV. 

The circularly polarized photons were generated from longitudinally polarized electrons via incoherent 
brems\-strahlung on a thin radiator foil (20 $\mu$m Vacoflux50). The radiator foil was magnetized 
and allowed for constant monitoring of the electron polarization during data taking with the help of 
M\o{}ller scattering. For the current experiments, the polarization of the electrons was in the range 
$P_e=$60-65\% and was determined with an uncertainty of 2\% \cite{Kammer_09}.

The circular photon polarization  $P_{\gamma}$ was deduced from the longitudinal electron polarization 
with the following polarization transfer equation \cite{Olsen_59}:
\begin{equation}
\frac{P_{\gamma}}{P_{e}} = \frac{3+(1-x)}{3 + 3(1-x)^2 -2(1-x)}\cdot x\, ,
\label{eq:olsen}
\end{equation}
where $x=E_{\gamma}/E_0$, and $E_{\gamma}$ is the energy of the photon. The photon energies were 
determined by the tagging spectrometer, which is mainly a dipole magnet and analyzes the momentum
of the scattered electrons. The tagger
consists of 480 fibers of 2 mm diameter and 96 scintillating bars with a thickness of 1.4-5 cm. 
This setup enabled the energy tagging of the photons in the energy range from 0.442~GeV to 2.3~GeV 
with a resolution of 2-25 MeV \cite{Elsner_09}. Typical tagger rates were 6.7~MHz for the measurement 
of the unpolarized cross sections and 10.5~MHz for the measurement of the polarization observable $E$.

The energy tagged photons impinged on a target, which was centered in the Crystal Barrel (CBB) detector, 
as shown in Fig.\ \ref{fig:Setup}. The target for the measurement of the unpolarized cross section was
similar to earlier measurements with a liquid deuterium target \cite{Jaegle_11}. The target cell was 
a capton cylinder (0.125 mm foil thickness) with a diameter of 3~cm and a length of 5.26~cm.
It was filled with LD$_2$ with a density of 0.169 g/cm$^3$ corresponding to a surface thickness
of 0.26 nuclei/barn. For the double polarization experiment, a 1.88~cm long deuterated butanol 
(C$_4$D$_9$OD) target with an effective density of 0.65~g/cm$^3$ was used. The deuterated butanol was 
polarized with Dynamic Nuclear Polarization (DNP) \cite{Bradtke_99}. The target polarization was  
50-60\% for the current work and was measured with an NMR coil every two days with an overall relative 
uncertainty of 5\%. The polarization values between the measurements were interpolated using an exponential 
function with a decay time of approximately 340 hours. Data from each single run of data taking 
(approximately 30 min) were normalized individually to the beam and target polarization degrees determined 
for this run.

The main part of the solid angle 
($30^{\circ}\leq \theta \leq 156^{\circ}$ and $0^{\circ}\leq \phi \leq 360^{\circ}$) was covered by the 
CBB calorimeter \cite{Aker_92}, which consisted of 21 azimuthally symmetric rings of 60 CsI(Tl) crystals 
each. Hence, an angular coverage of $\Delta\phi=\Delta\theta=6^{\circ}$ (for the 21st ring 
$\Delta\phi=12^{\circ}$ ) was provided and an angular resolution of $1.5^{\circ}$ was achieved by using 
the center of gravity of a shower. The energy resolution was given by $\sigma_E/E = 1.5/\sqrt[4]{E [GeV]}$ 
\cite{Aker_92}. In the CBB, charged particles were detected using the inner detector \cite{Suft_05}, which 
was made of three layers of 513 scintillating fibers. By requiring at least two intersecting hit fibers, 
an angular resolution of $\Delta\theta=0.4^{\circ}$ and $\Delta\phi=0.1^{\circ}$ was reached.

Up to an angle of $\theta=11.2^{\circ}$, the CBB was complemented with the Forward Plug (FP) detector 
made of the same CsI(Tl) crystals as the CBB (actually these are the modules of the old configuration
of the CBB \cite{Aker_92}, which had been removed from the detector when it was used together with a
larger forward wall from the TAPS detector as in \cite{Jaegle_11}). In contrast to the CBB, the FP detector 
provided time information due to the photomultiplier (PMT) readout (CBB was read out by photodiodes). 
In front of the FP crystals, two layers of plastic scintillators were mounted to identify charged particles.

The most forward region ($2^{\circ}\leq\theta\leq12^{\circ}$) of the experiment was covered by the 
MiniTAPS (MT) detector, which is a subunit of the TAPS detector \cite{Novotny_91,Gabler_94} made of 
216 BaF$_2$ crystals read out via PMTs. Each crystal had its own plastic veto to identify 
charged particles. A gas Cerenkov detector with $n=1.00043$, mounted between the CBB and the MT detectors, 
was used as an online veto to reduce electromagnetic background in the trigger.

This detector arrangement covered almost the complete solid angle for the detection of photons and 
recoil nucleons, but not for triggering on them, because the missing time information from the 
CBB did not allow to use it in the first-level trigger. This was not a problem for reactions with recoil
protons because in that case, the inner scintillating fiber detector can trigger on the protons. But for 
final states composed of photons and neutrons, only the FP and the MT could trigger on the photons.
Therefore, it was not possible to use the $\eta\rightarrow\gamma\gamma$ decay to study the 
$\gamma n\rightarrow n\eta$ reaction, but the $\eta\rightarrow 3\pi^0\rightarrow 6\gamma$ decay was
used. For the latter decay, photon hits in the FP and/or in the MT had a reasonable probability (in the 
meantime the CBB has been upgraded for timing information, so that this restriction does not any more 
apply for future measurements). In order to avoid systematic uncertainties in the comparison of reactions
off quasifree neutrons and protons, the trigger generation for $\eta$ production in coincidence with 
recoil neutrons and recoil protons was done identically, i.e. signals from the scintillating fiber 
detector were not used in the trigger decision. 

\begin{table}[!t]
\centering
\begin{tabular}{|cc|cc|}
\hline
\multicolumn{2}{|c|}{First Level Trigger} & \multicolumn{2}{|c|}{Second Level Trigger} \\
MiniTAPS & FP & \multicolumn{2}{|c|}{FACE} \\
 \hline 
1 & - & $\geq 2$& ($\geq 3$) \\
- & 1 & $\geq 2$& ($\geq 3$)\\
1 & 1 & $\geq 1$ &($\geq 2$)\\
- & $\geq 2$ & $\geq 1$ & ($\geq 2$)\\
$\geq 2$ & - & $\geq 1$& ($\geq 2$)\\
1 & $\geq 2$ & - &($\geq 1$)\\
$\geq 2$ & 1 & - & ($\geq 1$)\\
$\geq 2$ & $\geq 2$ &  - & -  \\
 \hline
\end{tabular}
\caption{Three (four) particle hardware trigger, which was used for the present work. For MiniTAPS, 
the numbers represent the number of hit sectors, and for the FP and FACE, the number is the cluster 
multiplicity. For the first level trigger, the two columns are connected with a logical `and',
for the second level trigger the first column is for the measurement with the unpolarized
target and the second column (in brackets) for the double polarization experiment.}
\label{tab:Trigger}
\end{table} 
 
The trigger was composed of two levels. Details are shown in Table~\ref{tab:Trigger}. The first level
used only signals from the detectors with timing information, i.e. FP and MT, which were available
in less than 400~ns. For that purpose, the MT was subdivided into four logical sectors, each comprising
54 BaF$_2$ modules, as shown in Fig.~\ref{fig:Setup} (right hand side). Each sector for which at least 
one module responded with a signal above the threshold of the discriminators (leading edge type, LED) was 
considered as activated. The first ring of modules closest to the beam pipe was not allowed to trigger 
because of the high count rates in these modules. For events with only one segment hit, the LED threshold 
was 100~MeV for all rings and for events with two or more segments responding it was 120~MeV for the 
second ring and 80~MeV for all other rings. The FP was equipped with a Cluster-Finder (CF) algorithm which 
combines neighboring groups of responding crystals into a cluster and delivered the number of clusters 
per event. Only modules with signals above a threshold of 30~MeV were considered. The first level 
trigger was then derived from the cluster multiplicity in the FP and the sector multiplicity in the MT as 
summarized in Table~\ref{tab:Trigger}. One should note, that the LED thresholds were only relevant
for the generation of the trigger, not for the readout of detector elements. When a valid trigger had been 
generated MT modules were read out when the additionally available constant fraction discriminator (CFD) 
had a signal above threshold. These thresholds were set below 17~MeV for the two inner-most rings and below
10~MeV for the others.

The second level of the trigger included the cluster multiplicity in the CBB, which was determined with
the Fast Cluster Encoder (FACE) based on cellular logic. It delivers the number of clusters in the CBB
in a time range of $\approx 6$ $\mu$s, which is not fast enough for the first level trigger. The threshold 
energy for clusters was set to 20~MeV. The required CBB multiplicities in dependence of the first level
trigger are given in the right hand side of Table~\ref{tab:Trigger}. The combined multiplicity from the FP,
MT, and CBB was required to be three for the measurements with the unpolarized target and
four for the butanol target. A total multiplicity of four also allows to investigate events from double 
$\pi^0$ production, which were measured simultaneously. A cluster multiplicity of three allows to also
include four-photon events where two photons hit the same sector in TAPS. For the measurement
with the polarized target, a stricter trigger was used in order to increase the $\eta$ count rate.
During the data analysis, an additional software trigger was imposed, which required that the hardware 
trigger condition was already satisfied by hits assigned as photons (no trigger contribution from
recoil nucleons). This was done in order to minimize systematic effects in the comparison of reactions 
off quasifree protons and off quasifree neutrons. Recoil protons are much more likely to generate 
clusters above the respective trigger thresholds than recoil neutrons and it is difficult to
simulate the trigger probability of neutrons.

\section{Data Analysis}
\label{sec:Ana}
The calibration procedures for the different detector components are discussed in much detail for a 
slightly different setup (no FP, forward region covered by a larger TAPS wall) in \cite{Jaegle_11,Gutz_14}.
The higher level analysis (reaction identification, extraction of cross sections, elimination of
effects from Fermi smearing, etc.) was done analogously to the one reported in 
\cite{Witthauer_13,Werthmueller_14} for the data measured at MAMI. All analysis steps will be briefly 
summarized in this section. 

\subsection{Event Selection}
\label{sec:Event}

\begin{figure}[ht]
\centerline{
\resizebox{\columnwidth}{!}{\includegraphics{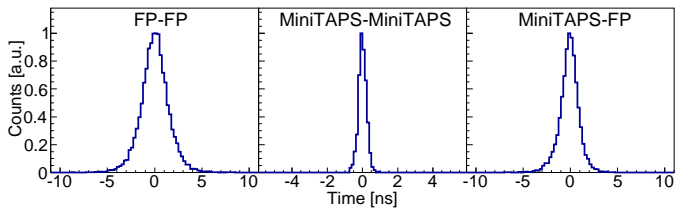}}}
\caption{Coincidence time spectra: time difference between two photon candidates in the Forward Plug 
(FP-FP), in MiniTAPS, and one photon candidate in MiniTAPS and one in the Forward Plug.}
\label{fig:Coinc}       
\end{figure}

\begin{figure}[h]
\centerline{
\resizebox{\columnwidth}{!}{\includegraphics{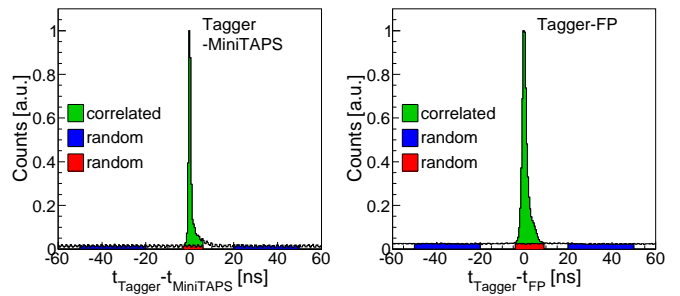}}}
\caption{Coincidence time between a hit in the tagger and the calorimeter (MiniTAPS or FP). 
The random hits (red) below the peak (green) were subtracted by a sideband (blue) analysis 
(colored areas are not actual size).}
\label{fig:Rand}  
     
\end{figure}

A coincidence analysis for the FP and MT was performed for all events and the corresponding time spectra are shown 
in Fig.\ \ref{fig:Coinc}. Non-coincident events were rejected by cuts on these time spectra. However, the 
background level of random coincidences was very low and not visible on a linear scale. The coincidence-time 
resolution for two hits in the MT detector was 0.51 ns (FWHM), whereas, due to the slow rise-time of the 
signals from the CsI crystals, the time resolution for two hits in the FP detector was only 3.28 ns (FWHM).
The tagger scintillators contributed $\approx$1~ns to the time resolution between production detector and tagger 
so that the total time resolution was $\approx$1.25~ns for coincidences between MT and tagger and $\approx$2.5~ns 
between FP and tagger (see Fig.~\ref{fig:Rand}). 

For each event, all electron hits in the tagger were registered during the coincidence window.
The corresponding time spectra for coincidences between the MT and tagger and the FP and tagger are shown in
Fig.~\ref{fig:Rand}. Random coincidences were first suppressed by a cut on the prompt peak
 and residual background below the prompt peak was removed by a sideband subtraction. 
 The colored areas shown in Fig.~\ref{fig:Rand} 
do not show the actual size of the windows. For this analysis, the prompt peak region ($\Delta p$) was chosen between 
-3 and 6 ns for MiniTAPS-Tagger and -4 and 9 ns for FP-Tagger. Events in that region had a weight of 1, 
whereas hits in the sidebands (between -400 and -100 ns or between 100 and 400 ns) had a weight of:
\begin{equation}
 w = \frac{\Delta p}{\Delta R_1 + \Delta R_2}\, ,
\end{equation}
where $\Delta R_1$ and $\Delta R_2$ were the widths of the two background windows (blue).
For the analysis with at least one hit in the MT, the timing from the MT was used. For other events, the 
timing from the FP was used. The slightly asymmetric shape of the peaks shown in 
Fig.\ \ref{fig:Rand} is caused by massive  particles misidentified as photons and disappears after the 
subsequent analysis steps of particle identification. The 2 ns modulation visible in the random background 
is due to the 500 MHz acceleration field of ELSA. The average background level in Fig.~\ref{fig:Rand} is
10\% for FP-tagger coincidences and 7\% for MT-tagger coincidences and this is further reduced by the
subsequent analysis cuts. 

\subsection{Particle Identification}
\label{sec:PartID}

In the first step of the particle identification, clusters (i.e. connected groups of activated detector 
modules) in the calorimeters were classified as `charged' or `neutral' with the help of the charge-sensitive 
detectors. In the CBB, clusters were identified as `charged' when an intersection point of at least two 
fibers in the inner detector could be reconstructed, and the angular difference between the cluster in the CBB 
and the reconstructed hit in the inner detector was less than $12^{\circ}$ in azimuthal and $30^{\circ}$ 
in polar angle (the angular cuts were motivated by the construction and properties of the inner detector 
\cite{Suft_05}). 
The energy threshold per fiber was set to approximately 150 keV. Clusters in the FP were 
regarded as `charged' when at least two overlapping vetoes in the angular range of 
$\Delta \phi \leq 14^{\circ}$ and $\Delta \theta \leq 15^{\circ}$ were hit. No energy threshold could 
be set in the analysis for the FP vetoes since the energy information was not read out. In the MT, a cluster 
was `charged' when at least one veto in front of all crystals belonging to the cluster was hit and the energy 
threshold was approximately 100 keV. Clusters with a single hit in the FP veto or a hit in only one of the
layers of the inner detector were regarded as `neutral', but rejected for the analysis of recoil neutrons.

\begin{figure*}[!t]
\centerline{\resizebox{0.99\textwidth}{!}{\includegraphics{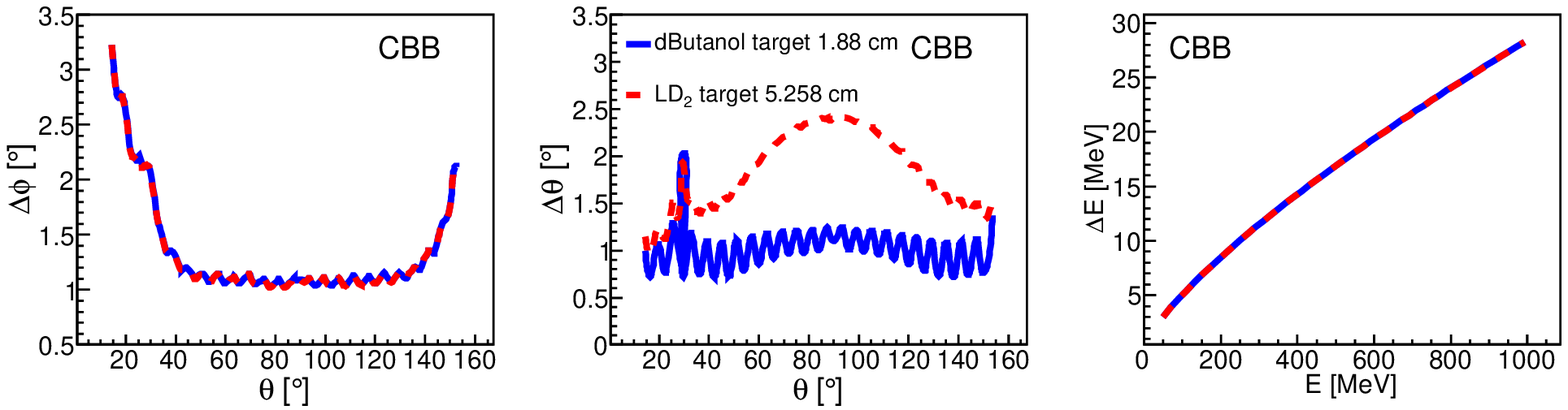}}}
\centerline{\resizebox{0.99\textwidth}{!}{\includegraphics{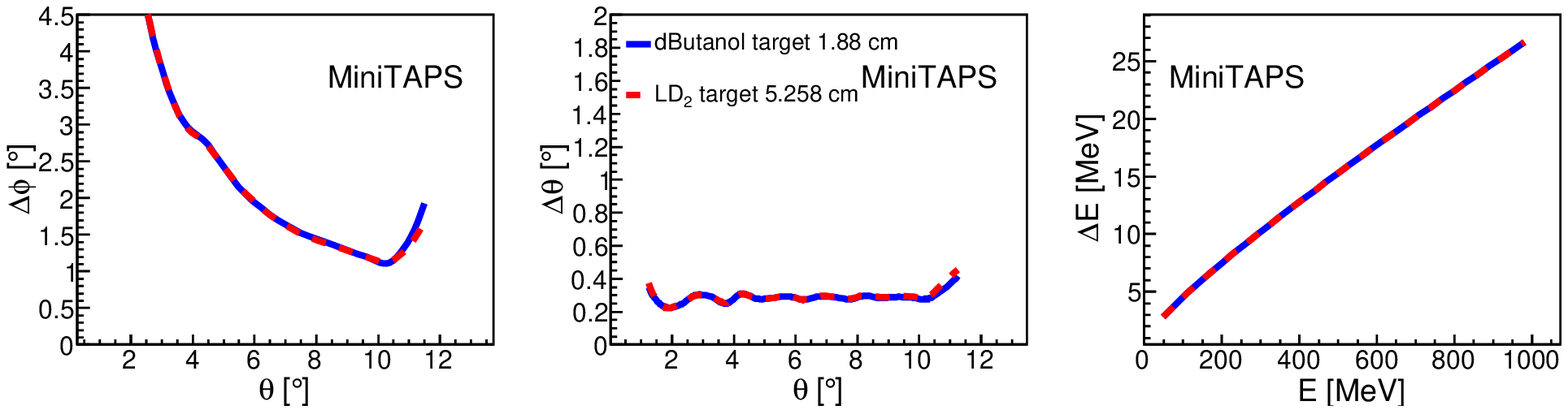}}}
\caption{Energy and angular resolutions ($\sigma$) for photons using the deuterium (red) and deuterated butanol 
(blue) target. The resolutions are shown for the CBB including the FP (top row) and the MT detector 
(bottom row).}
\label{fig:Res}       
\end{figure*}

The next step in the analysis procedure was the assignment of the events to event classes according to the 
number of charged (c) and neutral (n) hits. Three different event classes were defined since the 
$\eta\to3\pi^0\to6\gamma$ decay was detected either in coincidence with the recoil proton ($\sigma_p$), 
the recoil neutron ($\sigma_n$), or without any condition on the recoil nucleon ($\sigma_{\rm incl}$):
\begin{equation}
\label{eq:eselection}
\begin{tabular}{ll}
$\sigma_p$: & 6n and 1c \\
$\sigma_n$: & 7n  \\
$\sigma_{\rm incl}$: & 6n or 7n or (6n and 1c)\, .  \\
\end{tabular}
\end{equation}
The inclusive channel was only analyzed for the unpolarized data.

To combine the six decay photons to three $\pi^0$, a $\chi^2$ analysis was performed by comparing the 
invariant mass of two photons, $m_{k} (2 \gamma )$, to the nominal $\pi^0$ mass, $m_{\pi^0}$:
\begin{equation}
\chi^2=  \sum_{k=1}^{3} \frac{\left( m_{k} (2 \gamma )
-m_{\pi^0} \right) ^2}{\left(\Delta m_k(2 \gamma )\right)^2}\, ,
\label{eq:Chi6g}
\end{equation}
where $\Delta m_k(2 \gamma )$ is the uncertainty of the invariant mass, which depends on the uncertainty
of the deposited energy  $\Delta E$, and the uncertainties of the azimuthal and the polar angle of the 
two photons, $\Delta \phi$ and $\Delta \theta$, respectively. The latter were determined using 
Monte Carlo simulations of isotropically distributed photons and are shown in Fig. \ref{fig:Res} for 
the two different targets. It is clearly visible that the polar angle resolution for the CBB is significantly 
better for the short (1.88~cm) deuterated butanol target than for the longer ($\approx 5.3$~cm) deuterium 
target. The poor resolution around $\theta = 30^{\circ}$ was caused by the transition region between the CBB 
and the FP. For the MT, polar-angle resolution is identical for both targets (target extension in $z$-direction
does not matter) and better than for the CBB (due to the larger distance). The energy resolution is similar for
both calorimeters.

For events with six neutral hits or six neutral and one charged hit, the $\chi^2$ analysis was only
used to combine the six neutral hits to the three most probable pairs corresponding to the 
intermediate $\pi^0$ mesons. The event was accepted when the combination with the best $\chi^2$
has invariant masses of all three pairs of photons between 115 - 156~MeV (the mass of the $\pi^0$ is 135~MeV).
The same selection was used for events with seven neutral hits and in this case, the neutral hit, which was 
not assigned to a $\pi^0$ decay photon, was treated as a neutron candidate. 
The $\chi^{2}$ values were only calculated to find the best combination, no cuts were applied based on the 
$\chi^{2}$ values.

\begin{figure}[b]
\centerline{
\resizebox{1.0\columnwidth}{!}{\includegraphics{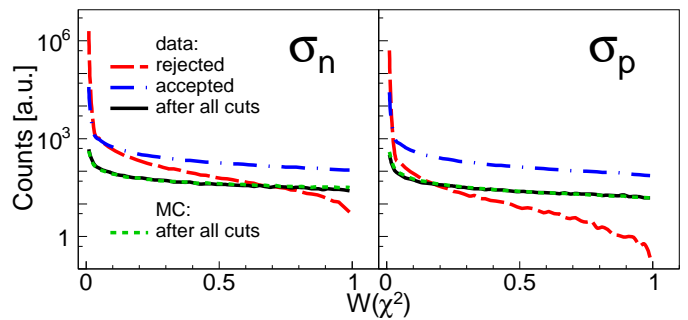}}}
\caption{Confidence level from the $\chi^2$ test for the event classes $\sigma_n$ (left) and 
$\sigma_p$ (right). The distributions for the accepted (blue dash-dotted) and rejected (red long dashed) 
events are shown together with the histograms for the final events selection (black solid) for 
experimental data and MC events (green short dashed).}
\label{fig:Conf}       
\end{figure}

The quality of the reconstruction is reflected in the resulting confidence level distributions, which 
are shown in Fig.\ \ref{fig:Conf}. The results for the liquid deuterium target (results from the butanol 
target are very similar) are shown separately for events in coincidence with neutrons and protons.
The confidence levels for the final event selection after all cuts are essentially flat and in good agreement 
with the Monte Carlo simulations, which indicates that the resolutions for the $\chi^{2}$ test were determined 
realistically. 
    
\begin{figure}[h]
\centerline{
\resizebox{0.8\columnwidth}{!}{\includegraphics{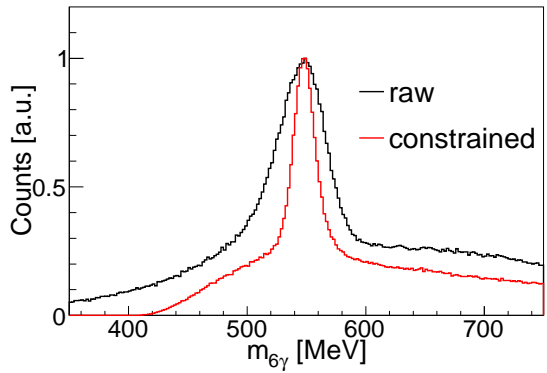}}}
\caption{Raw invariant mass $m{6\gamma}$ (black) and invariant mass after constraining (red) as shown in 
Eq. \ref{eq:const}.}
\label{fig:constr}       
\end{figure}

Subsequent to the assignment of the photon pairs to pion decays, the mass of the $\pi^0$ meson
was used to improve the experimental resolution. This was done by replacing the measured energies 
$E_{\gamma_1,\gamma_2}$ by $E'_{\gamma_1,\gamma_2}$ with:
\begin{equation}
E'_{\gamma_1,\gamma_2} = \frac{m_{\pi^0}}{m_{\gamma_1\gamma_2}}E_{\gamma_1,\gamma_2},
\label{eq:const}
\end{equation}
where $m_{\pi^0}$ is the nominal mass of the pion and $m_{\gamma_1\gamma_2}$ is the invariant mass of
the photon pair. Since the relative angular resolution for photons is better than the energy resolution 
(see Fig.~\ref{fig:Res}), the angular uncertainties were neglected. This reconstruction improved the resolution 
for the subsequent invariant and missing mass analyses of the $\eta$ mesons significantly 
(see Fig.\ \ref{fig:constr}). This improvement is in particular important for the rejection of background
in the missing mass analysis.

Particle identification up to this stage used only the information from the charged particle detectors 
and the $\chi^2$ analysis. Additional information was available for the MT detector from a pulse-shape 
analysis (PSA), time-of-flight (ToF) versus energy, and $E-\Delta E$ analyses which helped to distinguish
between photons, neutrons, and protons in the angular range where most recoil nucleons were detected.
These analyses are discussed in Sec.~\ref{sec:Checks}, because most of them only confirmed the correct 
assignment of particle types and were not used for additional cuts.
   
\subsection{Reaction Identification}
\label{sec:ReacID}
Quasifree $\eta$ production was identified by subsequent cuts on the invariant mass of the
six decay photons and on the reaction kinematics. For the latter, the coplanarity of the $\eta$ - nucleon
system and the missing mass of the reaction was investigated. 

The coplanarity analysis was only possible for events for which the recoil nucleon was detected
(not for the inclusive analysis). The coplanarity angle is the azimuthal angle difference between the 
recoil nucleon, $\phi_{N}$, and the $\eta$ meson, $\phi_{\eta}$. It was 
calculated with the following equation:
\begin{equation}
\Delta \phi = \begin{cases} \phi_{\eta}-\phi_{N}, 
& \mbox{if } \phi_{\eta}-\phi_{N} \geq 0 \\ 2\pi - |\phi_{\eta}-\phi_{N}|, & \mbox{if }  
\phi_{\eta}-\phi_{N} < 0\, . \end{cases}
\label{eq:Cop}
\end{equation}
\begin{figure}[b!]
\centerline{
\resizebox{0.92\columnwidth}{!}{\includegraphics{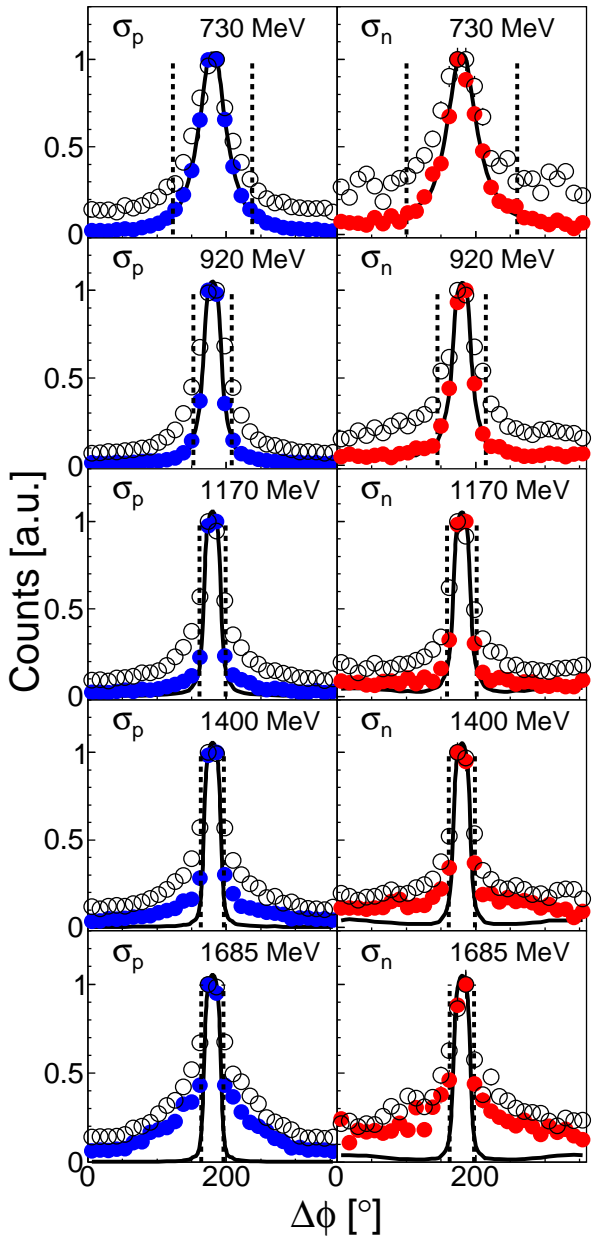}}}
\caption{Azimuthal angle difference between the $\eta$ meson and the nucleon for the $\sigma_p$ and 
$\sigma_n$ event classes for five different bins of incident photon energy (energy indicated in the 
figure). Shown are the distributions for the deuterium target (colored solid symbols) and for the 
deuterated butanol target (black open) symbols. The spectra were normalized in the peak maximum. 
The line shape from the simulation with the deuterium target is shown as a black solid line and the 
cut positions ($\pm 2 \sigma$) are indicated by the dashed vertical lines.}
\label{fig:Cop}       
\end{figure}

\begin{figure}[b!]
\centerline{
\resizebox{\columnwidth}{!}{\includegraphics{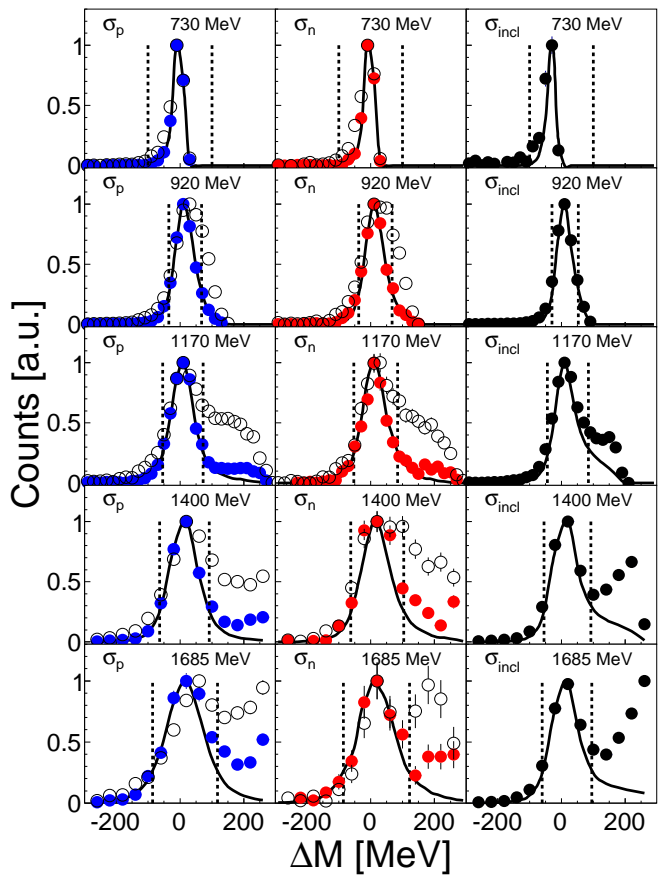}}}
\caption{Missing mass of the recoil nucleon $\Delta M$ for the event classes $\sigma_p$, $\sigma_n$ and 
$\sigma_{\rm incl}$ for different incident photon energies (indicated in the figure). Same labeling 
as in Fig. \ref{fig:Cop}.}
\label{fig:MM}       
\end{figure}

Due to four-momentum conservation, the angular difference has to be equal to $180^{\circ}$. Since 
the $\phi$ angle is independent of a boost in the $z$-direction, this is valid in the center of mass 
(cm) frame, as well as in the laboratory frame. However, for quasifree nucleons, the Fermi motion 
causes a smearing of the angle and hence, a broadening of the peak. The corresponding spectra filled 
right after the $\chi^2$ selection and the PSA cut are shown in Fig.\ \ref{fig:Cop} for the deuterium 
target (colored symbols) and the deuterated butanol target (black symbols). The peaks for the data with 
the deuterated butanol target are broader than for the deuterium target due to the larger Fermi 
motion in the carbon nuclei. Since the kinematics has a strong influence on the width of the peak, 
the cut position of $\pm 2 \sigma$ with respect to the mean position was determined for different bins 
of incident photon energy and $\cos{(\theta_{\eta}^{\ast})}$ ($\theta_{\eta}^{\ast}$ is the polar 
angle of the $\eta$ meson in the beam-target cm system assuming the initial state nucleon at rest). 
The same cuts were applied to the data taken with the deuterium and the deuterated butanol target.  

Within the selected cuts, the experimental data from the deuterium target and the simulated spectra 
(solid line) are in agreement. The deviations at higher energies are mainly caused by  $\eta\pi$ 
background. Problematic are in particular the $n\pi^+\eta$ and $p\pi^-\eta$ final states when the single 
charged pion escapes detection. Most of this background was later removed by the missing mass cut, 
which was not applied to these spectra.

\begin{figure}[b]
\centerline{
\resizebox{\columnwidth}{!}{\includegraphics{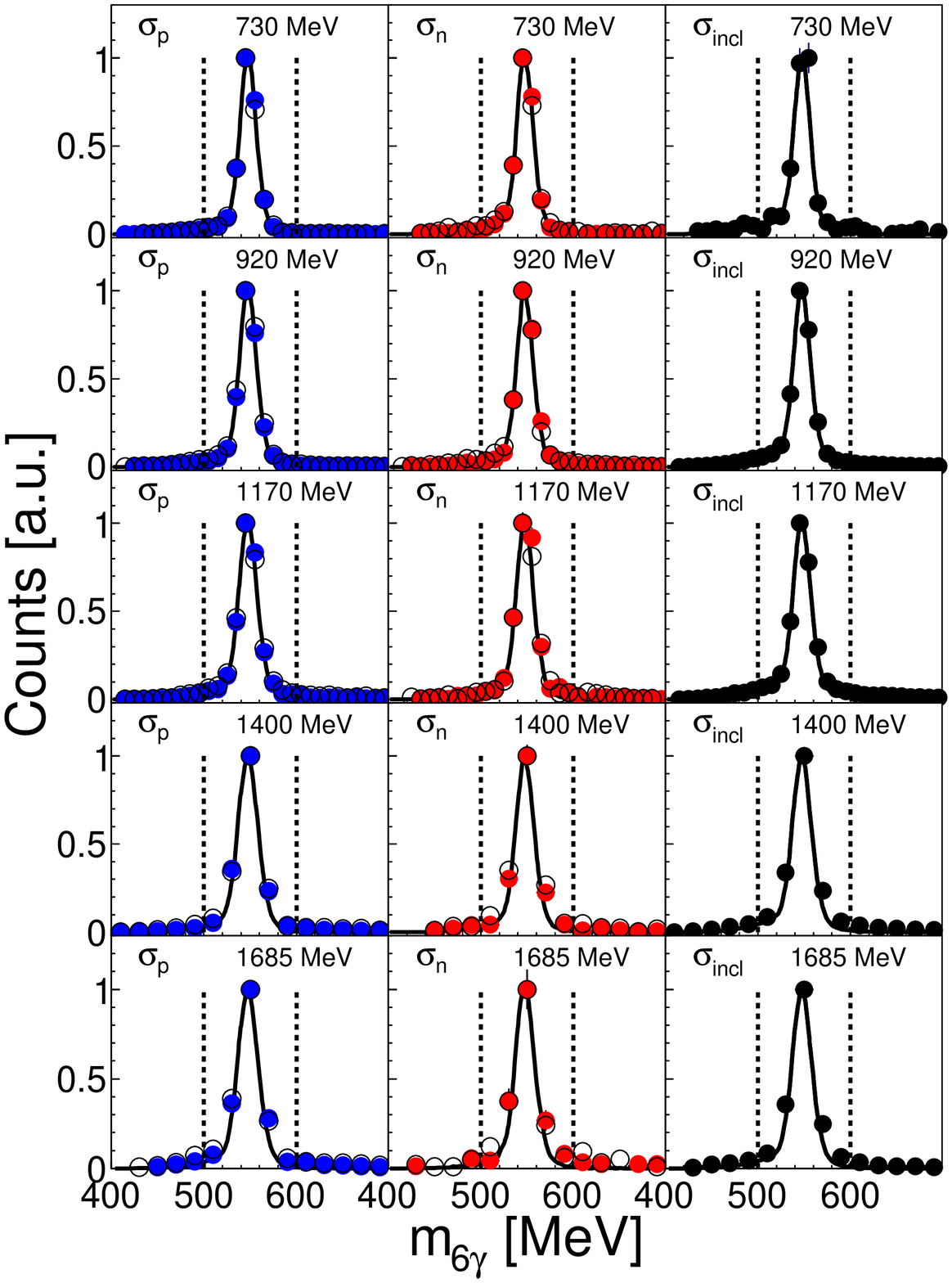}}}
\caption{Invariant mass for the event classes $\sigma_p$, $\sigma_n$, and $\sigma_{\rm incl}$ for different 
incident photon energies (indicated in the figure). Same labeling as in Fig. \ref{fig:Cop}.}
\label{fig:IM}       
\end{figure}

For further identification of the reaction, the mass $M$ of the recoil nucleon (treated as missing 
particle also when detected) was calculated from:
\begin{equation}
M = \sqrt{\left(E_{\gamma}+ m_N -E_{\eta}\right)^2-\left(\vec{p}_\gamma - \vec{p}_\eta \right)^2}\, ,
\label{eq:MM}
\end{equation}
where $E_{\gamma}$ and $\vec{p}_{\gamma}$ are the energy and the momentum of the incident photon beam, 
respectively, $E_{\eta}$ and $\vec{p_{\eta}}$ are the energy and the momentum of the $\eta$ meson, 
respectively, and $m_N$ is the nucleon mass. Subtracting the nominal nucleon mass from Eq.\ \ref{eq:MM}, 
i.e. $\Delta M = M-m_{N}$, should result in a peak at zero, which is visible in Fig.\ \ref{fig:MM} for 
different incident photon energies. Shown are the missing mass spectra for the deuterium target 
(red and blue symbols), the corresponding simulated line shape (black line), and spectra for the 
deuterated butanol target (black circles).

In Eq.\ \ref{eq:MM}, it is assumed that the initial state nucleon is at rest, thus, Fermi motion has a 
large influence on the missing mass spectra. The Fermi motion not only induces a broadening of the peak, 
but also causes a slightly asymmetric shape of the peak at low incident photon energies. This is caused 
by the fact that close to threshold, Fermi momenta antiparallel to the photon beam are favored since they 
correspond to larger cm energies. 

Above $E_{\gamma}\simeq 800$ MeV, the background on the right hand side of the peak was mainly caused by
$\gamma p \to \eta \pi^{+}n$ and $\gamma n \to \eta \pi^{-}p$ reactions, when the charged 
pion was not registered. Also, the $\eta\pi^0$ final state contributed to the background when both
$\pi^0$ decay photons were not identified, which, however, is rare. This background was rejected by applying 
an energy and $\cos{(\theta_{\eta}^{\ast})}$ dependent missing mass cut of $\pm1.5\sigma$ with respect 
to the mean position of the peak (indicated by the vertical lines in the figure). 
For incident photon energies below 800 MeV, the events were accepted in the range between $-100$ MeV 
and $+100$ MeV. Here, one should note that in this missing-mass range no significant background is visible
for the liquid deuterium target, the simulated line shapes agree with the experimental data. Additional
background in this range for the butanol target is due to reactions with nucleons bound in the carbon (oxygen)
nuclei, but this is later on subtracted (see Fig.~\ref{fig:MMFit}).  

Finally, the invariant mass of the $\eta$ meson was reconstructed from the six decay photons:
\begin{equation}
m_{6\gamma} = \sqrt{\left(\sum\limits_{i=1}^{6} E_{\gamma_i}\right)^2 -\left(\sum\limits_{i=1}^{6} 
\vec{p}_{\gamma_i}\right)^2 }\, ,
\end{equation}
where $E_{\gamma_i}$ is the energy of the $i$-th photon and $\vec{p}_{\gamma_i}$ is the corresponding 
momentum. 

Typical spectra are shown in Fig.\ \ref{fig:IM} for the different event classes. All measured data are 
in good agreement with the simulated line shapes and almost free of background. Since the width and position 
of the invariant mass peak is essentially independent of the incident photon energy, the spectra were  
integrated in the invariant mass range between 500 and 600 MeV, as indicated by the vertical lines in 
the figure.

The spectra shown in Figs.\ \ref{fig:Cop} , \ref{fig:MM}, and \ref{fig:IM} are integrated over the whole 
angular range, however, the actual analysis was done separately for each energy and 
$\cos{(\theta_{\eta}^{\ast})}$ bin.

\subsection{Additional Checks}
\label{sec:Checks}
Additional detector information was used to check the particle and reaction identification, however, 
only very soft or no cuts were applied to these spectra.

For the forward angular range covered by the MT wall, where a large fraction of the recoil nucleons was 
detected, further methods of particle identification were exploited. A clean separation of neutrons from photons 
was achieved in the MT by a Pulse Shape Analysis (PSA). Crystals of BaF$_2$ have two different scintillation 
light components \cite{Novotny_91,Gabler_94}, one with a slow decay time ($\tau=650$ ns) and another with a fast 
decay time ($\tau=0.9$ ns). The relative intensity of these two light components is different for photons and 
nucleons due to their different interaction mechanisms in matter (the fast component is more intense for photons). 
In the experiment, the signals from the BaF$_2$ crystals were integrated over a short (40 ns) and a long (2 $\mu$s) 
gate. For the analysis, it is convenient to define a PSA radius $r_{\rm PSA}$ and a PSA angle 
$\phi_{\rm PSA}$ using:
\begin{align}
r_{\rm PSA} &= \sqrt{E_l^2 + E_s^2} \label{eq:PSA1}\\
\phi_{\rm PSA}&=\arctan{\frac{E_s}{E_l}} \label{eq:PSA2}\, .
\end{align} 

\begin{figure}[h]
\centerline{
\resizebox{\columnwidth}{!}{\includegraphics{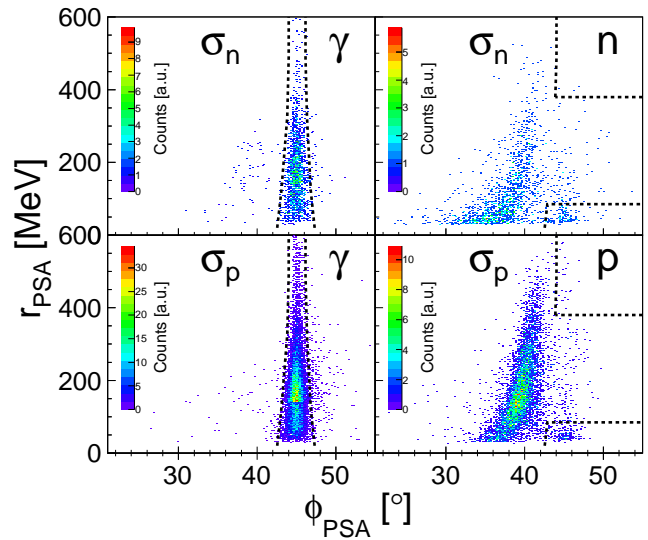}}}
\caption{Pulse shape analysis plots for the event classes $\sigma_n$ (top row) and $\sigma_p$ (bottom row). 
Shown are the PSA spectra for photon candidates (left hand side) and nucleon candidates (right hand side) 
after all analysis cuts were applied. The cut positions are indicated by the dashed lines. Photons are required 
to lie in a band of $\pm3\sigma$ with respect to the mean position of the photon band. Nucleon candidates 
with $R_{\rm PSA} <85$ MeV or $R_{\rm PSA} >380$ MeV are cut away when lying closer than the $3\sigma$ line to the 
photon band.}
\label{fig:PSA}       
\end{figure}

Plotting the PSA angle versus the PSA radius results in the spectra shown in Fig.\ \ref{fig:PSA}. 
The signals were calibrated such that the short-gate ($E_s$) and long-gate ($E_l$) responses were 
the same for photons, i.e. in a representation short-versus-long-gate signal photons appear on the 45$^{\circ}$ 
line. In contrast to the photons, the nucleons deposit less of their energy in the short gate and thus, are 
located at lower angles. 

Having applied all other analysis cuts (i.e. $\chi^2$ analysis, coplanarity, missing mass, and invariant 
mass cut) the spectra were almost free of background and soft cuts (dashed lines) were chosen to eliminate 
the small residual background. Photons were required to lie in a band of $\pm3\sigma$ 
with respect to the mean position of the photon band. Nucleon candidates with a PSA radius smaller than 
85 MeV or larger than 380 MeV were cut away when lying closer than the $3\sigma$ line to the photon band. 
The background in the nucleon spectra at low PSA radii was mainly caused by electrons that did not activate 
the veto detectors. The cut was not applied for PSA radii between 85 MeV and 380 MeV because in this region, 
the nucleon band (punch-through nucleons not stopped in the MT) slightly overlaps with the photon band and 
real nucleons could get lost.

\begin{figure}[h]
\centerline{
\resizebox{\columnwidth}{!}{\includegraphics{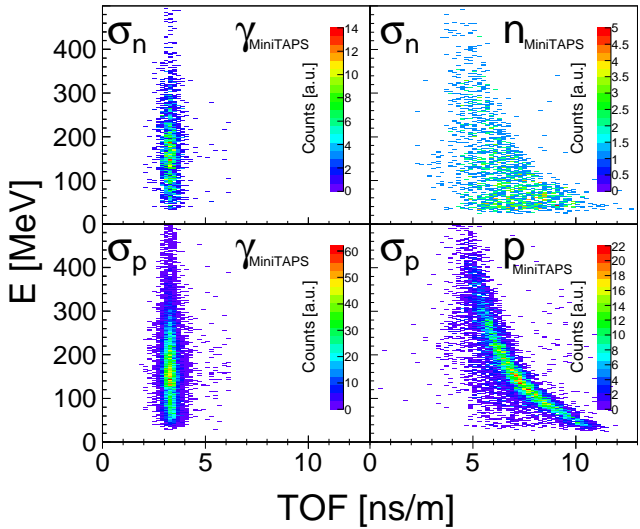}}}
\caption{TOF versus deposited energy for photons (first column), neutrons (top right), and protons (left bottom) 
in MiniTAPS for the event classes $\sigma_n$ (top row) and $\sigma_p$ (bottom row). }
\label{fig:TOF}       
\end{figure}

Since the MT provided a good time resolution, the time of flight (TOF) versus energy spectra allowed 
a redundant identification of photons (see Fig.~\ref{fig:TOF}, left hand side) which appear at a band of
approximately 3.3~ns/m independent of energy. These spectra also distinguish between recoil protons 
and neutrons. Most protons deposit their total energy in the MT and thus appear in a band corresponding to
the kinematic TOF versus energy relation for recoil nucleons, while neutrons deposit a 
random amount of energy in the MT. The important result is that in the TOF-versus-energy spectra
for neutron candidates, no residual structure of the proton band appears, which would indicate 
misidentification of protons as neutrons due to inefficient charge particle counters.    

\begin{figure}[h]
\centerline{
\resizebox{\columnwidth}{!}{\includegraphics{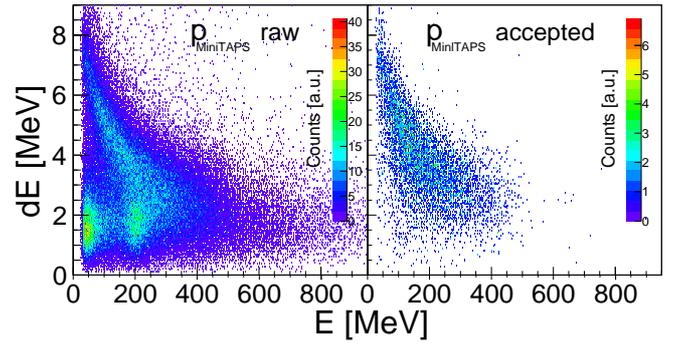}}}
\caption{$\Delta$E versus E for proton candidates in MiniTAPS right after the event selection 
(left hand side) and after all subsequent analysis steps (right hand side).}
\label{fig:dEE}       
\end{figure}

Protons could be further identified by plotting the energy $\Delta$E deposited in the MT vetoes 
versus the energy $E$ deposited in the corresponding crystals. The $\Delta$E versus E plot is shown in
Fig.\ \ref{fig:dEE} for proton candidates. 
%The spectrum on the left hand side shows the result after
%primary selection of the events (Eq.~\ref{eq:eselection}) and the spectrum at the right hand side show the
%result after all further cuts. 
The energy resolution for these spectra was not good because the
veto detectors were read out by long wave-length shifting fibers, which resulted in low photon statistics.
However, they demonstrate that the background structures from electrons and charged pions 
(minimum ionizing peak at $E=200$ MeV and $\Delta E=1.5$ MeV in Fig.\ \ref{fig:dEE}) were 
eliminated by the other analysis cuts. 

\begin{figure}[t]
\centerline{
\resizebox{\columnwidth}{!}{\includegraphics{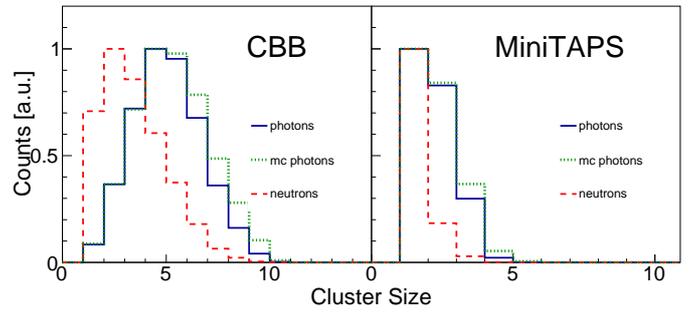}}}
\caption{Cluster multiplicity of photon and neutron candidates in the CBB (including FP) and MT. 
The distribution for photons and neutrons clearly differs in the CBB. For the photons, distributions from 
experimental data and simulation (mc) are compared.}
\label{fig:ClustMult}       
\end{figure}

In the CBB (including the FP), neutrons and photons could not be directly separated. However, the 
identification by the $\chi^2$ analysis and further steps of reaction identification
was cross-checked by the comparison of the cluster multiplicity, i.e. the number of 
crystals per cluster of neutron and photon candidates. The corresponding distributions are shown in 
Fig.\ \ref{fig:ClustMult} for neutron (red) and photon (blue) candidates in the CBB and the MT. 
As expected, the distribution for neutrons in the CBB peaks at lower values than for photons, 
indicating a good separation of photons from neutrons for hits with cluster multiplicity one. 
However, for all other cluster multiplicities, the distributions are not well enough
separated for an analysis on an event-by-event basis. The two distributions for the MT detector are more 
similar; however, in this case, photons and neutrons can be additionally distinguished by PSA. 

\subsection{Kinematic Reconstruction of the Final State}
\label{sec:KinRec}

Cross sections were extracted as a function of the incident photon energy $E_{\gamma}$ and bins of  
$\cos{(\theta_{\eta}^{\ast})}$ in the beam-target cm system assuming the initial state nucleon at rest.
For the two exclusive channels, where the recoil nucleon was detected, 
cross sections were also determined as a function of the final state invariant mass $W$ in the $\eta$-nucleon 
cm system. The main advantage when using the final state invariant mass is that structures are not 
broadened due to Fermi motion. However, the measurement as a function of $W$ requires the full 
kinematic reconstruction of the reaction, i.e. the extraction of the four-momenta of the final state 
nucleon and the $\eta$ meson. As discussed in \cite{Jaegle_11,Witthauer_13,Werthmueller_14}, this 
reconstruction is possible when the four momentum of the $\eta$-meson and the polar and azimuthal 
angles (direction of momentum) of the recoil nucleon are measured. The kinetic energy of the recoil 
nucleon is then determined by energy and momentum conservation. For nucleons in the MT, the kinetic 
energy can also be determined using a TOF reconstruction, however, this results in a poor $W$ 
resolution \cite{Witthauer_13,Werthmueller_14}. 

\begin{figure}[t]
\centerline{
\resizebox{0.5\columnwidth}{!}{\includegraphics{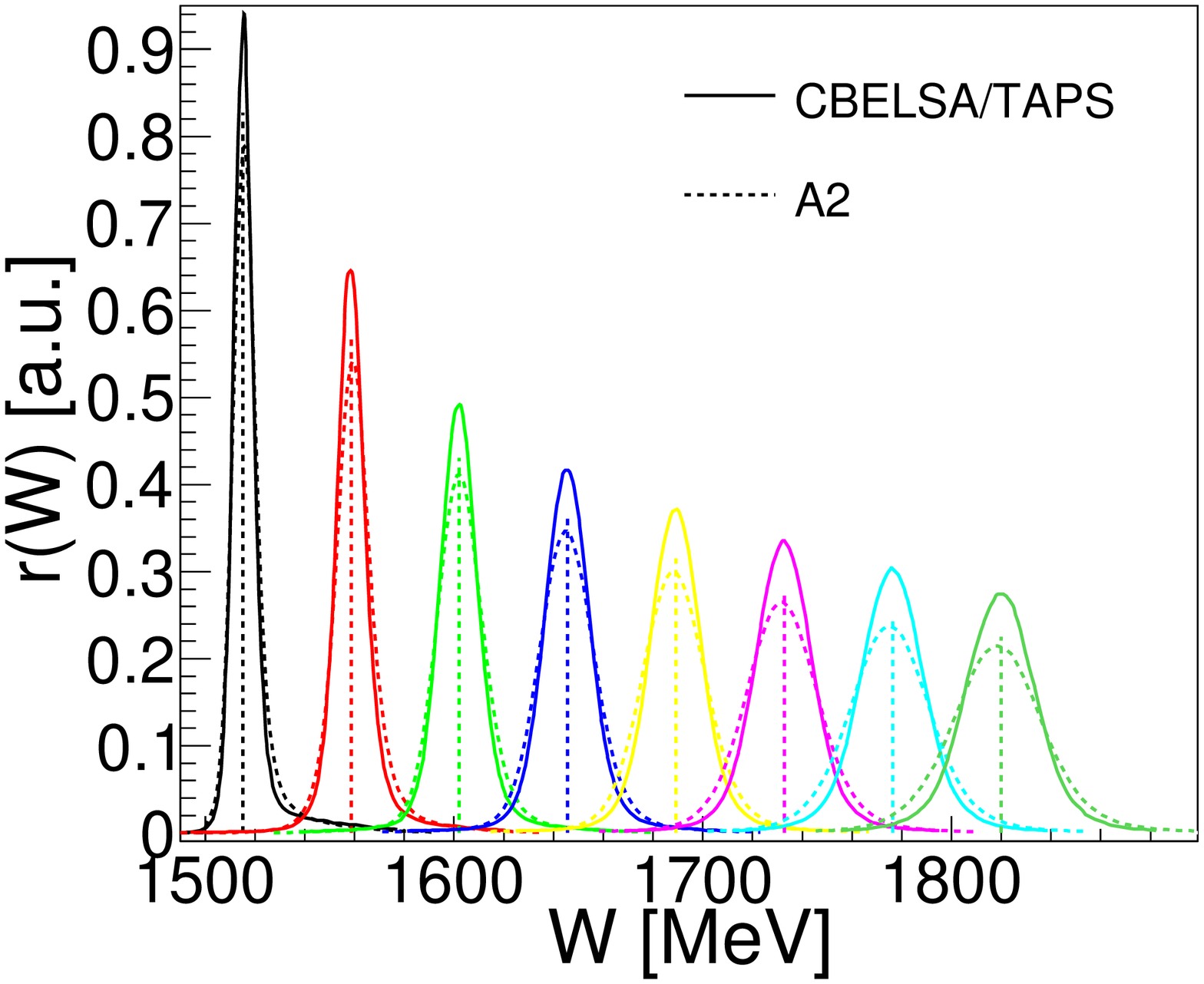}}
\resizebox{0.5\columnwidth}{!}{\includegraphics{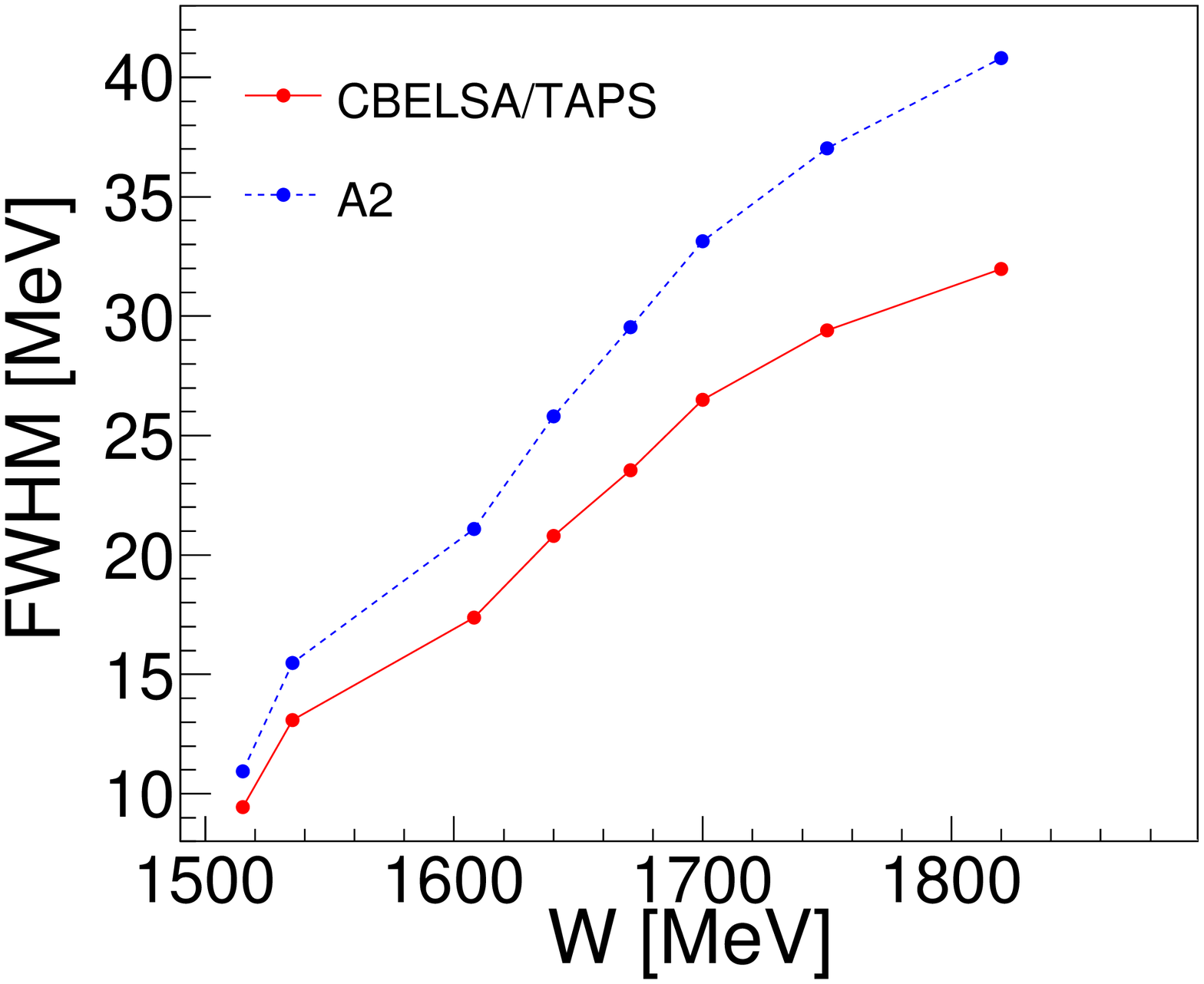}}}
\caption{Left hand side: response of the kinematic reconstruction of the final state obtained from 
Monte Carlo simulations with a fixed $W$ of 1515, 1535, 1608, 1640, 1671, 1700, and 1750 MeV. 
Right hand side:  $W$ resolution (FWHM) for the CBELSA/TAPS setup of the current work (red dots) 
compared to the resolution of the A2 setup (blue dots), which was used for the work in 
\cite{Witthauer_13,Werthmueller_14}.}
\label{fig:Wres}       
\end{figure}

The $W$ resolution achieved with the kinematic reconstruction of the final state is shown in 
Fig.\ \ref{fig:Wres}. It was determined by a Monte Carlo simulation of $\eta$ photoproduction 
reactions with a fixed invariant mass $W$ of 1515, 1535, 1608, 1640, 1671, 1700, and 1750 MeV. 
The resolution obtained with the CBELSA/TAPS setup is somewhat better than the resolution of the A2 setup 
\cite{Werthmueller_14} at the Mainz MAMI accelerator, which is shown for comparison in Fig~\ref{fig:Wres}.
This is mainly due to a better angular resolution for the recoil nucleon because the CBB has higher
granularity than the Crystal Ball in Mainz. In the region of the narrow structure in $\eta$ photoproduction 
of the neutron, the resolution at ELSA is approximately 24 MeV FWHM, compared to the 30 MeV FWHM for the 
A2 setup.

\begin{figure}[h]
\centerline{
\resizebox{\columnwidth}{!}{\includegraphics{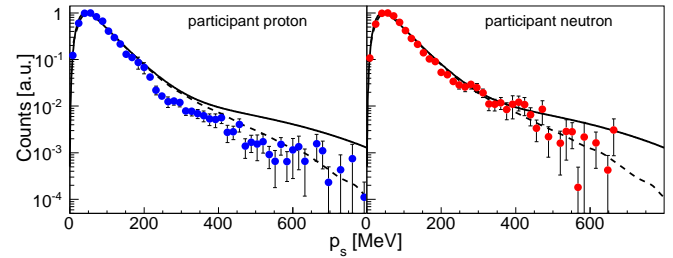}}}
\caption{Spectator momentum distribution for the participant proton (left hand side) and neutron 
(right hand side) in deuterium. The experimental distributions are compared to calculations using 
the Paris N-N potential (solid line) \cite{Lacombe_81} and the corresponding kinematically allowed 
distribution for $\eta$ photoproduction (dashed line). The $y$-axis is shown logarithmically.}
\label{fig:Fermi}       
\end{figure}

The kinematic reconstruction of the final state was cross-checked by a comparison of the final-state
momenta of the spectator nucleons, which also follow from the reconstruction, to the
expectation from the deuteron wave function. In the approximation of a spectator - participant model without
final state interactions, the final state momentum of the spectator nucleon should be equal to its initial 
Fermi momentum. The measured spectator momentum distributions for participant protons and neutrons are
compared in Fig.~\ref{fig:Fermi} to the Fermi momentum distribution of nucleons bound in the deuteron
extracted from the deuteron wave function of the Paris nucleon-nucleon potential \cite{Lacombe_81}.
For large Fermi momenta, $\eta$ photoproduction can be kinematically forbidden due to overall energy
conservation. Results from a simulation based on the deuteron wave function taking into account the
kinematic constraints for $\eta$ production are shown as dashed lines in Fig.~\ref{fig:Fermi}
and agree quite well with the measured data. The agreement is better for participant neutrons than for 
protons because for the latter energy dependent detection efficiency effects are larger and not 
considered in the simulation (in contrast to neutrons low energy protons do not reach the detector).  

\subsection{Absolute Normalization of the Cross Sections}
\label{sec:Extraction}
The cross sections were computed from the yields obtained by the integration of the final invariant mass 
distributions after application of all other analysis cuts. The yields were normalized to the target surface 
density, the $\eta\to3\pi^0$ decay ratio $\Gamma_i / \Gamma= 32.68\%$ \cite{PDG_16}, the incident photon flux, 
and the detection efficiency. The different steps were done separately for each energy and 
$\cos{(\theta_{\eta}^{\ast})}$ bin and are summarized in this section. 

The target surface densities were 0.266 deuterons per barn for the liquid deuterium target 
and 0.8727 deuterons per barn (corresponding to 0.08727 butanol molecules per barn) for the deuterated 
butanol target. The incident photon flux was calculated from the number of electrons in the photon 
tagger and the tagging efficiency. The tagging efficiency is the fraction of the photons which pass through 
the collimator and impinge on the target. The number of photons on the target was constantly monitored 
by a dedicated detector system, consisting of the Gamma Intensity Monitor (GIM) and the Flux Monitor 
(FluMo). The rate in the GIM was $5.1-6.9$ MHz and its efficiency was $40-80\%$ (energy dependent). 
The efficiency was mainly reduced by dead time effects and discriminator thresholds and was corrected 
with the FluMo, which was calibrated at low intensities where the latter had an efficiency of 100\%.

\begin{figure}[t]
\centerline{
\resizebox{\columnwidth}{!}{\includegraphics{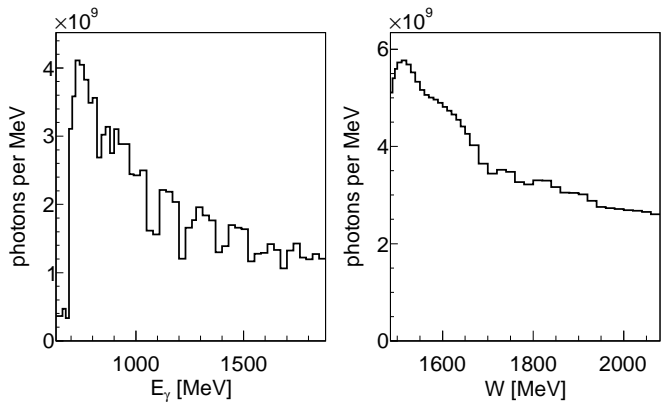}}}
\caption{Typical photon flux on the target as a function of the incident photon energy (left hand side) 
and the final state invariant mass (right hand side). }
\label{fig:flux}       
\end{figure}

A typical spectrum for the resulting photon flux as a function of incident photon energy is shown 
in Fig.\ \ref{fig:flux}. The photon flux as a function of the final state energy $W$ was calculated by 
folding the incident photon energy distribution with the Fermi momentum distribution of nucleons bound
in the deuteron taken from the deuteron wave function parameterized by the Paris N-N potential 
\cite{Lacombe_81}.

\begin{figure}[t]
\centerline{
\resizebox{\columnwidth}{!}{\includegraphics{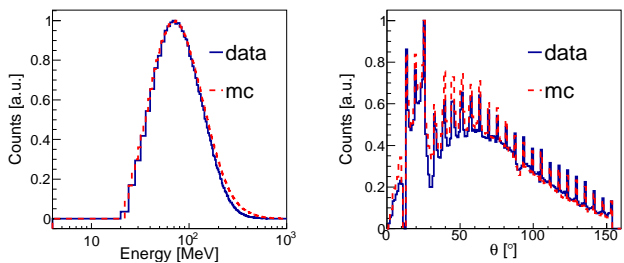}}}
\caption{Energy (left hand side) and polar angle (right hand side) distribution of the $\eta$ decay photons. 
Compared are the spectra from experimental data and simulation (mc). }
\label{fig:CompareMC}       
\end{figure}

\begin{figure}[b]
\centerline{
\resizebox{\columnwidth}{!}{
    \includegraphics[height=0.5\textheight]{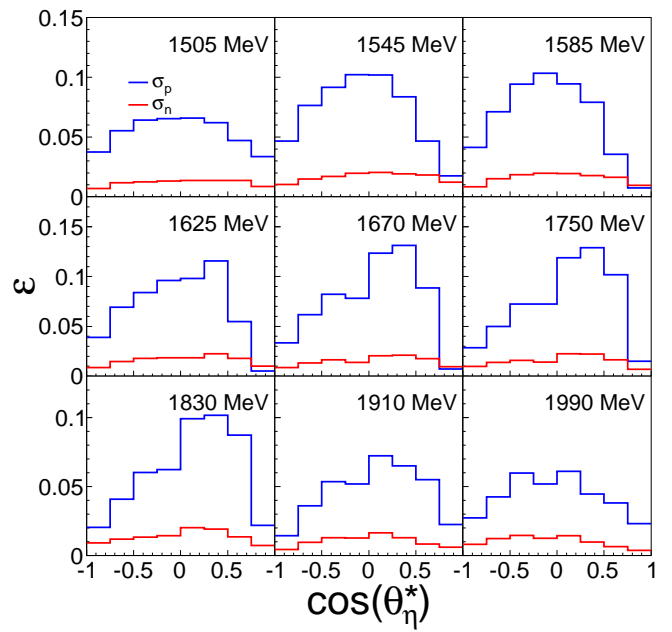}
   }} 
  \caption{Angular detection efficiencies for different bins of the final state invariant mass 
  $W$ (energy indicated in the figure) for the reaction off the proton (blue) and off the neutron (red).}
\label{fig:Eff} 
\end{figure}

The detection efficiency was determined from Monte Carlo simulations with the GEANT3 package  
\cite{Brun_86}. Events were generated with the Pluto event generator \cite{Froehlich_07}, including
the effects of Fermi motion. They were weighted with a $1/E_{\gamma}$ bremsstrahlung distribution
for the photon flux and cross sections for quasifree $\eta$ photoproduction from the deuteron
taken from \cite{Jaegle_11,Werthmueller_14}. The efficiencies were determined for each event class 
separately and for every energy and $\cos{(\theta_{\eta}^{\ast})}$ bin. Typical results for nine bins 
of final state invariant $W$ are shown in Fig.\ \ref{fig:Eff}. Shown are the $\cos{(\theta_{\eta}^{\ast})}$ 
distributions for the event classes $\sigma_p$ (blue) and $\sigma_n$ (red) for the deuterium target. 
The efficiencies are between $2.5\%$ and $14\%$ for the reaction off the proton and $1\%$  to $2\%$ 
for the reaction off the neutron. 

The simulations are very reliable and precise for the detection of the $\eta$-decay photons in the 
detector. As an example Fig.~\ref{fig:CompareMC} compares the measured and simulated energy and polar-angle 
distributions of data and simulation for the $\eta\rightarrow 6\gamma$ final state. Simulated cluster 
multiplicities (i.e. the number of crystals responding in one electromagnetic shower) are compared to 
experimental data in  Fig.\ \ref{fig:ClustMult}. The simulations include also the correct implementation 
of the experimental trigger, which was only based on photon hits as discussed in Sec.\ \ref{sec:1}. 
In the analysis, all trigger thresholds (for measured data and simulated responses)
were set by software safely above the used hardware thresholds so that no bias could arise.

\begin{figure}[h]
\centerline{
\resizebox{\columnwidth}{!}{
    \includegraphics[height=0.3\textheight]{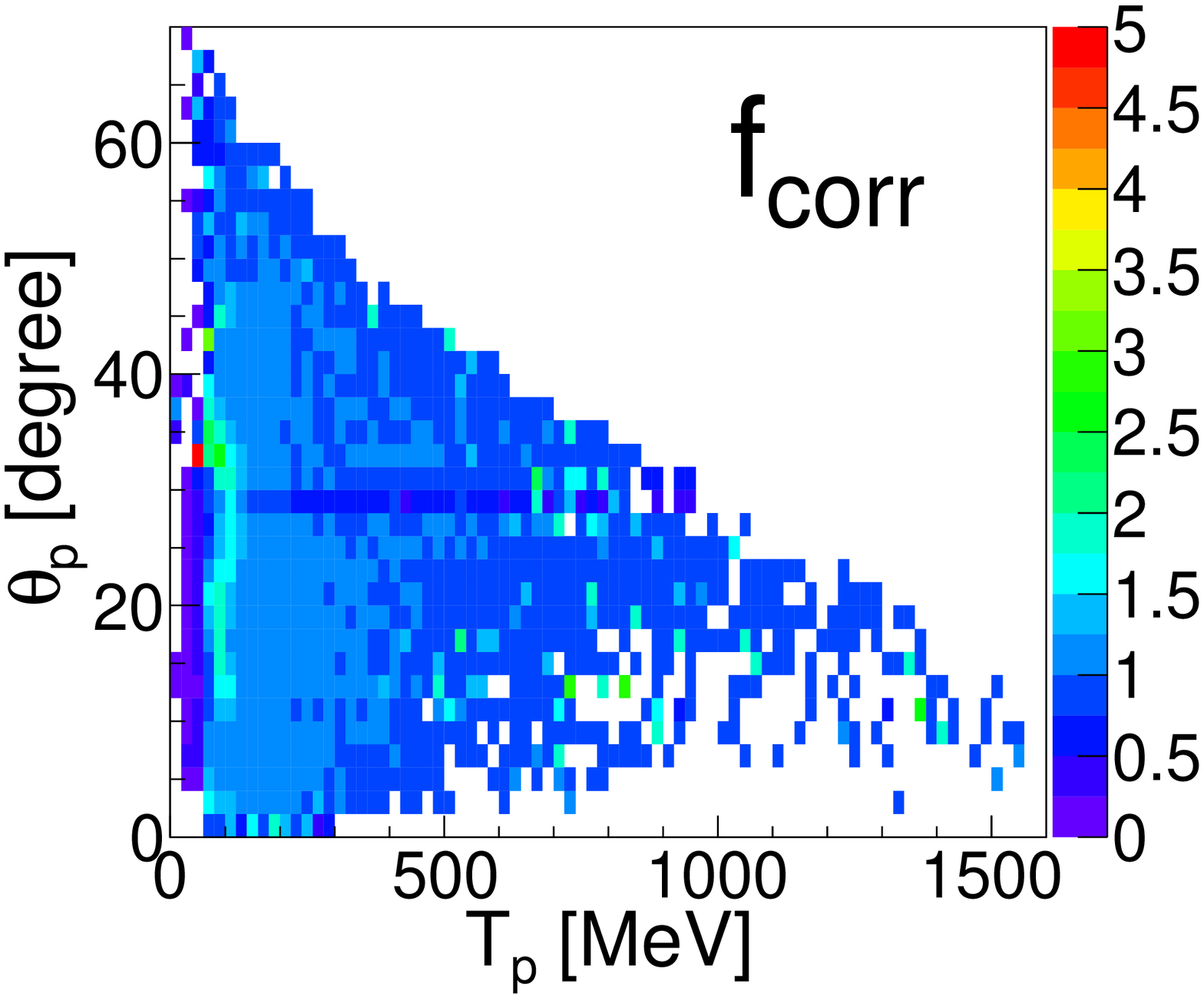}
    \includegraphics[height=0.3\textheight]{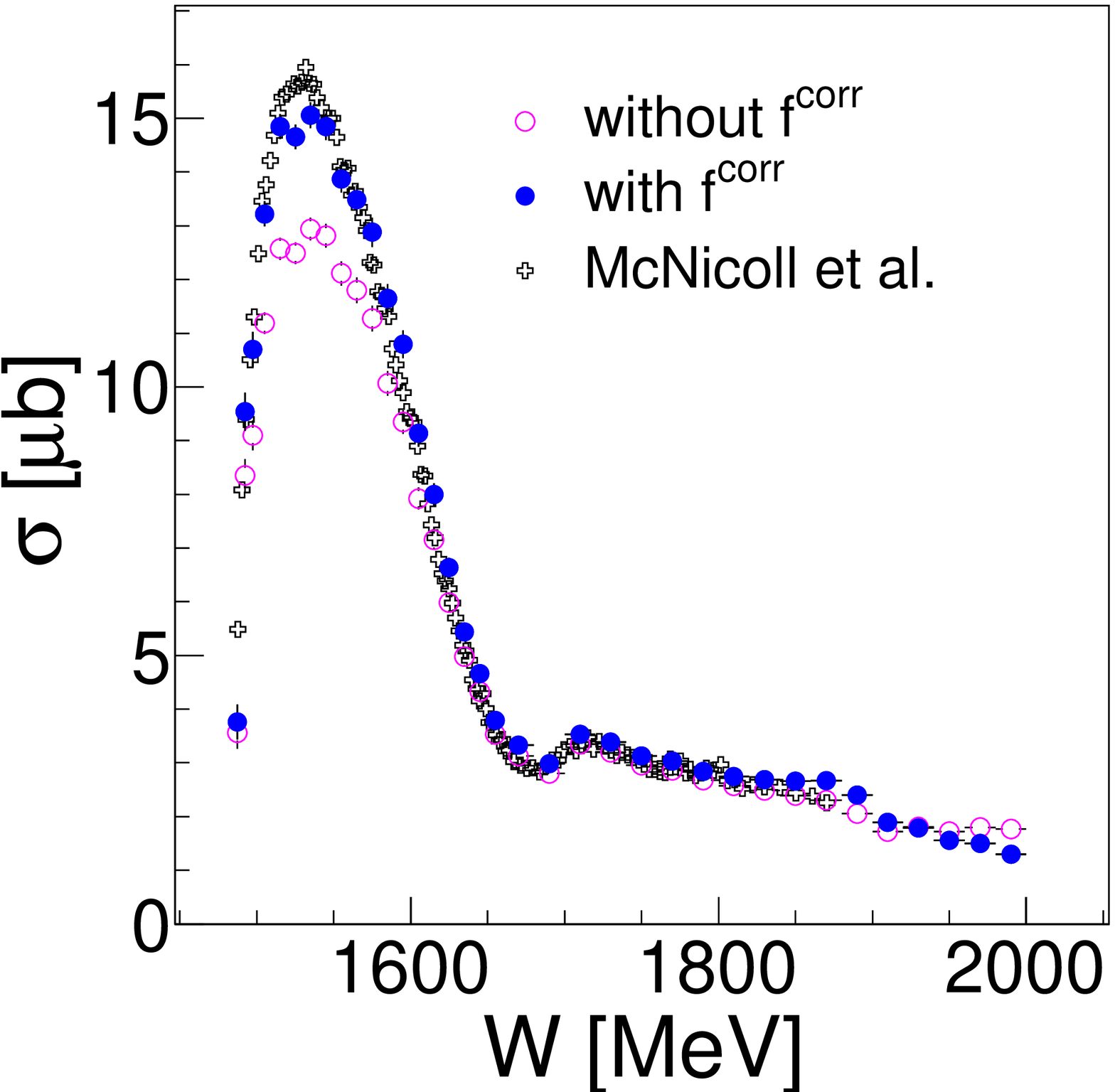}
   }} 
  \caption{Left hand side: proton efficiency correction factor $f_{\rm corr}$ as a function of the 
  laboratory angle of the proton $\theta_{p}$ and its kinetic energy $T_p$. 
  Right hand side: effect of $f_{corr}$ on the total cross section. The cross sections from the 
  current experiment with (blue dots) and without correction factor (magenta open circles) are compared 
  to results on the free proton target from \cite{McNicoll_10} (black stars).}
\label{fig:NuclE} 
\end{figure}

The precision of the simulation is worse for recoil nucleons, in particular for charged protons detected 
close to the transition regions from the CBB to FP and from the FP to MT, where the results are very 
sensitive to small inaccuracies in the modeling of insensitive support materials and detector geometry. 
Therefore the detection efficiency for recoil protons was independently studied using $\eta$ photoproduction 
off the free proton measured with a liquid hydrogen target. The number of events with detection of the 
recoil proton was compared to the total number of events with an identified $\eta$ meson. The ratio of 
these two event rates corresponds directly to the proton detection efficiency. This was done for the 
measured data and also for the results of the Monte Carlo simulation of the hydrogen experiment. 
The comparison of the detection efficiency derived from the measurement and the simulated one, both as 
a function of proton laboratory angle $\theta_p$ and proton kinematic laboratory energy $T_p$, was then 
used to extract correction factors for the simulation of the deuterium experiment. Fig.\ \ref{fig:NuclE} 
shows the relative proton efficiency correction as a function of $\theta_p$ and $T_p$. The largest deviations 
from unity occur in the region of the FP detector ($11.8^{\circ}\leq\theta_p\leq 27.5^{\circ}$) and the 
transition from the CBB to the FP around $\theta_p \simeq 30^{\circ}$. They are most probably caused 
by geometrical discrepancies between simulation and experiment and the fact that no precise energy 
thresholds could be determined for the FP vetoes and the inner detector, since they did not provide 
energy information. It is shown at the right hand side of Fig.\ \ref{fig:NuclE} that the correction 
significantly improves the agreement of the present results with previous measurements \cite{McNicoll_10}
of the free proton cross section in the energy range of the $N(1535)$ resonance. This effects are negligible 
at larger $W$.

A correction of the neutron efficiency (as in \cite{Witthauer_13,Werthmueller_14}) was not possible, 
since the analysis would require the unambiguous identification of the charged pion in the reaction 
$\gamma p \to \pi^{+}n$ or $\gamma p \to \pi^{0} \pi^{+}n$. This was only possible in the MT 
detector via a $\Delta E-E$ analysis, but not in the CBB detector or in the FP detector due to the 
missing energy information of the charge sensitive detectors. However, analyses of data from a
similar detector setup at the MAMI accelerator \cite{Witthauer_13,Werthmueller_14}, for which 
identification of charged pions was possible, have shown that the effects for neutron detection
are smaller because low energy neutrons are much less affected by passing through inactive structural
material of the detector.

The determination of the unpolarized cross sections also included the subtraction of the contribution 
from the target cell ($2\times 125$ $\mu$m capton foil). Since no empty target measurement was available, 
this contribution was estimated from a measurement with a carbon target. The yields were normalized 
to the number of nuclei in the target windows and subtracted to obtain the final cross section values. 
The empty target contribution was up to 5\% for the exclusive cross sections $\sigma_p$, $\sigma_n$
and up to 10\% for the quasifree inclusive reaction.

\subsection{Extraction of the Double Polarization Observable $\bm{E}$ and the Helicity Dependent 
Cross Sections}
\label{sec:ExDPE}

The analysis of the polarized data was done analogously to unpolarized data. The yields with 
helicity 1/2 and 3/2 were normalized with nucleon numbers, branching ratio, photon flux, and detection 
efficiency. 

\begin{figure}[t]
\centerline{
\resizebox{0.9\columnwidth}{!}{\includegraphics{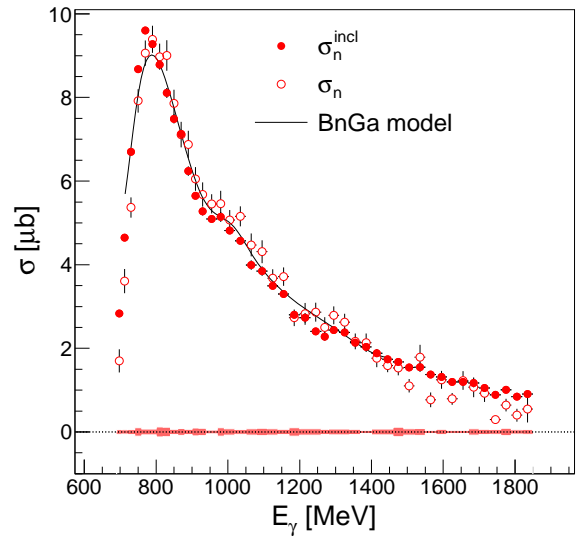}}}
\caption{Comparison of the exclusive ($\sigma_n$, open symbols) and semi-inclusive ($\sigma_n^{incl}$, 
closed symbols) analysis of the reaction $\gamma n\rightarrow n\eta$. Curve: Fermi folded model 
calculation by BnGa \cite{Anisovich_15}. The systematic uncertainty of the inclusive cross section is given 
by the shaded area on the bottom, except 10\% total normalization uncertainty.}
\label{fig:Incl}       
\end{figure}

The statistical quality of the data measured with the polarized target was limited and not sufficient
for an exclusive analysis of $\eta$ production in coincidence with recoil neutrons. However, due to the
low neutron detection efficiency, statistics for the measurement of the $\gamma n\rightarrow n\eta$ 
reaction is increased by roughly a factor of three when requiring that no recoil proton has been detected 
instead of requiring the detection of a coincident neutron. Since the proton detection efficiency 
is typically above 95\%, this type of analysis does not introduce substantial background. The analysis
accepted only events for which none of the charge sensitive detectors responded. This condition
enhances proton suppression because in contrast to the analysis of the exclusive proton data,
hits where the proton was stopped in the charged particle detectors were also assigned to protons. 
The method was tested with the unpolarized data for which also the exclusive $\gamma n\rightarrow n\eta$
reaction with detection of the recoil neutron could be studied. In Fig.\ \ref{fig:Incl}, total cross sections
of the exclusive ($\sigma_n$, coincident detection of recoil neutron) and semi-inclusive ($\sigma_n^{incl}$, 
vetoing of events with charged hits) analyses for the $\gamma n\rightarrow n\eta$ reaction are compared. 
They are in quite good agreement, but the statistical
quality of the inclusive data is much better. The drawback of this analysis method is that without detection
of the recoil neutron, the final state kinematics cannot be reconstructed so that the effects from
nuclear Fermi motion are not removed and all structures in the total cross section are significantly
broadened by it. 

\begin{figure}[t]
\centerline{
\resizebox{\columnwidth}{!}{\includegraphics{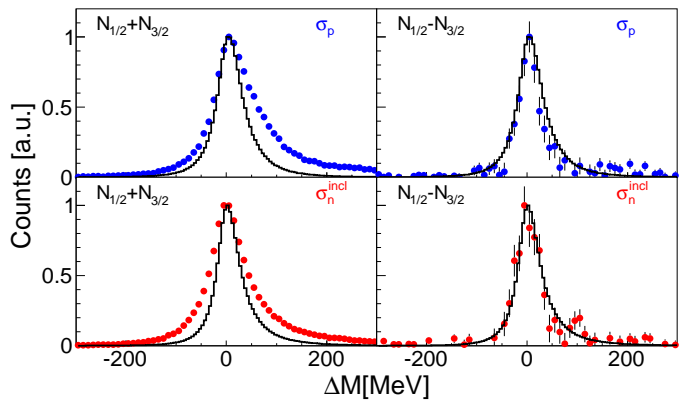}}}
\caption{Missing mass for the sum (left) and the difference (right) of two helicity states for the 
exclusive reaction off the proton (top) and the inclusive reaction off the neutron (bottom). 
The experimental spectra (dots) are compared to the simulated line shape (black line)
for nucleons bound in the deuteron.}
\label{fig:MMDiff}       
\end{figure}

The double polarization observable $E$ was deduced in two different ways, corresponding to the two
parts of Eq.\ \ref{eq:E}. In version (1) (first part of Eq.\ \ref{eq:E}), the difference of the two 
helicity dependent cross sections was divided by the sum of the two. This analysis required the  
subtraction of the unpolarized background from carbon and oxygen nuclei in deuterated butanol.
This background becomes clearly visible when one compares the missing mass spectra of the sum and
the difference of the two helicity states. This is shown in Fig.\ \ref{fig:MMDiff}. The spectra
of the helicity difference correspond only to reactions on nucleons bound in the deuteron. 
The background from the unpolarized nucleons bound in the heavier spin-zero nuclei drops out. 
These spectra agree very well with a MC simulation for the reaction on the deuteron, taking into 
account Fermi motion in the deuteron. The helicity sum spectra are much broader due to the unpolarized 
background from the heavier nuclei with larger Fermi momenta.  

\begin{figure}[t!]
\centerline{
\resizebox{\columnwidth}{!}{\includegraphics{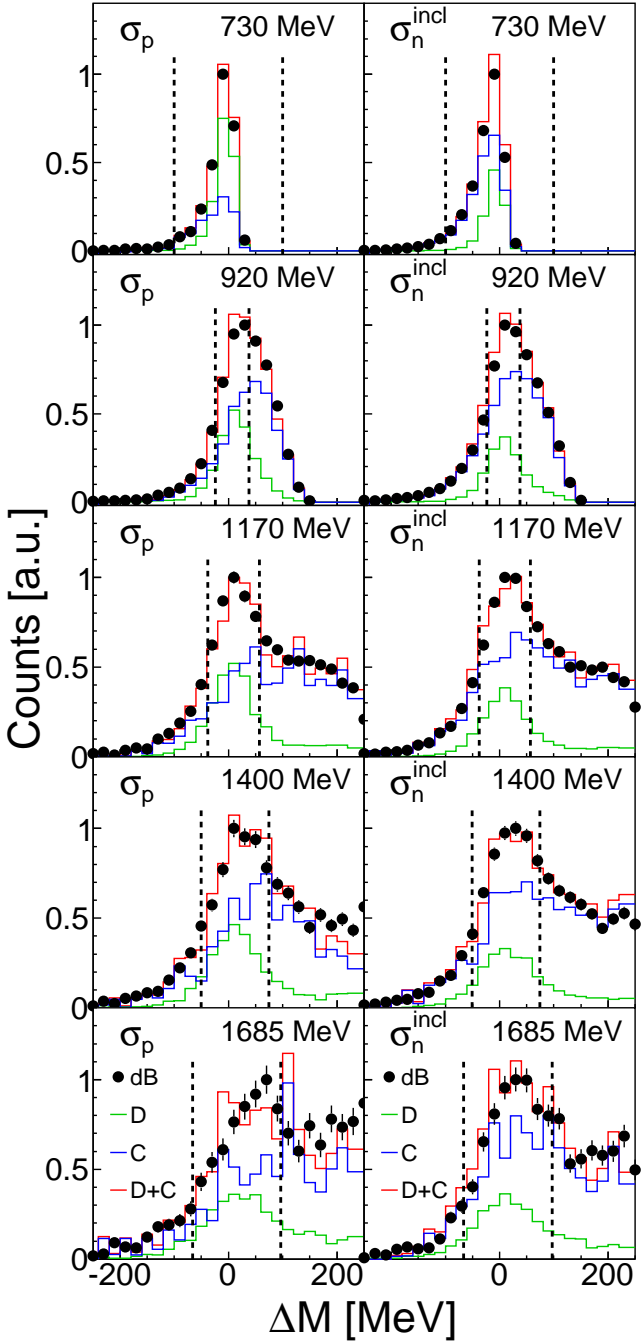}}
}
\caption{Normalized missing mass spectra for the reactions $\gamma p\rightarrow p\eta$ (left hand side) 
and $\gamma n\rightarrow (n)\eta$ (right hand side) for five bins of incident photon energy (mean energy 
indicated). The data points from the measurement with the deuterated butanol target are shown in black, 
and the line shapes from the measurement with the carbon and deuterium  target are shown in blue and green, 
respectively. The red line is the sum of the deuterium and the carbon data. The vertical lines are the 
missing mass cut positions.}
\label{fig:MMFit}       
\end{figure}

The contribution of the unpolarized carbon and oxygen nuclei was extracted from a dedicated measurement 
with a carbon foam target of a similar density as the deuterated butanol target. The data were analyzed 
with the same procedure as the data from the deuterated butanol and the yields were extracted and normalized 
with nucleon numbers, branching ratios, photon flux, and efficiencies. The normalized yields from the 
carbon measurement were then directly subtracted from the ones of the measurement with the deuterated 
butanol target. This procedure was cross-checked with a comparison of the yields from the butanol, carbon,
and the liquid deuterium target, which is shown in Fig.\ \ref{fig:MMFit}. The yields from all three targets 
were normalized absolutely as discussed above (no free parameter). The results for the deuterium
target and the carbon target were added and agree very well with the measurements of the butanol target. 
For the carbon subtracted spectra (not shown), the cuts indicated in Fig.\ \ref{fig:MMFit} were applied.

In a second analysis version (version (2)), the difference of the two helicity states was normalized 
to the unpolarized cross section measured with the liquid deuterium target, corresponding to the second 
part of Eq.\ \ref{eq:E}.

The helicity dependent cross sections $\sigma_{1/2}$ and $\sigma_{3/2}$ follow from the $E$ asymmetry via:
\begin{align}
\sigma_{1/2} &= \sigma_0(1+E) \nonumber\\
\sigma_{3/2} &= \sigma_0(1-E)\, ,
\label{eq:helsplit}
\end{align}
where $\sigma_0$ is again the unpolarized cross section measured with the liquid deuterium target.
The two different extraction methods of $E$ correspond to two solutions for $\sigma_{1/2}$ and $\sigma_{3/2}$,
which both use the unpolarized cross section $\sigma_0$ (see Eq.~\ref{eq:helsplit}) for the absolute normalization. 
A third extraction method (version (3)) for $\sigma_{1/2}$ and $\sigma_{3/2}$ 
uses directly the two cross sections measured with the butanol target for the parallel and antiparallel spin 
orientations with subtraction of the carbon background and is thus completely independent of the liquid 
deuterium data. All versions are statistically correlated, but have a different systematic error and hence 
are ideally suited to crosscheck the extraction of the observables.

\subsection{Systematic Uncertainties}
\label{sec:Sys}
For the unpolarized cross sections, the systematic uncertainty is composed of overall normalization 
uncertainties and energy and $\cos{(\theta_{\eta}^{\ast})}$ dependent effects. 

The normalization uncertainties come from the target surface density (7\%, estimated from uncertainties 
in the measurement of the target pressure and possible changes of target length between room temperature 
and cooled state), the $\eta$ decay branching ratio (0.8\% calculated from \cite{PDG_16}), the photon 
flux (8\%, mainly related to the precision of the GIM and the trigger electronics), and the empty target 
contribution (2.5\%, very conservatively estimated to be approximately half the contribution 
of the empty target measurement). These contributions were added in quadrature for a total uncertainty 
of approximately 10\%.

\begin{figure}[t]
\centerline{
\resizebox{\columnwidth}{!}{\includegraphics{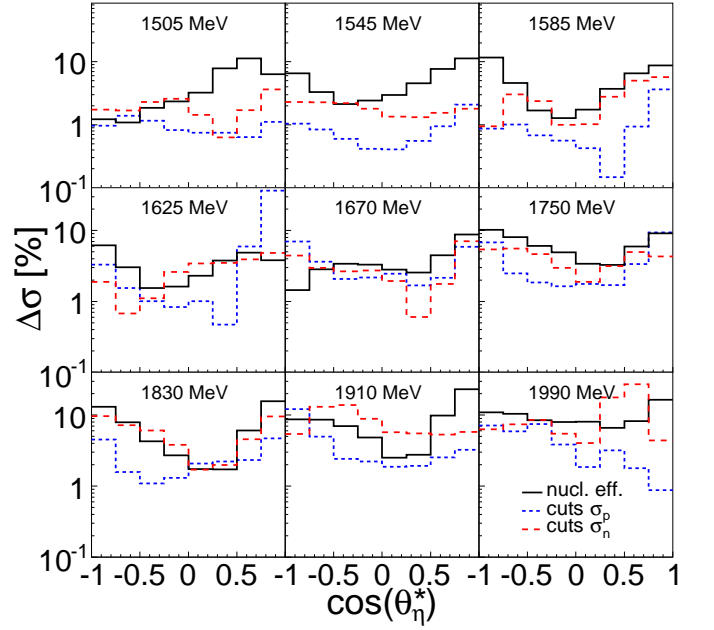}}}
\caption{Typical relative systematic uncertainties induced by the analysis cuts (proton: blue short dashed, 
neutron: red long dashed lines) and the nucleon efficiency (solid line, same for proton and neutron). }
\label{fig:Sys}    
\end{figure}

The energy and $\cos{(\theta_{\eta}^{\ast})}$ dependent effects result from the analysis cuts and the 
MC simulations, more precisely from how well the MC simulations reflect the analysis cuts. 
They were determined separately for each reaction channel and estimated by a comparison of the results 
from analyses with varying cuts on the coplanarity angle, the missing mass, and the invariant mass.
Only the precisely determined incident photon energies, the parameters (angle and energy) of the $\eta$-decay
photons, and the angles of the recoil nucleons enter into this cuts. In the comparison, all cuts were
simultaneously varied by $\pm3\%$.

The uncertainty of the nucleon efficiency correction was investigated similarly as in 
\cite{Jaegle_11,Witthauer_13,Werthmueller_14} by a comparison of the sum $\sigma_{p}+\sigma_{n}$ of the 
two exclusive cross sections $\sigma_p$, $\sigma_n$ and the total inclusive cross section 
$\sigma_{\rm incl}$ measured without any conditions for recoil nucleons. The latter is 
independent on nucleon detection efficiencies. Since contributions from coherent production of $\eta$ 
mesons are completely negligible \cite{Weiss_01}, any deviations between $\sigma_{p}+\sigma_{n}$ and 
$\sigma_{\rm incl}$ result from imperfect nucleon detection efficiencies. Half of the observed differences 
were assigned to the uncertainty of the proton detection efficiency and the other half was assigned to 
that of the neutron. The uncertainties induced by the analysis cuts and the detection efficiency are shown in 
Fig.\ \ref{fig:Sys} for different final state energy bins and as a function of $\cos{(\theta_{\eta}^{\ast})}$.

The systematic uncertainties of the double polarization observable $E$ and of the helicity dependent cross 
sections $\sigma_{1/2}$ and $\sigma_{3/2}$ were determined from the 
uncertainty of the target polarization (5\%), the photon polarization (3\%), and the smoothed difference 
between the analysis versions (in case of three versions, the maximum difference was taken). 
This estimation is reasonable since the systematics of all versions is independent even though 
the statistical errors are highly correlated.  
%The uncertainty of the helicity dependent cross sections
%$\sigma_{1/2}$ and $\sigma_{3/2}$ follows then from the uncertainty of $E$ and $\sigma_0$ for analysis 
%versions (1,2) and from the absolute uncertainty of the cross sections measure with the butanol target
%for analysis version (3).  

\section{Results}
\label{sec:Res}
The results presented in this section for the unpolarized total cross section and angular distributions
were obtained from the analysis of the liquid deuterium data. The analysis of the asymmetry $E$
and the helicity dependent cross sections $\sigma_{1/2}$, $\sigma_{3/2}$ used data from the polarized 
deuterated butanol target, the liquid deuterium target, and the solid carbon target. The event selection 
and particle identification was identical for all data sets. 

\subsection{Unpolarized Cross Sections}
\label{sec:UCS}

\begin{figure}[t]
\centerline{
\resizebox{0.50\textwidth}{!}{\includegraphics{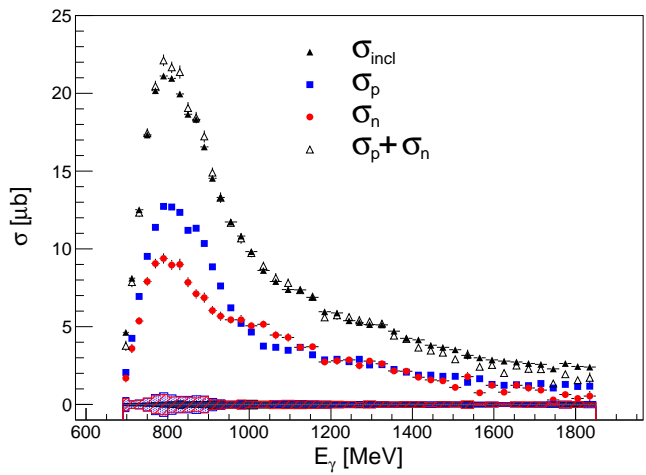}}}
\caption{Total exclusive cross sections $\sigma_p$ (blue squares), $\sigma_n$ (red dots), and the 
inclusive cross section $\sigma_{\rm incl}$ (black triangles) as a function of the incident photon energy. 
In addition, the sum of the two exclusive cross sections $\sigma_{p}+\sigma_{n}$ (black open triangles) 
is shown. The estimated systematic uncertainties are given at the bottom, except 10\% total normalization 
uncertainty.}
\label{fig:TXSE}       
\end{figure}

The angular distributions have been extracted as discussed above in small bins of energy $E$ 
(either incident photon energy $E_{\gamma}$ or final state energy $W$) and for eight equidistant bins of 
$\cos{(\theta_{\eta}^{\ast})}$. The total cross sections were deduced from the angular distributions via 
a fit of Legendre polynomials of third order:
\begin{equation}
\frac{d\sigma}{d\Omega}(E,\mbox{cos}(\theta_{\eta}^{\star})) = \frac{q_{\eta}^{*}(E)}{k_{\gamma}^{*}(E)}
\sum_{i=0}^{3} A_i(E) P_i(\mbox{cos}(\theta_{\eta}^{\star}))\,,
\label{eq:Leg}
\end{equation}
where $q_{\eta}^{*}$ and $k_{\gamma}^{*}$ are the $\eta$ and photon momenta in the center-of-mass frame, 
respectively, and  $A_i(E)$ are the Legendre coefficients. 
    
\begin{figure}[t]
\centerline{\resizebox{0.5\textwidth}{!}{
\includegraphics{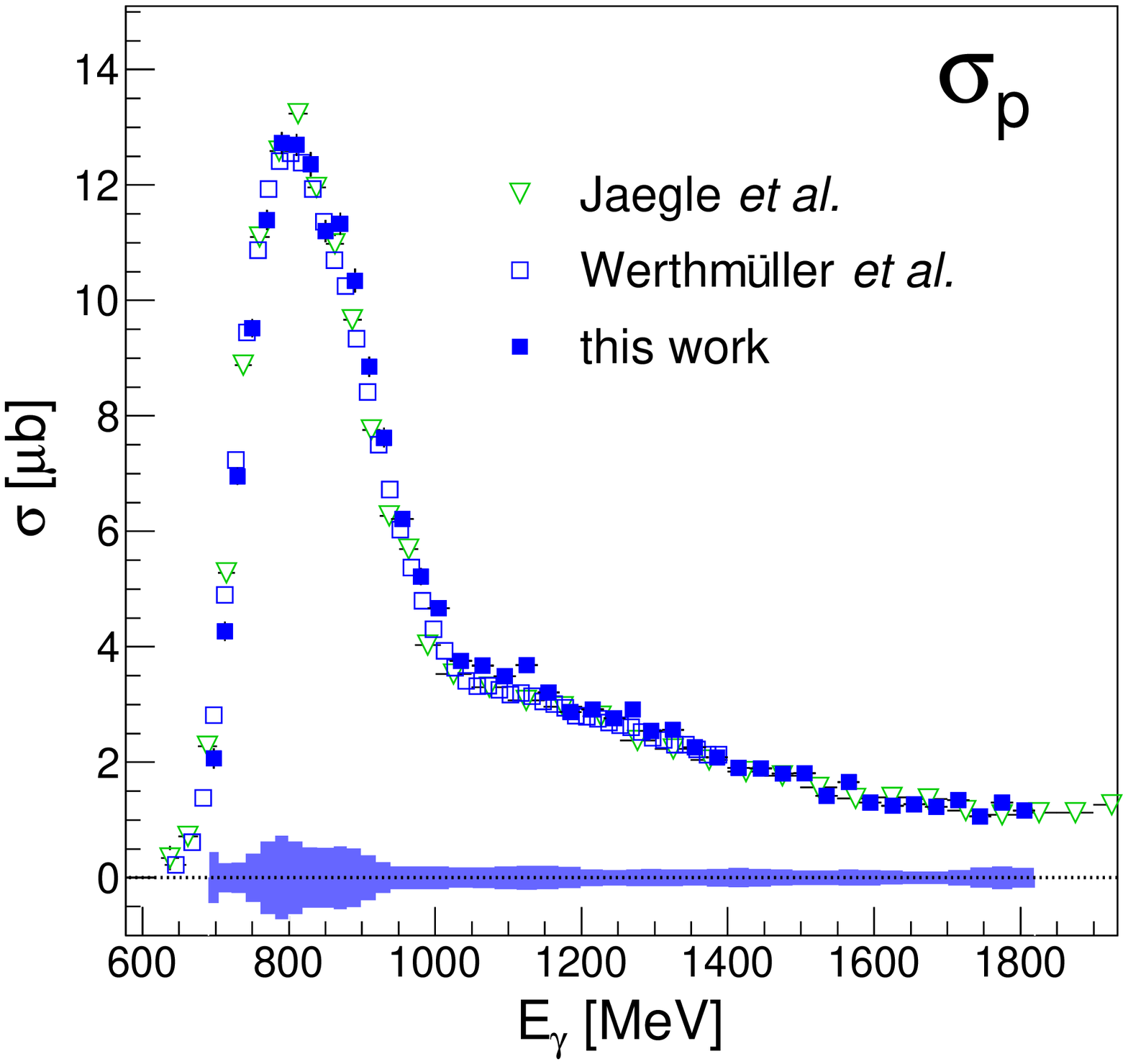}
}}
\centerline{\resizebox{0.5\textwidth}{!}{
\includegraphics{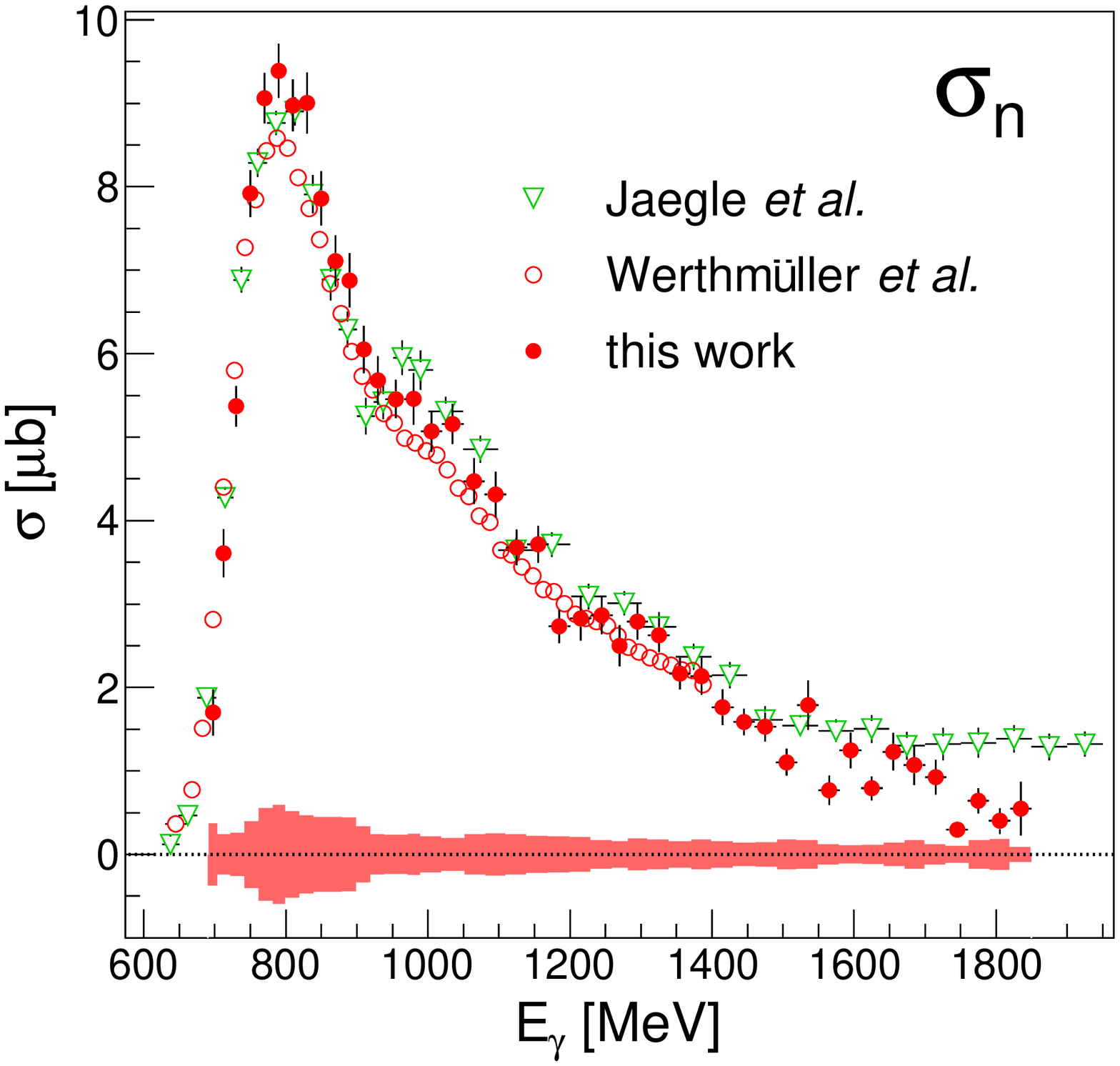}
}}
\caption{Total exclusive cross sections $\sigma_p$ (top) and $\sigma_n$ (bottom) from this work compared to 
the results from \cite{Jaegle_11,Werthmueller_14}. The estimated systematic uncertainties are given
at the bottom, except 10\% total normalization uncertainty.} 
\label{fig:TXS}       
\end{figure}

The total cross sections in the beam-target cm system assuming the initial state nucleon at rest are shown in 
Fig.~\ref{fig:TXSE} for the exclusive reaction on quasifree protons $\sigma_p$ and on quasifree neutrons 
$\sigma_n$, and the sum of these two cross sections $\sigma_{p}+\sigma_{n}$ is compared to the inclusive cross 
section $\sigma_{\rm incl}$. The excellent agreement between $\sigma_{p}+\sigma_{n}$ and 
$\sigma_{\rm incl}$ sets stringent limits for systematic uncertainties in the nucleon detection efficiencies, 
which would affect the proton/neutron cross section ratio.
    
\begin{figure*}[!t]
\centerline{
\resizebox{0.9\textwidth}{!}{\includegraphics{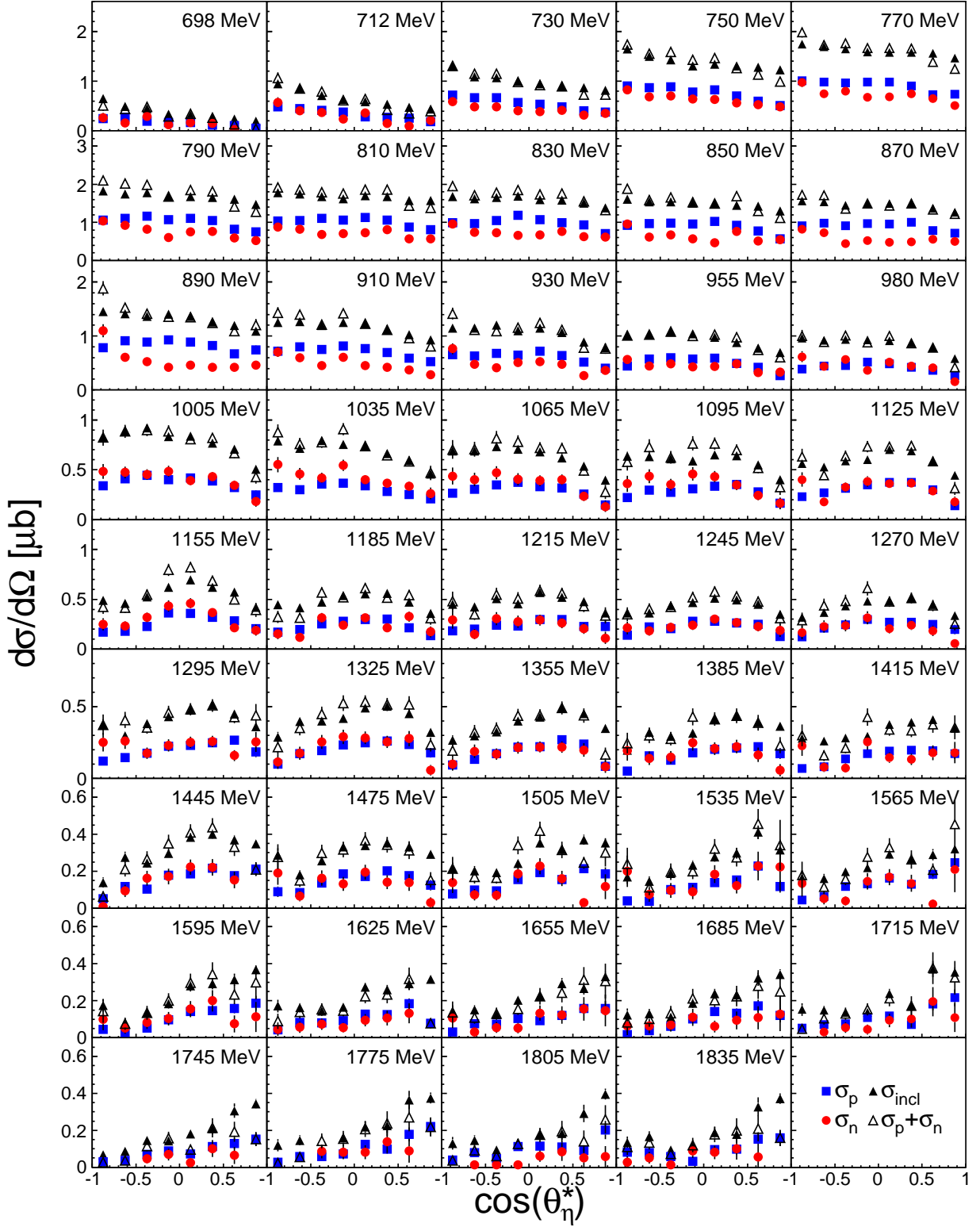}}}
\caption{Angular distributions $\sigma_p$, $\sigma_n$, and $\sigma_{\rm incl}$ for different bins of 
incident photon energy $E_{\gamma}$ (mean energy indicated in the figure) as a function of 
$\cos{(\theta_{\eta}^{\ast})}$ in the beam-target cm system assuming the initial state nucleon at rest. 
Same labeling as in Fig.~\ref{fig:TXSE}.}
\label{fig:XS}       
\end{figure*}

\begin{figure*}[t]
\centerline{
\resizebox{0.85\columnwidth}{!}{\includegraphics{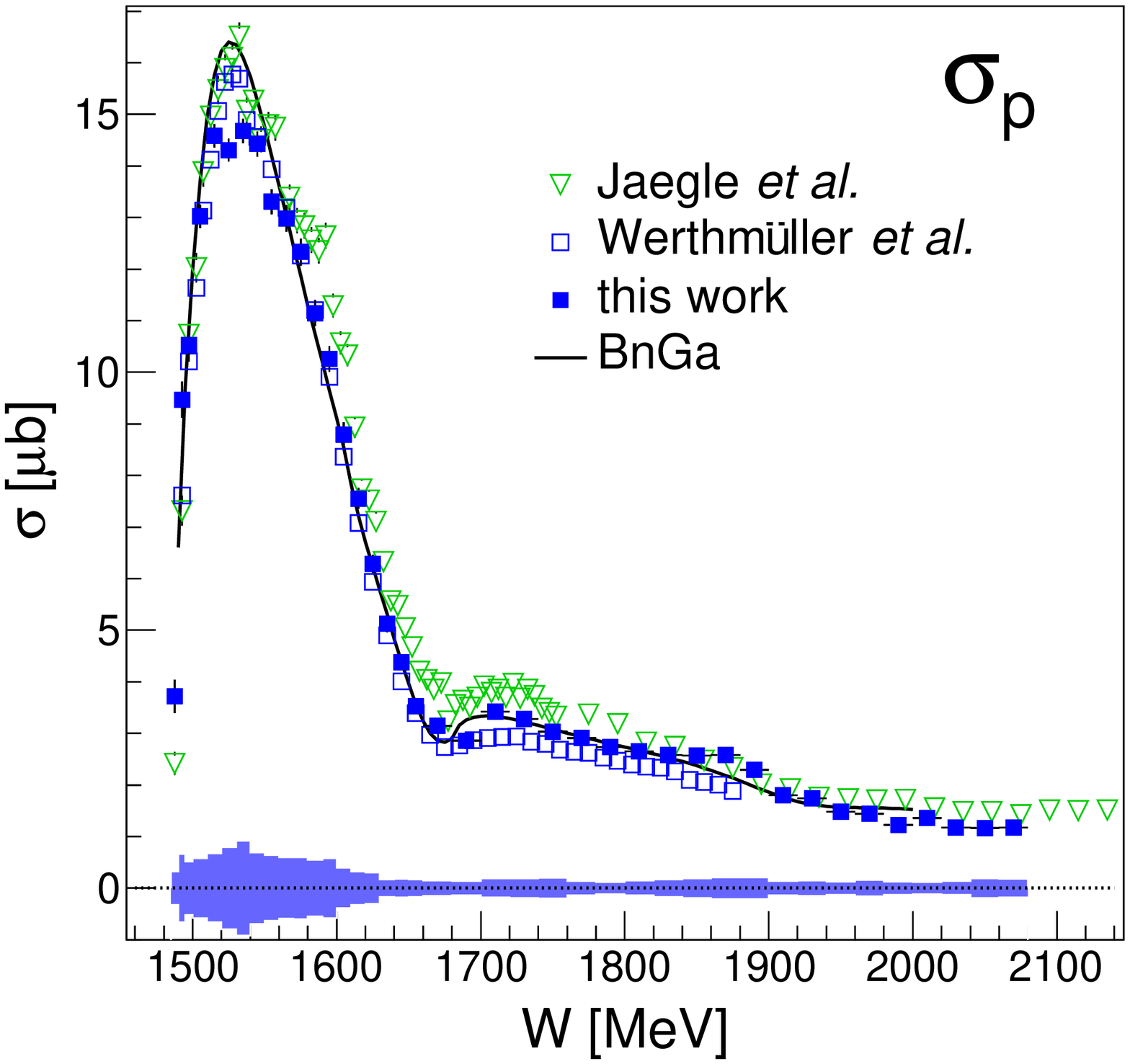}}
\resizebox{0.85\columnwidth}{!}{\includegraphics{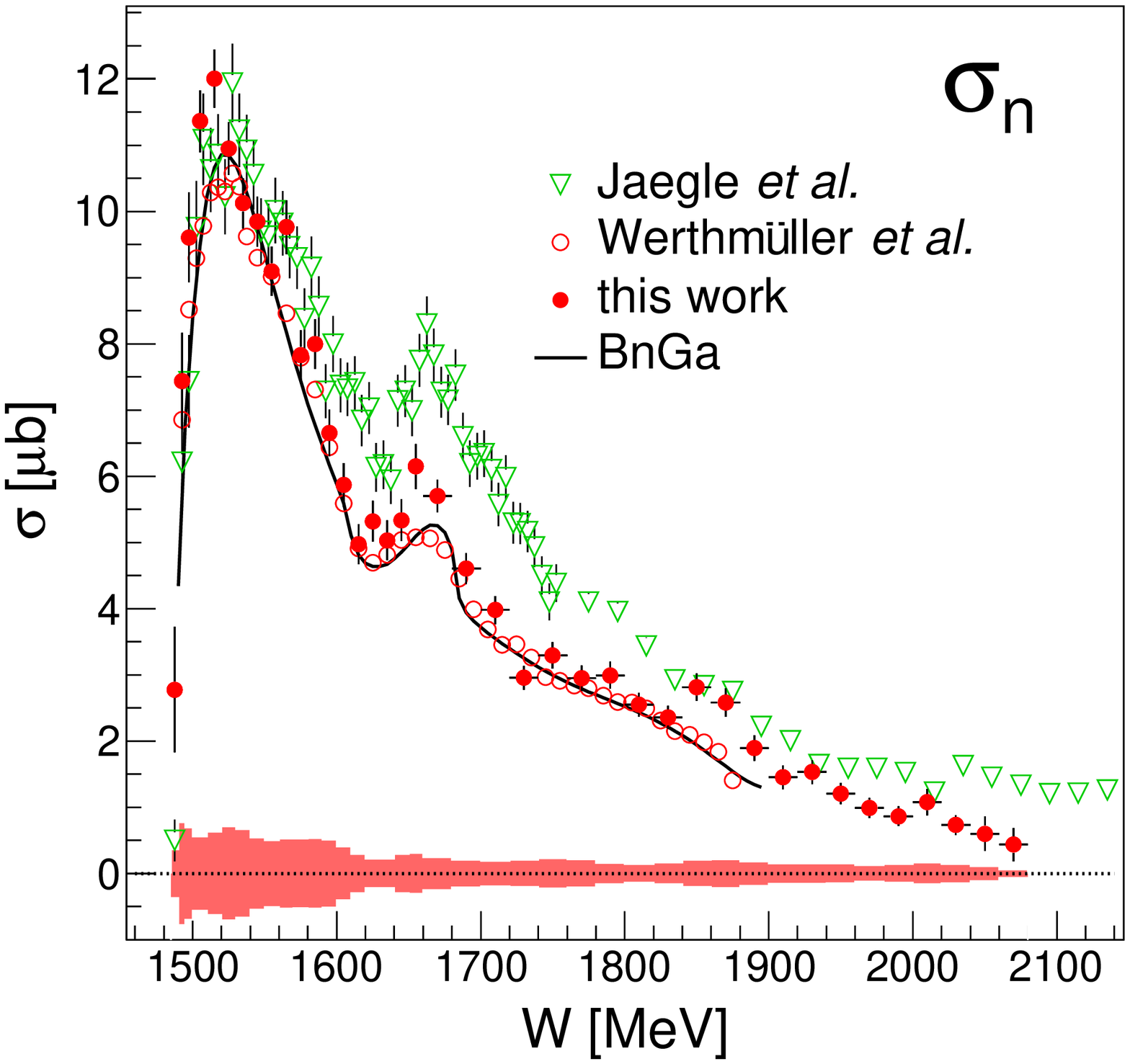}}
}
\caption{Total exclusive cross sections as a function of the final-state invariant mass $W$ for the reaction 
$\gamma p\rightarrow p\eta$ (left) and the reaction $\gamma n\rightarrow n\eta$ (right). Previous 
results from Werthm\"uller \textit{et al.} (MAMI) \cite{Werthmueller_14} and Jaegle \textit{et al.} (ELSA) 
\cite{Jaegle_11}, and BnGa model analyses \cite{Anisovich_15} are compared. The systematic uncertainties 
are indicated at the bottom, except 10\% total normalization uncertainty. }
\label{fig:TXSW}       
\end{figure*} 

\begin{figure}[b]
\centerline{
\resizebox{\columnwidth}{!}{\includegraphics{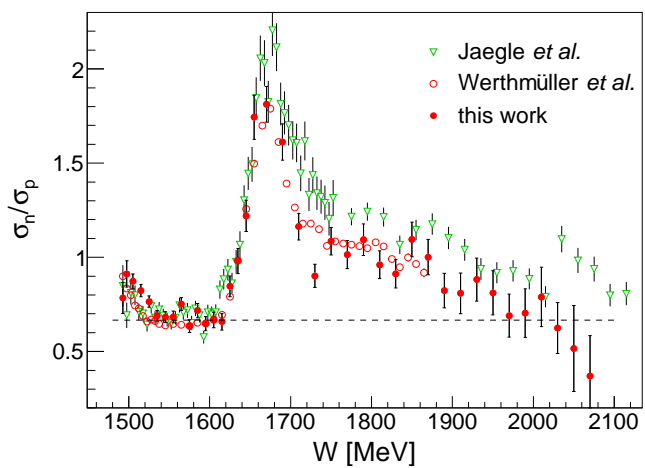}}
}
\caption{Ratio of neutron and proton cross section as a function of the final state energy $W$. 
The data of this work (red dots) are compared to previous results by Werthm\"uller \textit{et al.} 
(red circles) \cite{Werthmueller_14} and Jaegle \textit{et al.} (green triangles) \cite{Jaegle_11}. 
The dashed line marks the ratio of $\sigma_n/\sigma_p = 2/3$, which was already seen in earlier experiments 
\cite{Krusche_95a,Hoffmann_97,Hejny_99} and is expected from the excitation of the $N(1535) 1/2^{-}$ state.}
\label{fig:TXSWn}       
\end{figure}

\begin{figure*}[!t]
\centerline{
\resizebox{0.85\textwidth}{!}{\includegraphics{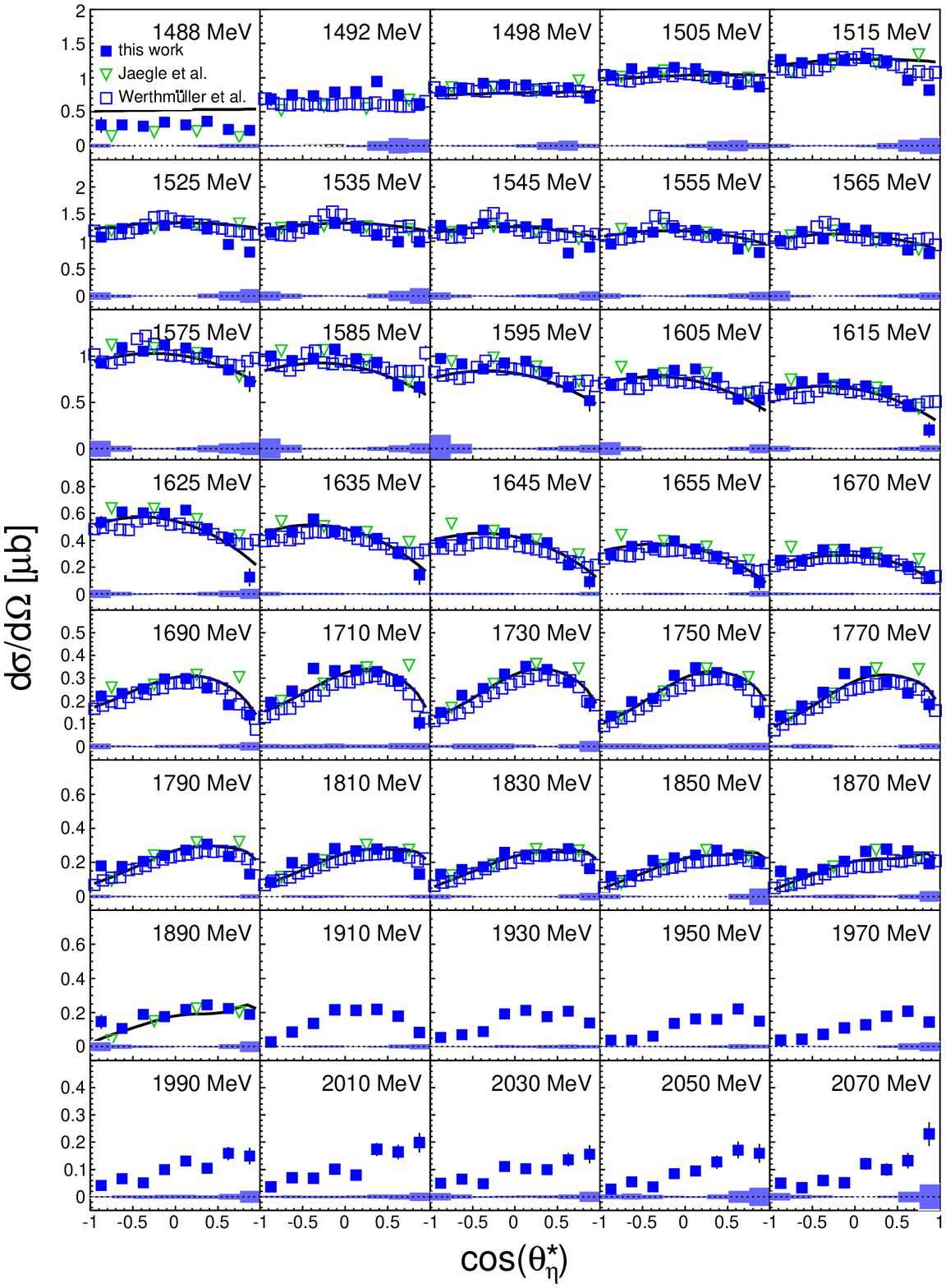}}}
\caption{Angular distributions in the $\eta$-nucleon cm system for the reaction $\gamma p\rightarrow p\eta$ for 
different bins of final state energy $W$ (mean energy indicated in the figure). Blue squares: current data, 
open squares: results by Werthm\"uller \textit{et al. }\cite{Werthmueller_14}, green triangles: Jaegle \textit{et al.} 
\cite{Jaegle_11}, black line: recent model analysis by BnGa \cite{Anisovich_15}. The systematic uncertainties 
are given in bottom of each histogram, except 10\% total normalization uncertainty.}
\label{fig:DXSWP}       
\end{figure*}

\begin{figure*}[!t]
\centerline{
\resizebox{0.85\textwidth}{!}{\includegraphics{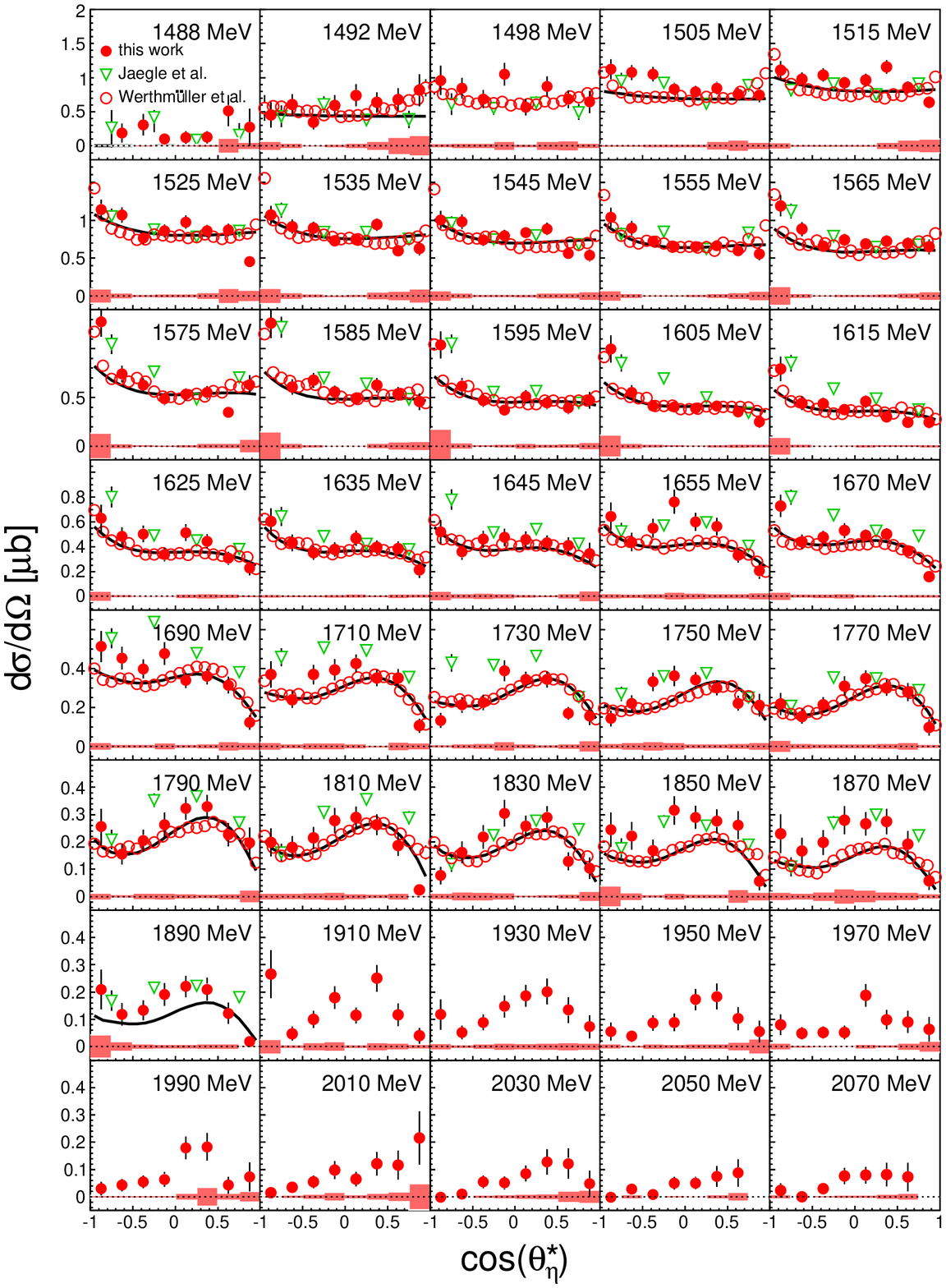}}}
\caption{Angular distributions in the $\eta$-nucleon cm system for the reaction $\gamma n\rightarrow n\eta$ 
for different bins of final state energy $W$ (mean energy indicated in the figure). Red dots: current data, 
red circles: results by Werthm\"uller et al. \cite{Werthmueller_14}, green triangles: Jaegle \textit{et al.} 
\cite{Jaegle_11}, black line: recent model analysis by BnGa \cite{Anisovich_15}. The systematic uncertainties 
are given in bottom of each histogram, except 10\% total normalization uncertainty.}
\label{fig:DXSWN}       
\end{figure*}

The present results for the exclusive cross sections $\sigma_p$ and $\sigma_n$ as a function of $E_{\gamma}$
are compared in Fig.~\ref{fig:TXS} to previous results from CBELSA/TAPS \cite{Jaegle_11} and Crystal Ball 
at MAMI \cite{Werthmueller_14}. The agreement for the proton data is excellent and for the neutron data it is
good. For the latter, in the region around the narrow bump at $E_{\gamma}\approx$1~GeV, the present data are
somewhat in between the two previous measurements. Note that for the present data, the results from the
exclusive analysis with detection of the coincident neutron are shown. The results from the analysis
based on rejection of events with proton candidates (see Fig.~\ref{fig:Incl}) have better statistical quality
(in particular above 1.4~GeV those results are certainly superior), however, for consistency we compare the 
data sets which have been analyzed in the same way.

The angular distributions in the beam-target cm system are summarized in Fig.\ \ref{fig:XS}. The angular 
dependence is in agreement with previous measurements \cite{Werthmueller_14}. Close to threshold, the 
distributions are almost flat and the slightly asymmetric shape is a Fermi motion effect. Around 
incident photon energies of 800 MeV, in the excitation maximum of the $N(1535) 1/2^{-}$, angular 
distributions for the proton target show a shallow maximum and those for the neutron target show a shallow 
minimum around polar angles of 90$^{\circ}$. This is a well understood effect from the interference 
between the contributions from the dominant $N(1535) 1/2^{-}$ resonance and the tiny contributions 
from the $N(1520) 3/2^{-}$ state \cite{Krusche_95,Weiss_03,Jaegle_11,Werthmueller_14}. 

The total cross section as a function of the final-state invariant mass are shown in Fig.\ \ref{fig:TXSW} 
for the proton (left) and the neutron (right). The systematic uncertainty (except 10\% total normalization 
uncertainty) is indicated by the shaded histograms. The present data agree quite well with the results from 
Werthm{\"u}ller {\it et al.} \cite{Werthmueller_14}. However, in contrast to the data as a function of incident 
photon energy without kinematic reconstruction of the final state (see Fig.~\ref{fig:TXS}), they disagree 
significantly with the CBELSA/TAPS results of Jaegle {\it et al.} \cite{Jaegle_11}. Already in 
\cite{Krusche_15} it was shown (by a comparison of the integrals) that the kinematically reconstructed results 
from \cite{Jaegle_11} are inconsistent with the Fermi-motion uncorrected data from the same work, 
which most likely is caused by an energy dependent normalization error (the effect is similar for proton 
and neutron data). 

The present results confirm this 
finding and the $W$ dependent CBELSA/TAPS data from \cite{Jaegle_11} are superseded by the present results.   
Although statistically not very significant, the structure in the neutron excitation function around
$W\approx 1.67$~GeV (and also the dip-like structure in the proton excitation function at the same $W$),
appear a bit sharper than in the previous MAMI data \cite{Werthmueller_14}. This is consistent
with the better $W$ resolution of the CBELSA/TAPS experiment due to the better angular resolution
of the CBB compared to the Crystal Ball. Agreement with the results from the BnGa model \cite{Anisovich_15}
is also good because they are based on fits to the previous MAMI data \cite{Werthmueller_14}.

The ratios of the neutron and proton cross sections from the present and previous 
\cite{Jaegle_11,Werthmueller_14} data are compared in Fig.\ \ref{fig:TXSWn}. They are in reasonable agreement
for all measurements and emphasize the rapid change above $W=1.6$~GeV. At smaller $W$, the ratio is dominated
by the excitation of the $N(1535) 1/2^{-}$ state for which a $\sigma_n/\sigma_p \simeq 2/3$ scaling 
\cite{Krusche_95a,Hoffmann_97,Hejny_99} was established by earlier experiments.

\begin{figure}[t]
\centerline{
\resizebox{\columnwidth}{!}{\includegraphics{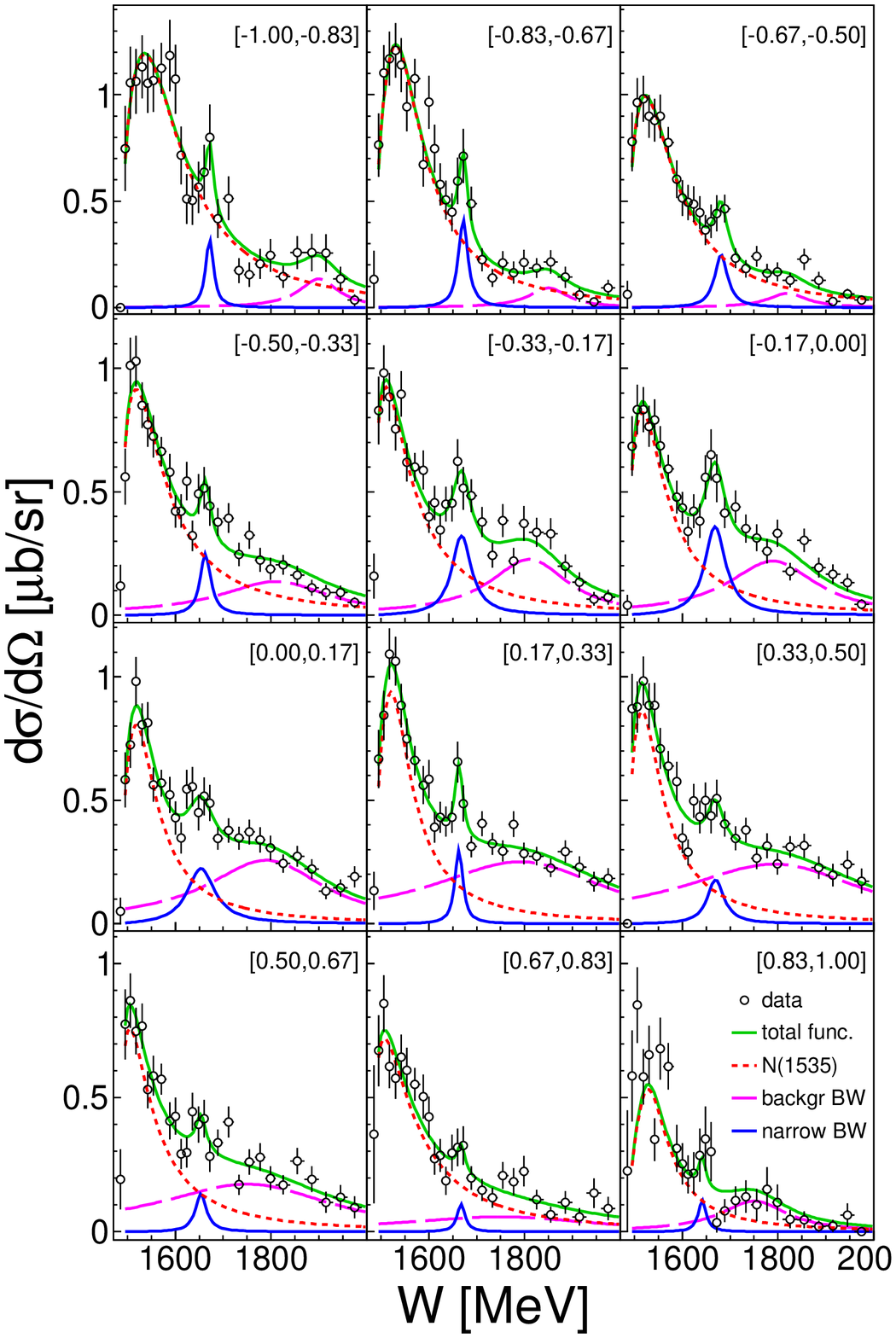}}
}
\caption{Differential cross sections in the eta-nucleon cm system for the reaction $\gamma n\rightarrow n\eta$ 
for 12 angular bins ($\cos{(\theta_{\eta}^{\ast})}$ range indicated in the figure). The total fit function 
(green line) is the sum of a energy dependent BW (red dashed line), a broad background BW (magenta long dashed), 
and a narrow BW (blue) function.}
\label{fig:AngFits}       
\end{figure}

The angular distributions for different bins of final state energy $W$ are shown and compared to previous 
results from \cite{Jaegle_11,Werthmueller_14} in Fig.\ \ref{fig:DXSWP} for the proton and in 
Fig.\ \ref{fig:DXSWN} for the neutron. Agreement between the present data and \cite{Werthmueller_14}
is on overall quite good. 

The previous data from \cite{Jaegle_11} were more coarsely binned (only four data
points per angular distribution) and, as discussed above, are affected by a normalization problem,
but even those data show a similar angular behavior. The present data extend the $W$ range for the angular
distributions by 160~MeV. Over most of the energy range, angular distributions for proton and neutron participant
nucleons are different and this is also reflected in the BnGa model results (which have been fit to the
data from \cite{Werthmueller_14}). 

\begin{figure}[t]
\centerline{
\resizebox{0.99\columnwidth}{!}{\includegraphics{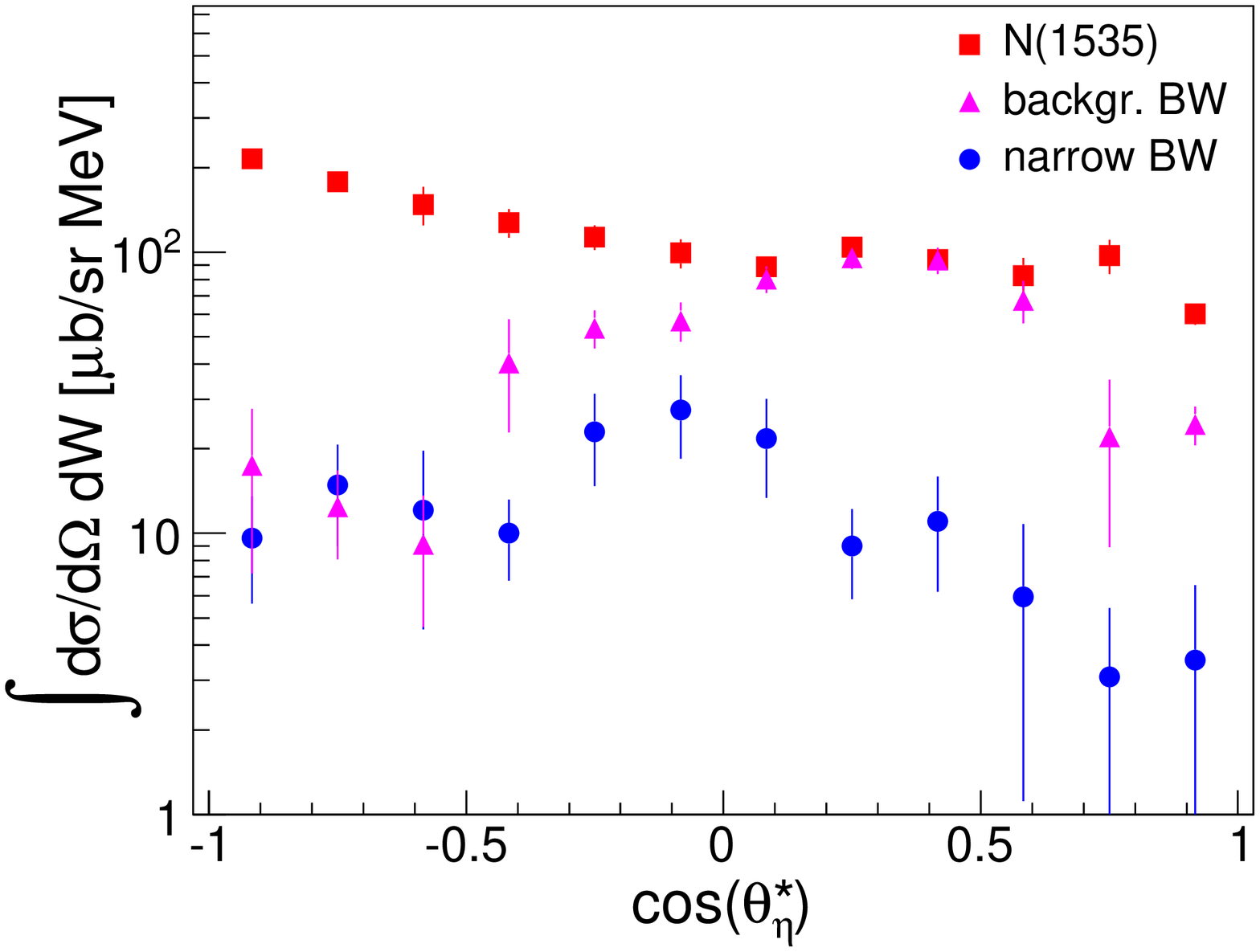}}
}
\caption{Integral of the angular fits to the differential cross sections from Fig.\ \ref{fig:AngFits}. 
Red squares: integral of the $N(1535)$ BW, magenta triangles: integral of the background BW, and 
blue dots: integral of the narrow BW.}
\label{fig:AngInt}       
\end{figure}

\begin{figure}[t]
\centerline{
\resizebox{\columnwidth}{!}{\includegraphics{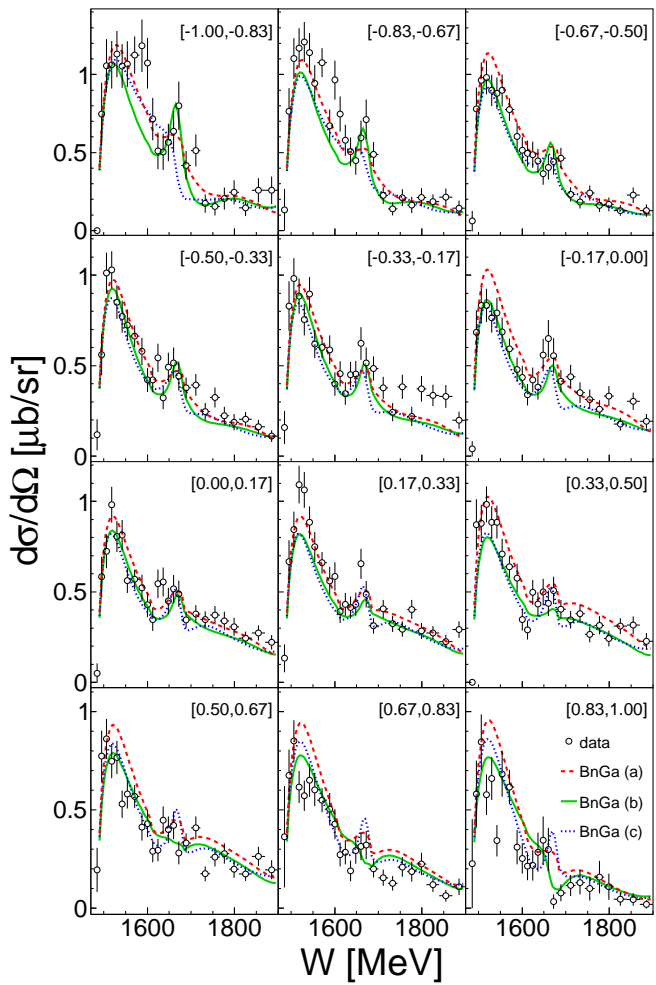}}
}
\caption{Differential cross sections in the eta-nucleon cm system for the reaction $\gamma n\rightarrow n\eta$ for 
12 angular bins ($\cos{(\theta_{\eta}^{\ast})}$ range indicated in the figure). The present data (black circles) are 
compared to model calculations from the BnGa group: BnGa (a) (model version with interference in $S_{11}$ 
wave, dashed red line) \cite{Anisovich_15}, BnGa (b) (model with narrow $P_{11}$ resonance with positive $A_{1/2}$ 
coupling, solid green line) \cite{Anisovich_15}, and BnGa (c) (narrow $P_{11}$ resonance with negative $A_{1/2}$ 
coupling, dotted blue line) \cite{Anisovich_15}. }
\label{fig:AngFitsBnGa}       
\end{figure}

A fit with the same phenomenological ansatz as in \cite{Werthmueller_14}, i.e. with three Breit-Wigner functions: one
for the $S_{11}$ wave, one for the narrow structure, and a third subsuming non-resonant backgrounds and
other partial waves, is used for a comparison to previous results. The fit yields a position of $1667 \pm 2$ MeV, 
a width of $35\pm3$ MeV, and an electromagnetic coupling of $\sqrt{b_{\eta}}A_{1/2}^n=13.4\pm 2$ $10^{-3}$GeV$^{1/2}$ 
for the narrow structure. Accounting for the experimental resolution, a width of only $23\pm2$ MeV is found. 
Table \ref{tab:Narrow} summarizes all results of the fit and compares them to previous results.
Note that the results for the $S_{11}$ wave are not corrected for experimental resolution so that in
particular the width $\Gamma$ deviates between the present results and \cite{Werthmueller_13,Werthmueller_14}. 

\begin{table}[t]
\centering
\resizebox{\columnwidth}{!}{
\begin{tabular}{|cc|ccc|}
\hline
& & $W_R$   & $\Gamma$  & $\sqrt{b_{\eta}}A_{1/2}^n$\\
& & [MeV] & [MeV] & [$10^{-3}$GeV$^{1/2}$] \\ \hline
\multirow{3}{*}{\parbox{1.3cm}{\centering narrow structure}} & this work & $1667\pm 3$ & $35\pm 3$ 
($23\pm 2$) & $ 13.4\pm 2$\\
& \cite{Werthmueller_13} & $1670\pm 1$& $50 \pm 2$ ($29\pm 3$)& $12.3 \pm 0.8$ \\
& \cite{Jaegle_11} & $1663\pm 3$& $25 \pm 12$ & - \\
 \hline 
\multirow{3}{*}{$N(1535)$} & this work & $1525\pm 2$ & $146\pm 13$ & $82\pm 4$\\
& \cite{Werthmueller_13} & $1529 \pm 1$ & $188 \pm 12$ & $90\pm 3$ \\
& \cite{Jaegle_11} & $1535 \pm 4$ & $166 \pm 23$ & $88\pm 6$ \\
 \hline
\end{tabular}
}

\caption{Fit parameters obtained from the fit of the total cross section. The position $W_R$, the width 
$\Gamma_R$, and the electromagnetic coupling $\sqrt{b_{\eta}}A_{1/2}^n$ are given for the narrow structure 
and the $N(1535)$ resonance. The width in the parentheses was extracted from a fit, which has been 
convoluted with experimental resolution. The indicated errors are statistical.}
\label{tab:Narrow}
\end{table}
    
In order to investigate the angular dependence of the narrow structure in more detail, differential cross 
sections were extracted for 12 angular bins. The distributions are shown in Fig.\ \ref{fig:AngFits}. 
The statistical quality is not as good as for the data from \cite{Werthmueller_14} but the energy 
resolution is better. Like the total cross section, they were fit as in \cite{Werthmueller_14} with 
a sum off three Breit-Wigner (BW) functions. A BW function 
with energy dependent width was used for the $N(1535)$ resonance, a standard BW for the narrow 
structure, and a much broader BW function was used to describe phenomenologically the remaining contributions 
from other resonances and non-resonant backgrounds. Due to the simple ansatz, the BW for the $N(1535)$ 
resonance accounts for the total $S_{11}$ partial wave in this range including contributions from the 
$N(1650)$ state, some non-resonant backgrounds, the $S_{11}$-$D_{13}$ interference, and also possible
effects from the interference of the $N(1535)$ with the $N(1440)1/2^+$ (Roper resonance). The corresponding 
integrals of the BW functions are plotted in Fig.\ \ref{fig:AngInt} as a function of $\cos{(\theta_{\eta}^{\ast})}$. 
The results are quite similar to \cite{Werthmueller_14}, but better cover the extreme forward 
and backward angles. As already discussed in \cite{Werthmueller_14}, the angular dependence of the narrow
structure is non-trivial. There is a general trend rising from forward to backward angles. This is what
would be expected for a $S_{11}$ - $P_{11}$ interference (the interference term is proportional to
$\cos{(\theta_{\eta}^{\ast})}$). In comparison to \cite{Werthmueller_14}, this trend is better visible 
in the present data due to the better angular coverage. However, like in \cite{Werthmueller_14}, there is also a 
maximum around polar angles of 90$^{\circ}$ degrees, which must have a different origin. This will need further 
investigation with reaction models such as the BnGa analysis.

As a first step, we compare in Fig.~\ref{fig:AngFitsBnGa} the present data to the calculations of the BnGa group 
\cite{Anisovich_15}, which were obtained by a fit of the data from \cite{Werthmueller_14}. Three different fits 
were discussed in \cite{Anisovich_15}. Fit (a) used only nucleon resonances included in the standard version 
of the BnGa model and tuned the interference pattern in the $S_{11}$ wave (contributions from $N(1535)$, $N(1650)$, 
and non-resonant backgrounds) to achieve best agreement with the experimental data. It turned out, that 
reasonable agreement requires a change in sign (with respect to the value quoted by RPP \cite{PDG_14,PDG_16}) 
of the electromagnetic $A_{1/2}$ coupling of the $N(1650)$ for the neutron. In the other two solutions 
(b), (c), an additional narrow $P_{11}$ resonance was introduced by hand in version (b) so that the 
interference term between this resonance and the $S_{11}$ partial wave had a positive sign and for (c) 
so that it had a negative sign. The conclusion was that the version (a) without an additional narrow 
$P_{11}$ gave the best overall $\chi^2$ and an upper limit was quoted for a contribution of a narrow 
$P_{11}$ state. However, the present results with better resolution show a clear forward - backward 
asymmetry of the narrow structure. It is more pronounced at backward angles. This is in better agreement 
with BnGa solution (b) with the positive sign of an $S_{11}$ - $P_{11}$ interference term than with 
solution (a) without any narrow state and it clearly contradicts solution (c) with the negative sign 
of the interference. Also the overall $\chi^2$ is slightly better for the BnGa solution (b) than (a) and (c) 
($2.43$ versus $2.85$ and $2.87$, respectively). Calculating the $\chi^2$ in the range of the narrow structure, 
i.e. between 1630 and 1690 MeV, yields values of (a) $2.70$, (b) $2.09$, and (c) $4.93$. Hence, the current 
results favor a $S_{11}$ - $P_{11}$ interference with positive sign.

A more detailed phenomenological analysis can be done by fitting the Legendre series Eq.~\ref{eq:Leg} to the
angular distributions. It is well known that also the angular distributions of the $\gamma p\rightarrow p\eta$ 
reaction change rapidly shape in this energy region. The ratio of the Legendre coefficients $A_1/A_0$ has an 
extremely steep slope around $W$=1.7~GeV and crosses zero shortly below this value \cite{Krusche_15}. 
This behavior could also be demonstrated over a significant range of $Q^2$ by electroproduction \cite{Denizli_07}
of $\eta$ mesons off protons. The most simple explanation is an $S_{11}$ - $P_{11}$ interference when one of the 
two partial waves passes rapidly through resonance as discussed in \cite{Denizli_07}. However, even for 
the proton this is certainly not the full picture, because also higher order terms show a strong energy dependence 
in this region. As summarized in \cite{Krusche_15} also $A_3/A_0$ changes rapidly sign at the same energy 
(with the inverse slope compared to $A_1/A_0$) while $A_2/A_0$ and $A_4/A_0$ show pronounced minima.

\begin{figure}[t]
\centerline{
\resizebox{0.99\columnwidth}{!}{\includegraphics{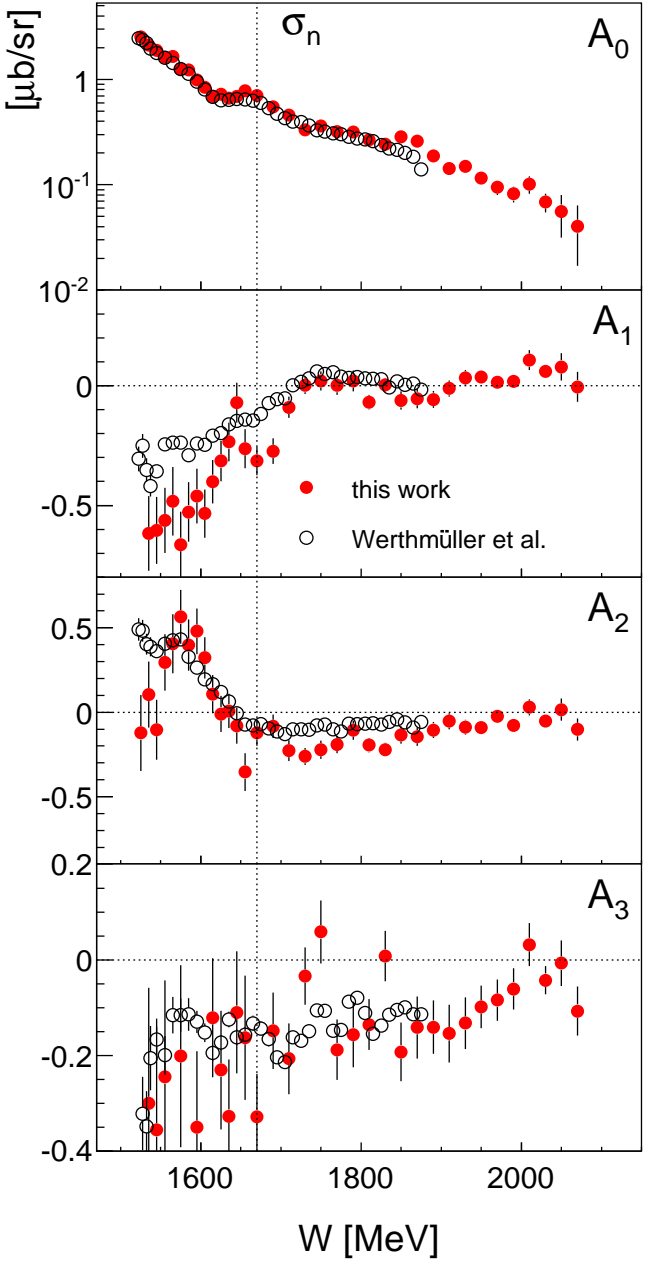}}
}
\caption{Legendre coefficients of the angular distributions for the reaction  $\gamma n\rightarrow n\eta$. 
The experimental results from this work (red dots) are compared to previous results from 
\cite{Werthmueller_14} (black circles). The dashed line indicates the position of the narrow structure at $W=1670$ MeV}
\label{fig:Leg}       
\end{figure}

The Legendre coefficients for $\gamma n\rightarrow n\eta$ up to $A_3$ from the present work are shown in 
Fig.~\ref{fig:Leg} and are compared to the published results from \cite{Werthmueller_14}. The narrow structure 
at $W=1670$ MeV (position indicated by the dashed line) is clearly visible in the $A_0$ coefficient and consistent 
with the previous results. However, the structure in the current results is slightly more pronounced, which must 
origin from the better $W$ resolution in the CBELSA/TAPS experiment. Another interesting finding is the dip 
structure in $A_1$ around the same energy. This is clearly more pronounced than in the previous unpolarized data
\cite{Werthmueller_14}, but a similar behavior was seen in the recently published results on the 
helicity dependent cross section $\sigma_{1/2}$ \cite{Witthauer_16}. The model calculation BnGa (b) with 
the narrow $P_{11}$ resonance with positive $A_{1/2}$ coupling \cite{Anisovich_15} agrees also with such a behavior. 
This points to the presence of some $S_{11}$ - $P_{11}$ interference, but like for the proton this is probably 
not the full explanation of the observed structures.  

\subsection{Double Polarization Observable $\bm{E}$ and Helicity Dependent Cross Sections}
\label{sec:DE}

\begin{figure}[h]
\centerline{
\resizebox{0.5\columnwidth}{!}{\includegraphics{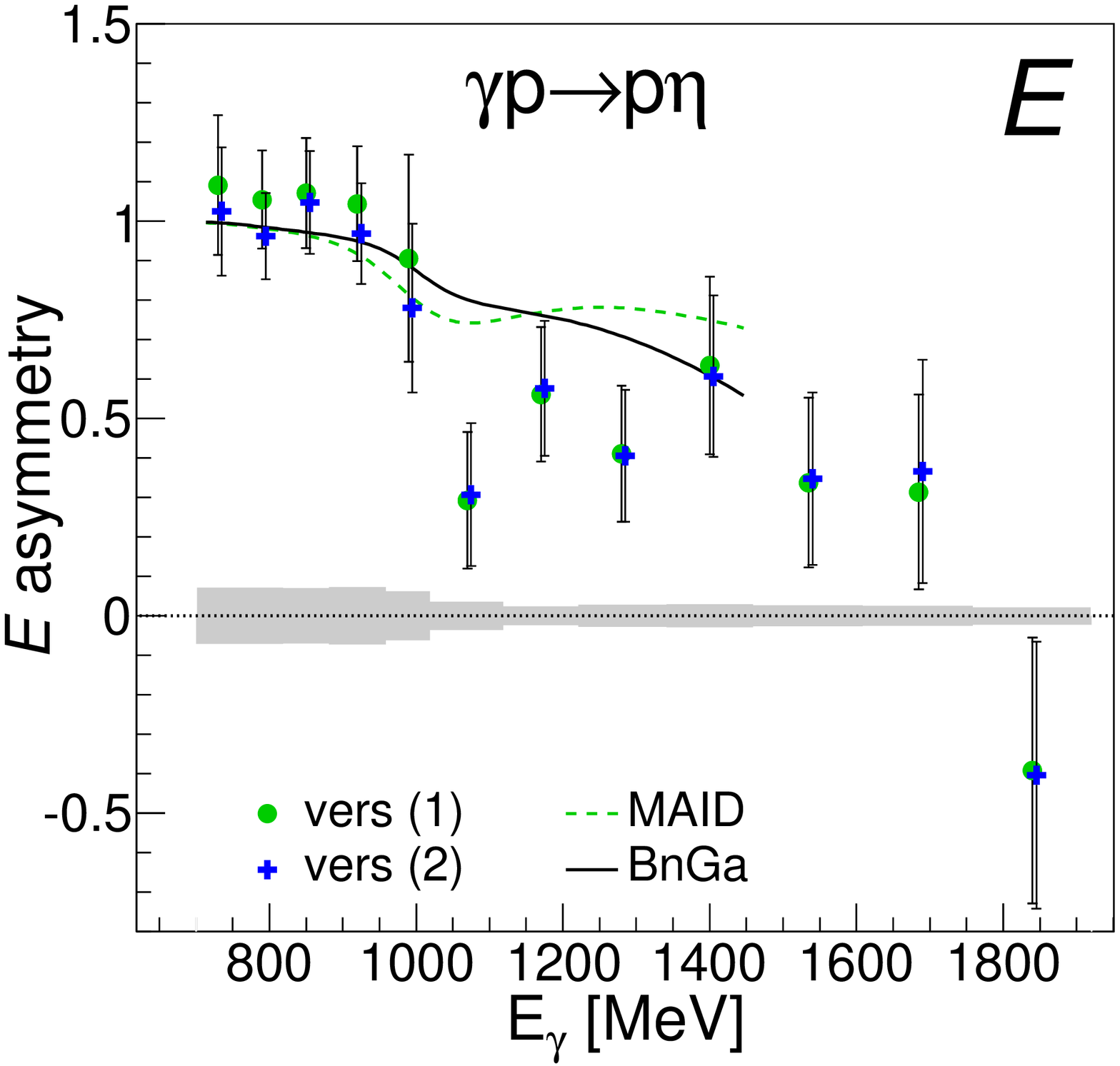}}
\resizebox{0.5\columnwidth}{!}{\includegraphics{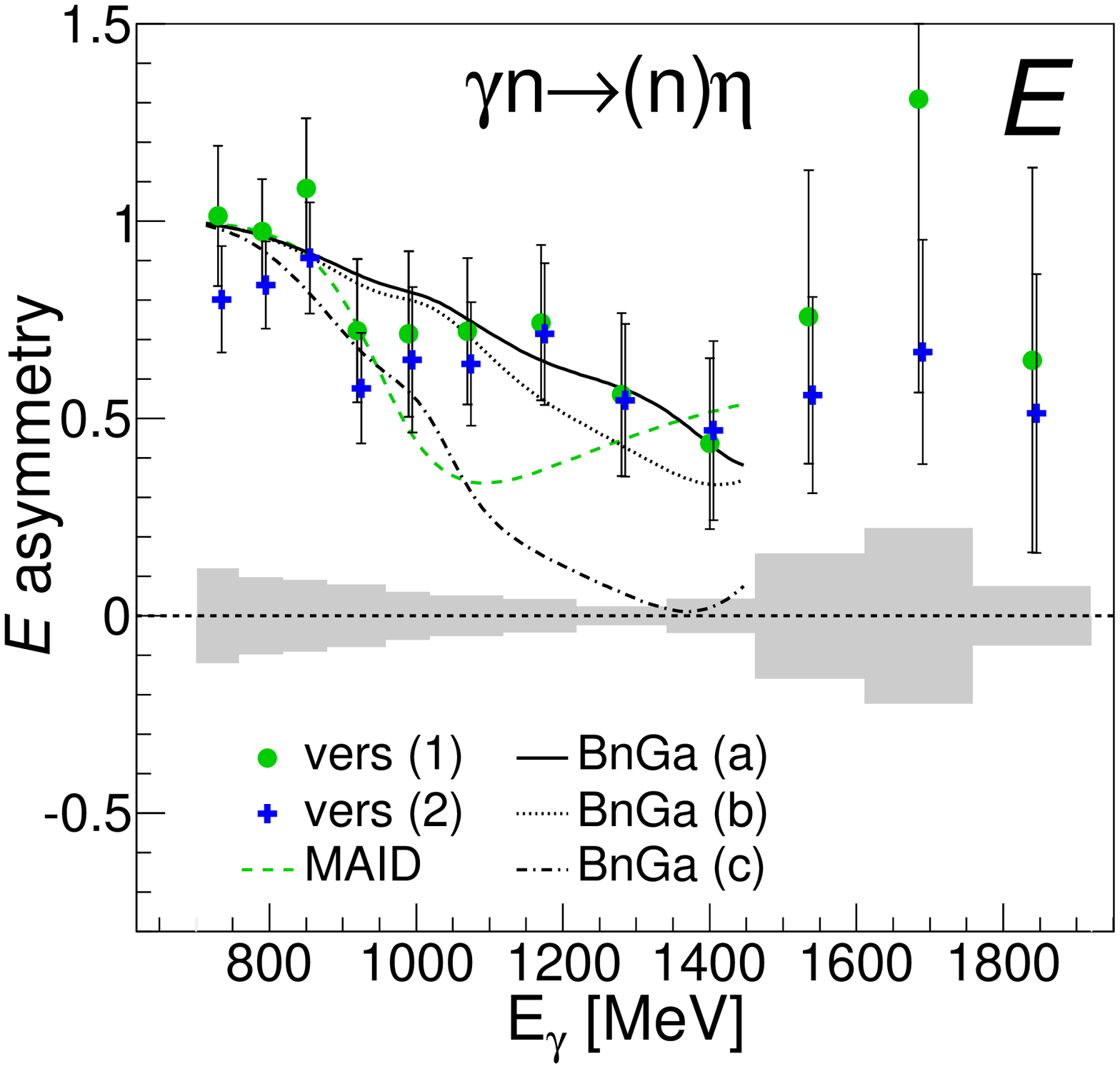}}
}
\caption{Double polarization observable $E$ for $\gamma p\rightarrow p\eta$ (left hand side) and 
$\gamma n\rightarrow (n)\eta$ (right hand side). Green dots: analysis version (1) (carbon subtraction), 
blue crosses: analysis version (2) (normalization to data from
unpolarized deuterium target). Gray shaded areas: systematic uncertainties. Curves: Fermi folded model 
predictions from MAID (green dashed) \cite{Chiang_02} and BnGa \cite{Anisovich_15}. For the neutron, three 
different BnGa calculations are given: BnGa (a) model version with interference in $S_{11}$ 
wave (solid) \cite{Anisovich_15}, BnGa (b) model with narrow $P_{11}$ resonance with positive $A_{1/2}$ 
coupling (dotted) \cite{Anisovich_15}, and BnGa (c) model version with narrow $P_{11}$ resonance with 
negative $A_{1/2}$ coupling (dashed-dotted) \cite{Anisovich_15}. For better visibility, the data points 
from version (2) were shifted by $+5$ MeV with 
respect to version (1).}
\label{fig:PolAll}       
\end{figure}

The double polarization observable $E$ for the reactions $\gamma p\rightarrow p\eta$ and 
$\gamma n\rightarrow (n)\eta$ is shown in Fig.\ \ref{fig:PolAll} as a function of the incident photon energy.
Apart from a small systematic deviation (version (1) is slightly higher than version (2)), the two different 
analysis versions are in good agreement. The experimental results are compared to model predictions from the 
BnGa analysis \cite{Anisovich_15} and the MAID model \cite{Chiang_02}. 
All models were
folded with the Fermi momentum distribution for nucleons bound in deuterium nuclei \cite{Lacombe_81}. The data
for the neutron were analyzed in the semi-inclusive way, i.e. all events without detection of any charged
particle were accepted as `neutron participant', while for reactions with `proton participant' detection and
identification of the recoil proton was required.  

Close to threshold, the extracted asymmetries are unity within statistical uncertainties as it is expected 
from the dominance of the $N(1535)1/2^{-}$ resonance and predicted by all models. At higher incident photon 
energies, $E$ drops slightly, indicating contributions from partial waves with $J\geq3/2$, although contributions
from $J=1/2$ partial waves dominate up to the highest energies. The contribution of $J\geq3/2$ partial waves
above incident photon energies of 1~GeV seems to be more important for the proton than for the neutron, but
this is at the edge of statistical significance. The model predictions from BnGa \cite{Anisovich_15}
and MAID \cite{Chiang_02} are similar and in reasonable agreement with the experimental data. For the neutron
target, the MAID prediction shows a significant effect from the $N(1675)5/2^-$ resonance at photon energies
around 1~GeV, which is not supported by the measurement. The polarization observable $E$ of the BnGa (c) calculation
exhibits a stronger fall-off towards higher energies than the other two BnGa models. This behavior is in clear 
contradiction to the current data.

\begin{figure}[t]
\centerline{
\resizebox{0.5\columnwidth}{!}{\includegraphics{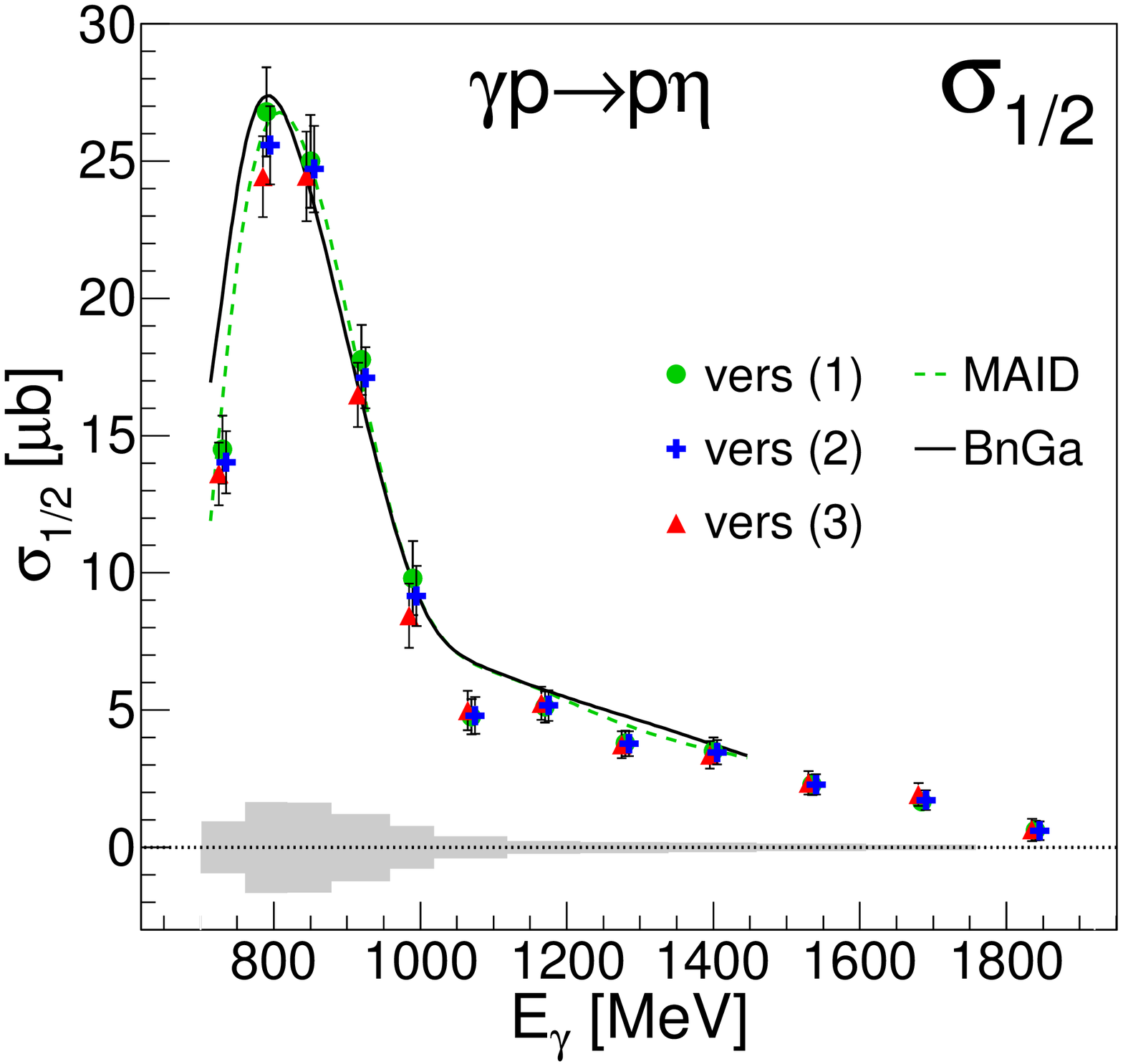}}
\resizebox{0.5\columnwidth}{!}{\includegraphics{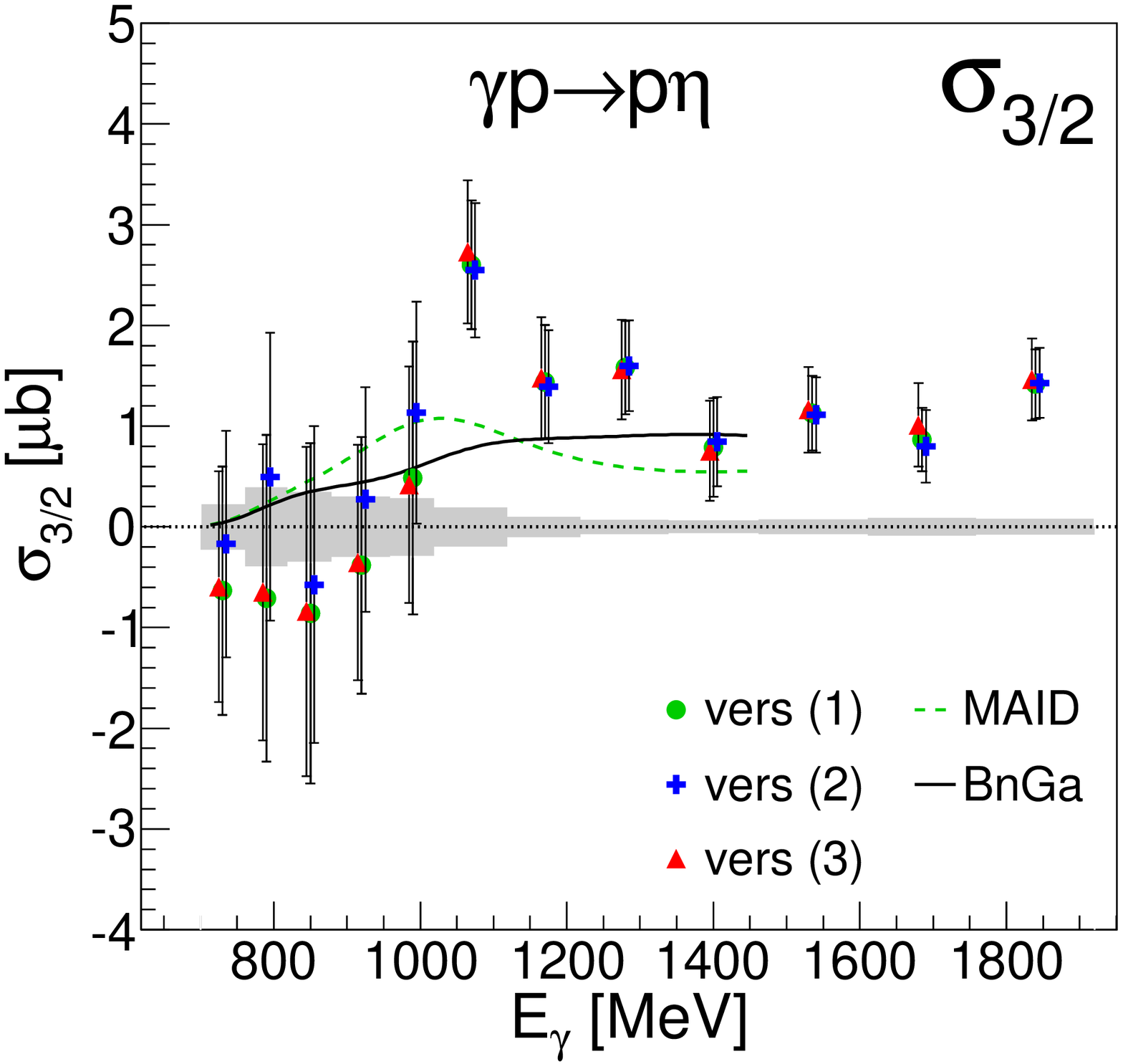}}}
\centerline{
\resizebox{0.5\columnwidth}{!}{\includegraphics{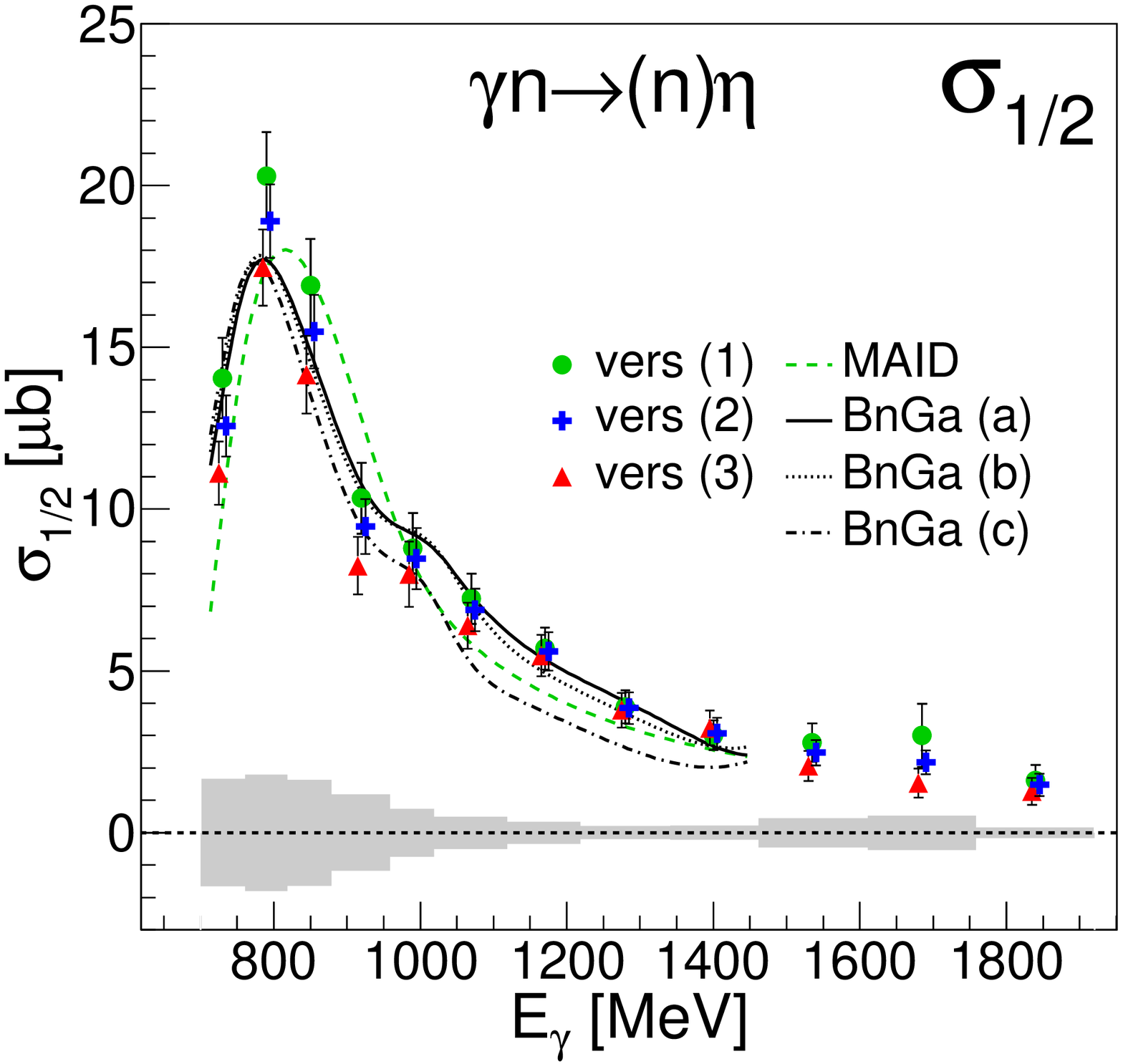}}
\resizebox{0.5\columnwidth}{!}{\includegraphics{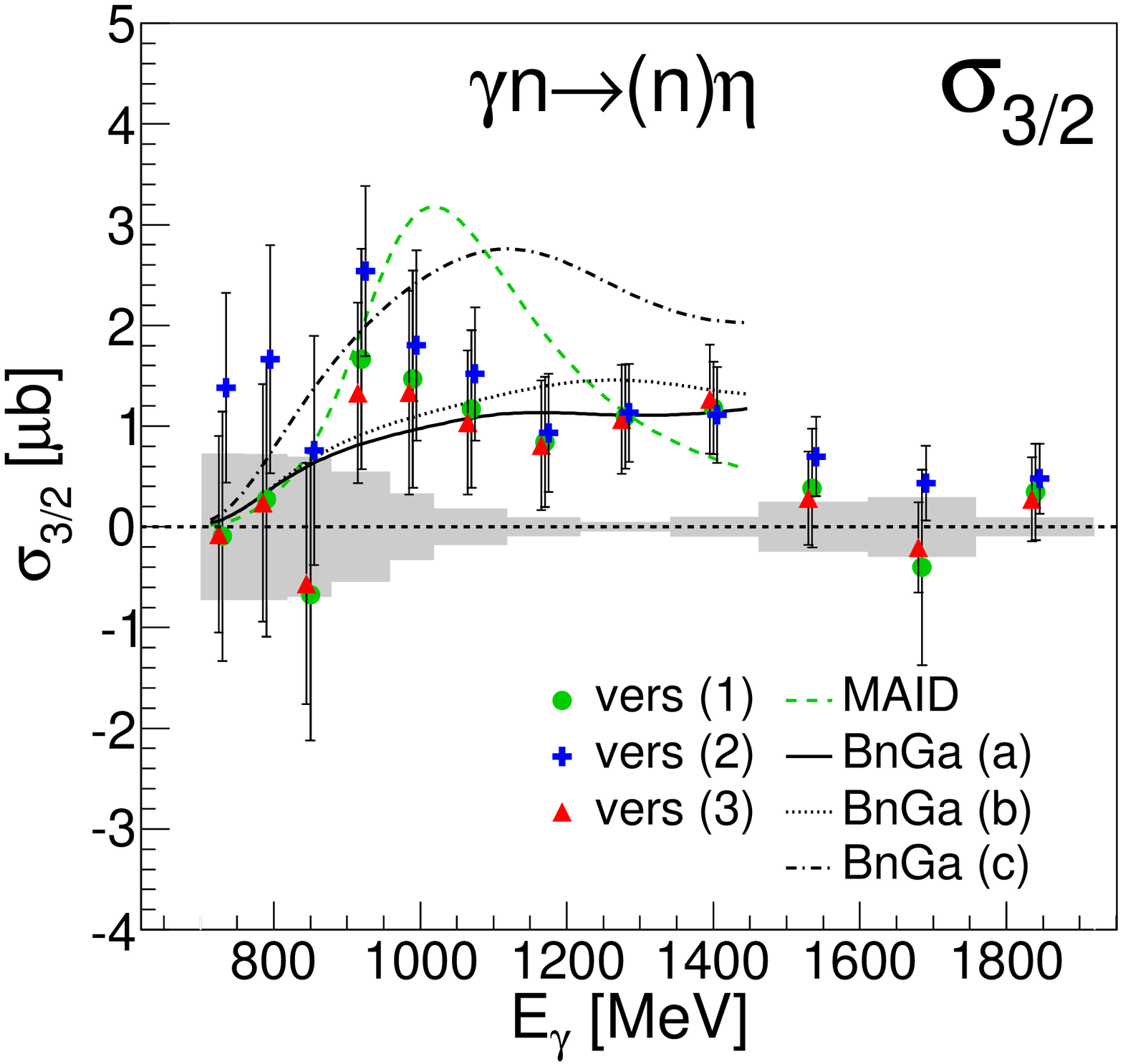}}
}
\caption{Helicity dependent cross sections $\sigma_{1/2}$ and $\sigma_{3/2}$ for 
$\gamma p\rightarrow p\eta$ and $\gamma n\rightarrow (n)\eta$ (reaction and helicity types indicated in figures)
from the three different analysis versions (see Sec.~\ref{sec:ExDPE}). Gray shaded areas: systematic uncertainties.
Curves: Fermi folded model predictions from MAID (green dashed) \cite{Chiang_02} and BnGa \cite{Anisovich_15}. 
For the neutron, three different BnGa calculations are given: BnGa (a) model version with interference in $S_{11}$
wave (solid) \cite{Anisovich_15}, BnGa (b) model with narrow $P_{11}$ resonance with positive $A_{1/2}$           
coupling (dotted) \cite{Anisovich_15}, and BnGa (c) model version with narrow $P_{11}$ resonance with negative 
$A_{1/2}$ coupling (dashed-dotted) \cite{Anisovich_15}. For better visibility, the data points from version (2) 
and version (3) were shifted by $\pm5$ MeV with respect to version (1).}
\label{fig:PolAllp}
\end{figure} 

Even more instructive is the discussion of the helicity dependent cross sections $\sigma_{1/2}$ and 
$\sigma_{3/2}$, which follow directly from the unpolarized cross section and the $E$ asymmetry.
They are summarized in Fig.~\ref{fig:PolAllp} and the results from the three different analysis methods,
which agree very well, are compared. The strong dominance of the $\sigma_{1/2}$ contribution for proton
and neutron is evident. The structure in the neutron excitation function around 1~GeV appears only
in the $\sigma_{1/2}$ part as a bump (broad due to the smearing of the Fermi motion). 
The $\sigma_{3/2}$ contribution is much smaller in this energy range (on the level
of approximately one microbarn) and does not show any peak-like structure as for example predicted by the 
MAID model due to the $N(1675)5/2^-$ contribution or the BnGa (c) model with the interference of a $P_{11}$ 
resonance with negative $A_{1/2}$ coupling. Note that all structures are much less narrow than for the above 
discussed unpolarized cross sections because the Fermi motion was not unfolded from the data. Due to the limited 
statistics, the neutron data had to be analyzed without a coincident detection of the recoil neutron and effects 
from Fermi motion could not be removed by an analysis of the final state kinematics. The results agree
with the MAMI data published in \cite{Witthauer_16} in so far as the narrow structure appears only in
$\sigma_{3/2}$. A direct comparison to the MAMI data is not yet possible because from MAMI only
kinematically reconstructed data have been published but not yet the Fermi smeared data.  

\begin{figure}[h!]
\centerline{
\resizebox{0.3\textwidth}{!}{\includegraphics{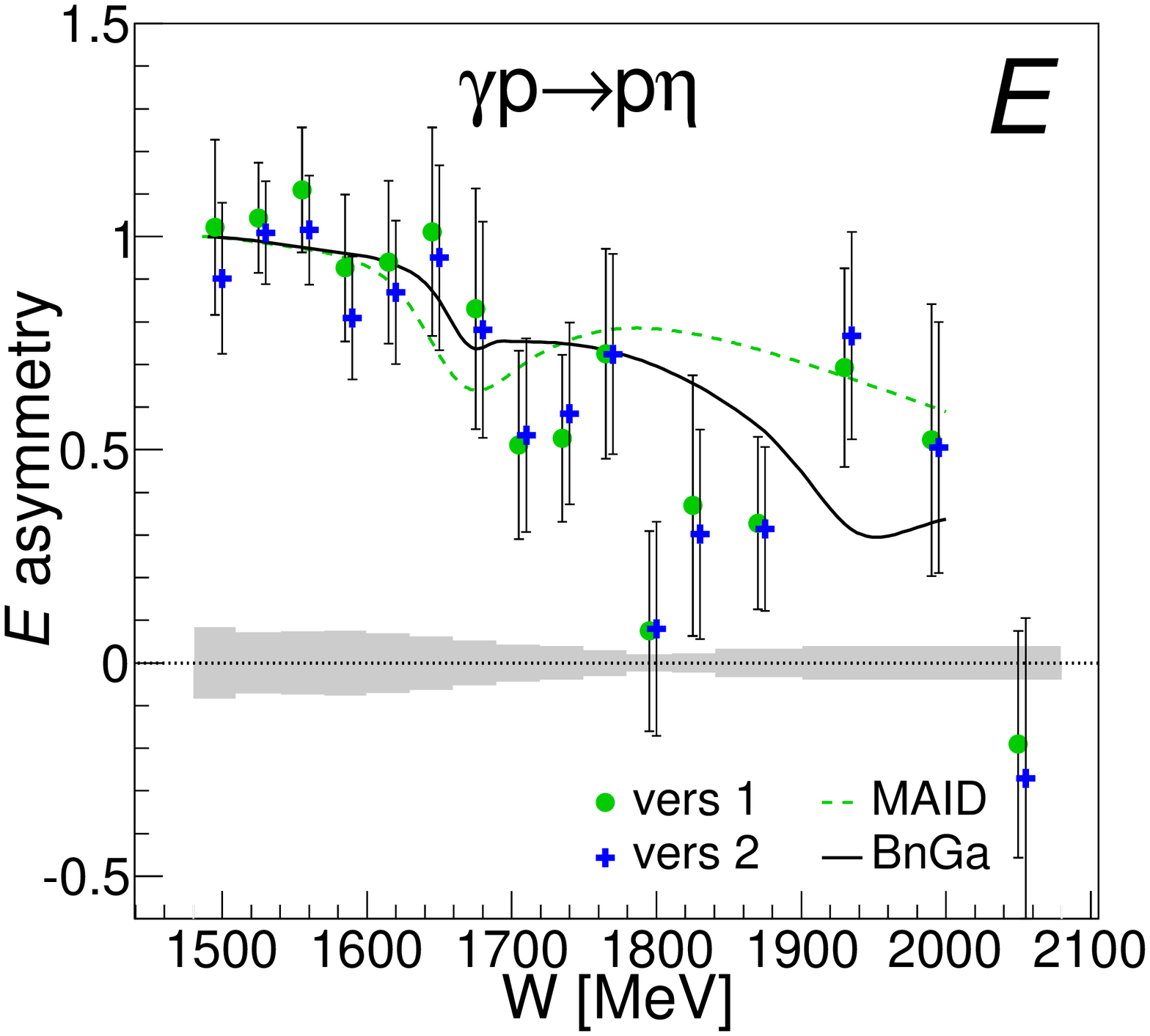}}}
\centerline{\resizebox{0.25\textwidth}{!}{\includegraphics{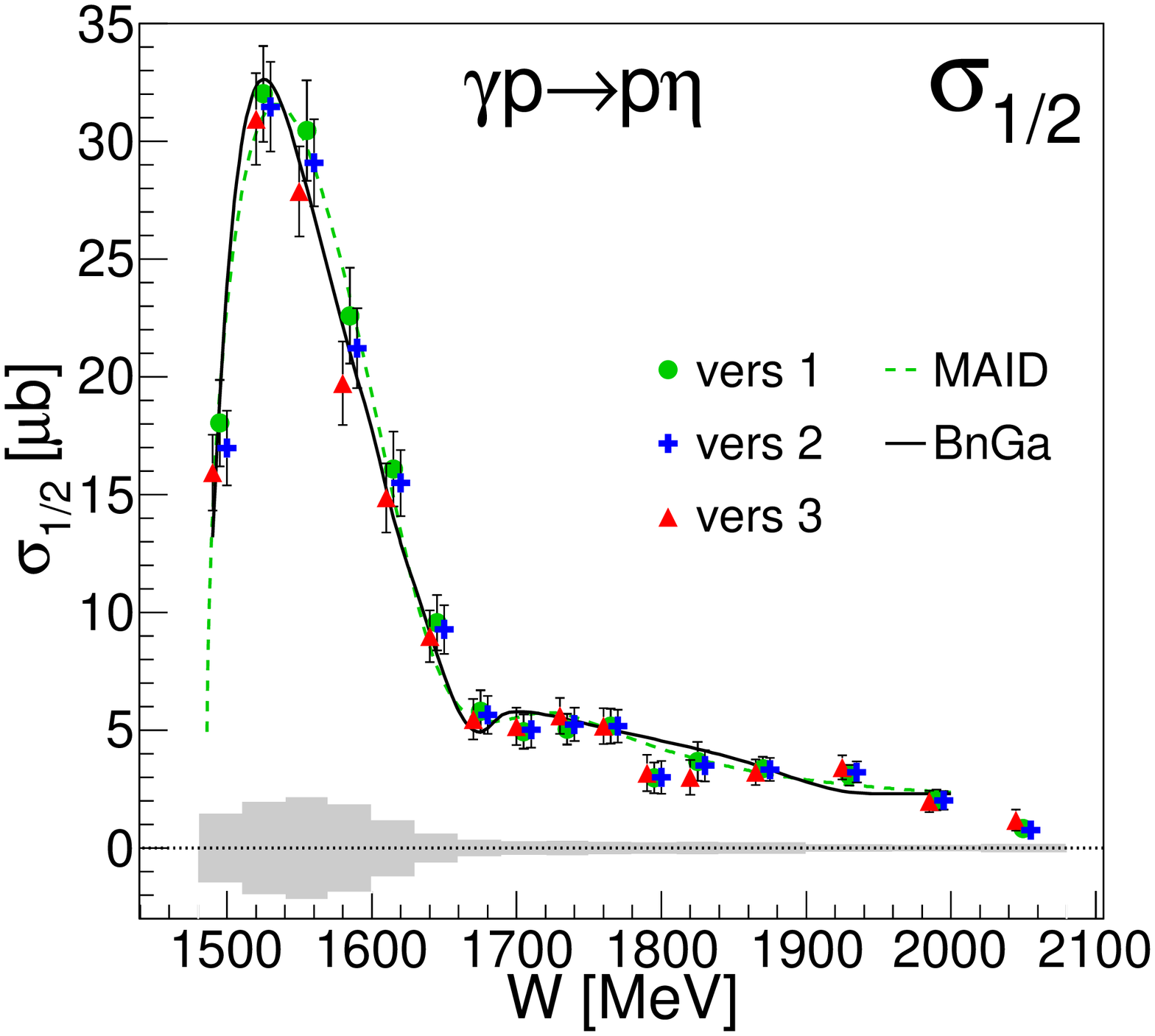}}
\resizebox{0.25\textwidth}{!}{\includegraphics{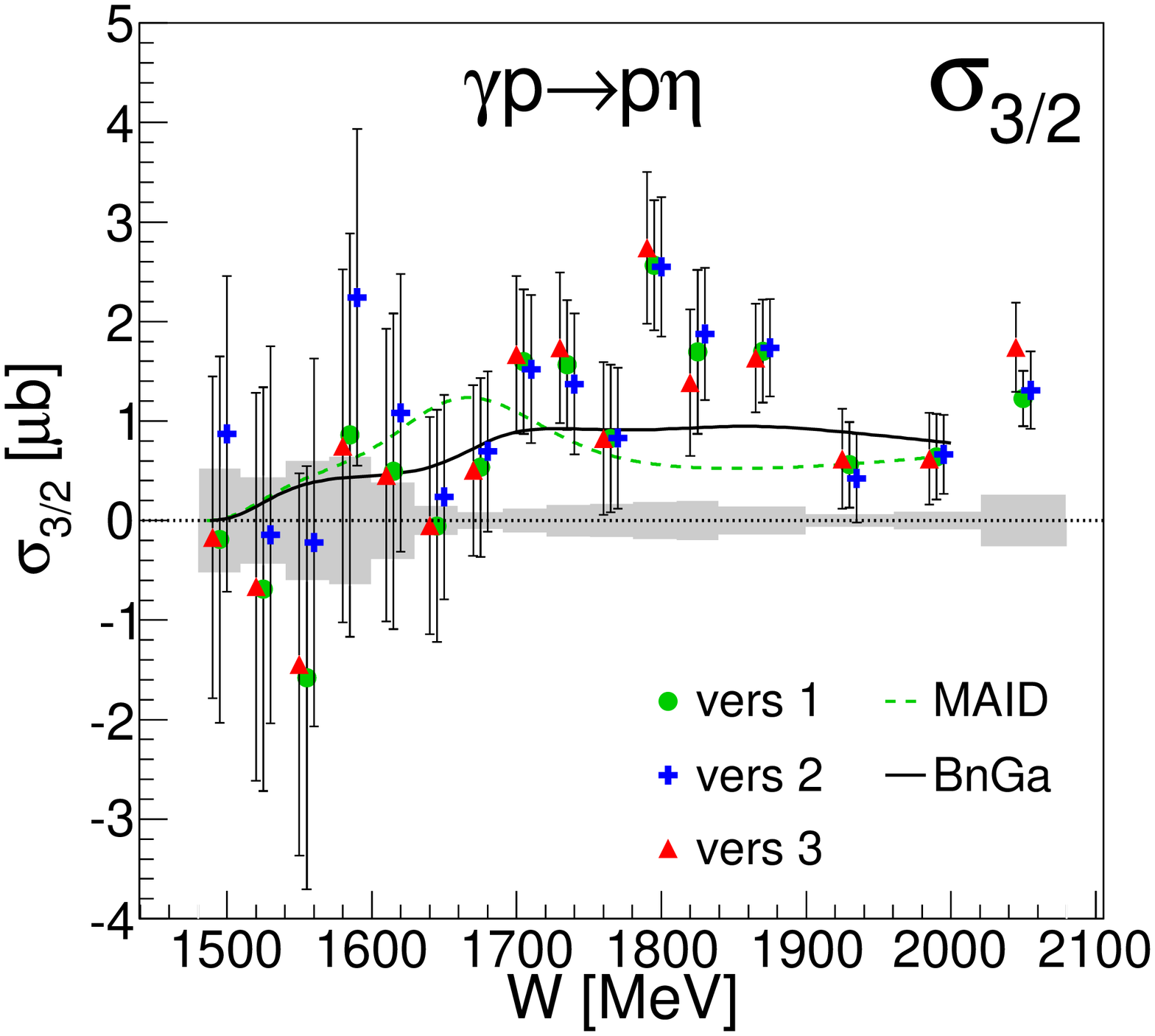}}
}
\caption{Double polarization observable $E$ and the helicity dependent cross sections  $\sigma_{1/2}$ and 
$\sigma_{3/2}$ for the reaction $\gamma p\rightarrow p\eta$ as a function of the final state invariant mass.
Curves: model predictions from MAID (green dashed) \cite{Chiang_02} and BnGa (model based on $S_{11}$ interference) 
\cite{Anisovich_15} for the free proton. For better visibility, the points from version (2) and version (3) 
were shifted by $\pm5$ MeV with respect to version (1).}
\label{fig:PolAllW}       
\end{figure}

The proton data shown in Fig.~\ref{fig:PolAllp} were analyzed in the same way for better comparison. However,
for the proton, the final state kinematics were completely determined because the recoil protons were detected.   
Therefore, this reaction was also analyzed analogously to the unpolarized cross section by kinematic
reconstruction of the final state. The corresponding results as a function of total cm energy $W$ are shown in
Fig.~\ref{fig:PolAllW}. They are compared to the original model predictions from MAID \cite{Chiang_02} and 
BnGa \cite{Anisovich_15} (without Fermi smearing). The agreement with both predictions for the $\sigma_{1/2}$
contribution is excellent and the very small $\sigma_{3/2}$ contribution is also well described in magnitude.
The data for $\sigma_{3/2}$ are within statistical errors not significantly different from zero up to 
$W\approx1.65$~GeV. At higher energies, small contributions, for example from the $N(1720)3/2^+$, are possible. 

\section{Summary and Conclusions}
\label{sec:Sum}
In this paper, total cross sections and angular distributions were presented for $\eta$ photoproduction 
from quasifree protons and neutrons. Total cross sections measured as a function of incident photon energy 
without correction for Fermi smearing agree with previous data \cite{Jaegle_11,Werthmueller_14}.
The data after correction for Fermi motion agree much better with \cite{Werthmueller_14} than with
\cite{Jaegle_11}. The position and width of the narrow structure seen in previous measurements of the 
$\gamma n\rightarrow n\eta$ reaction was also confirmed. Angular distributions have shown that the structure is 
more distinct for backward angles in the cm of the $\eta$ meson and the neutron and less pronounced in the 
forward direction. This is a behavior which would agree with an interference between a $P_{11}$ and 
the strongly dominant $S_{11}$ partial wave. However, the angular dependence is more complicated with a 
pronounced maximum around 90$^{\circ}$. Nevertheless, as a calculation of the $\chi^2$ values showed,
the data seem to be in better agreement with the BnGa model solution \cite{Anisovich_15} including an 
additional narrow $P_{11}$ state than without. This question needs further investigation. 

More stringent constraints on the model analyses can be obtained from the measurement of observables
exploiting the polarization degrees of freedom of the nucleons and the incident photons.
The present work includes results for the double polarization observable $E$ measured with a longitudinally
polarized target and a circularly polarized photon beam. This observable, together with the unpolarized cross
section $\sigma_0$, allows to split the reaction into contributions with antiparallel ($\sigma_{1/2}$)
and parallel ($\sigma_{3/2}$) orientation of nucleon and incident photon spin. The results show that for 
proton and neutron targets, $\eta$ photoproduction is strongly dominated by $\sigma_{1/2}$. This was expected 
for the threshold range, where excitation of the $N(1535)1/2^-$ resonance dominates, but contributions from 
states with $J\geq3/2$ also seem to be small at higher incident photon energies. The data for the neutron
show that the narrow structure around $E_{\gamma}\approx1$~GeV, observed in the unpolarized cross section,
is a feature of $\sigma_{1/2}$ and thus very likely related to the $S_{11}$ and/or $P_{11}$ partial
waves. 

\section{Acknowledgments}
We wish to acknowledge the outstanding support of the accelerator group
and operators of ELSA. This work was supported by Schweizerischer Nationalfonds (200020-156983, 132799, 121781, 117601) and
the Deutsche For\-schungs\-ge\-mein\-schaft (SFB/TR-16.)


\end{document}